\documentclass[aps,pra,groupedaddress,nofootinbib,notitlepage,showpacs,floatfix,twocolumn]{revtex4-1}

\pdfoutput=1

\usepackage{graphicx}
\usepackage{epstopdf}
\usepackage{amsmath}
\usepackage{amssymb}
\usepackage{hyperref}
\usepackage{bm}
\usepackage{threeparttable}
\usepackage{subfigure}
\usepackage{upgreek }
\usepackage{marginnote}

\usepackage[pdftex]{color}

\newcommand{\beq}{\begin{equation}}
\newcommand{\eeq}{\end{equation}}
\newcommand{\beqnn}{\begin{equation*}}
\newcommand{\eeqnn}{\end{equation*}}
\newcommand{\bea}{\begin{eqnarray}}
\newcommand{\eea}{\end{eqnarray}}
\newcommand{\beann}{\begin{eqnarray*}}
\newcommand{\eeann}{\end{eqnarray*}}
\newcommand{\bes} {\begin{subequations}}
\newcommand{\ees} {\end{subequations}}

\newcommand{\ket}[1]{ | #1\rangle}
\newcommand{\bra}[1]{\langle #1 | }
\newcommand{\ketbra}[2]{|#1\rangle\langle #2|}

\newcommand{\eps}{\varepsilon}
\newcommand{\bo}{\beta_{\textrm{opt}}}

\newcommand{\ignore}[1]{}

\begin{document}
\title{Error corrected quantum annealing with hundreds of qubits}
\author{Kristen L. Pudenz,$^{1,2,3}$ Tameem Albash,$^{2,4}$ \& Daniel A. Lidar$^{1,2,4,5}$}
\affiliation{$^{(1)}$Department of Electrical Engineering, $^{(2)}$Center for Quantum Information Science \& Technology, $^{(3)}$Information Sciences Institute, $^{(4)}$Department of Physics and Astronomy, $^{(5)}$Department of Chemistry\\
University of Southern California, Los Angeles, California 90089, USA}

\begin{abstract}

Quantum information processing offers dramatic speedups, yet is famously susceptible to decoherence, the process whereby quantum superpositions decay into mutually exclusive classical alternatives,
thus robbing quantum computers of their power. 
This has made the development of quantum error correction an essential and inescapable 
aspect of both theoretical and experimental quantum computing. So far little is known about protection against decoherence in the context of quantum annealing, a computational paradigm which aims to exploit ground state quantum dynamics to solve optimization problems more rapidly than is possible classically. Here we develop error correction for quantum annealing and provide an experimental demonstration using up to 344 superconducting flux qubits in processors which have recently been shown to physically implement programmable quantum annealing. 
We  demonstrate a substantial improvement over the performance of the processors in the absence of error correction. 
These results pave a path toward large scale noise-protected adiabatic quantum optimization devices.
\end{abstract}


\maketitle

\section{Introduction}
Combinatorial optimization problems are of great interest in both complexity theory and practical applications, and are also notoriously difficult to solve \cite{Papadimitriou:98-book}. Quantum computing harbors the promise of dramatic speedups over its classical counterpart \cite{Bacon:review}, yet it cannot function on a large scale without error correction \cite{Preskill:97a}. 
In this work we demonstrate how quantum annealing, a form of quantum computing tailored to optimization that can be more efficient than classical optimization \cite{Santoro,morita:125210,Somma:2012kx}, can be error corrected. Quantum annealing is a generalization of classical simulated annealing, an approach to optimization based on the observation that the cost function of an optimization problem can be viewed as the energy of a physical system, and that energy barriers can be crossed by thermal hopping \cite{Kirkpatrick13051983}. However, to escape local minima it can be advantageous to explore low energy configurations quantum mechanically by exploiting superpositions and tunneling. Quantum annealing \cite{finnila_quantum_1994,Kadowaki1998} and adiabatic quantum computation \cite{Farhi20042001,PhysRevLett.101.170503} are algorithms based on this idea, and a programmable quantum annealer (PQA) is its physical realization \cite{Brooke1999,2002quant.ph.11152K,Dwave}. 

\begin{figure}[t]
\begin{center}
\subfigure[]{\includegraphics[width=0.12\textwidth]{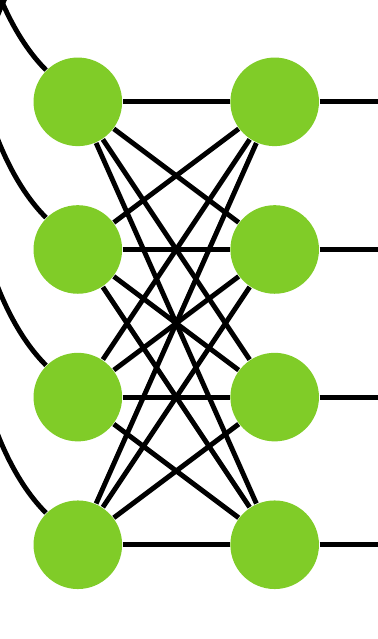}\label{fig:1a}} 
\subfigure[]{\includegraphics[width=0.12\textwidth]{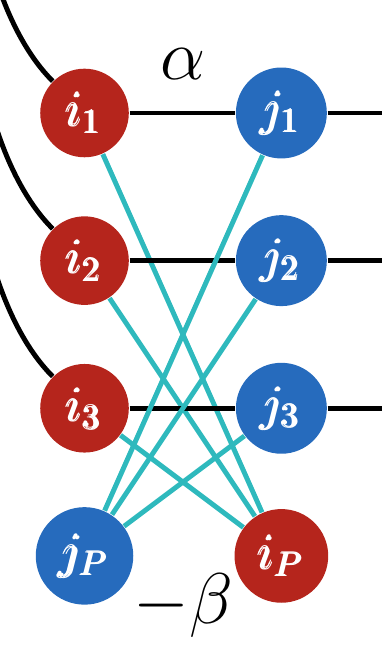}\label{fig:1b}} 
\subfigure[]{\includegraphics[width=0.48\textwidth]{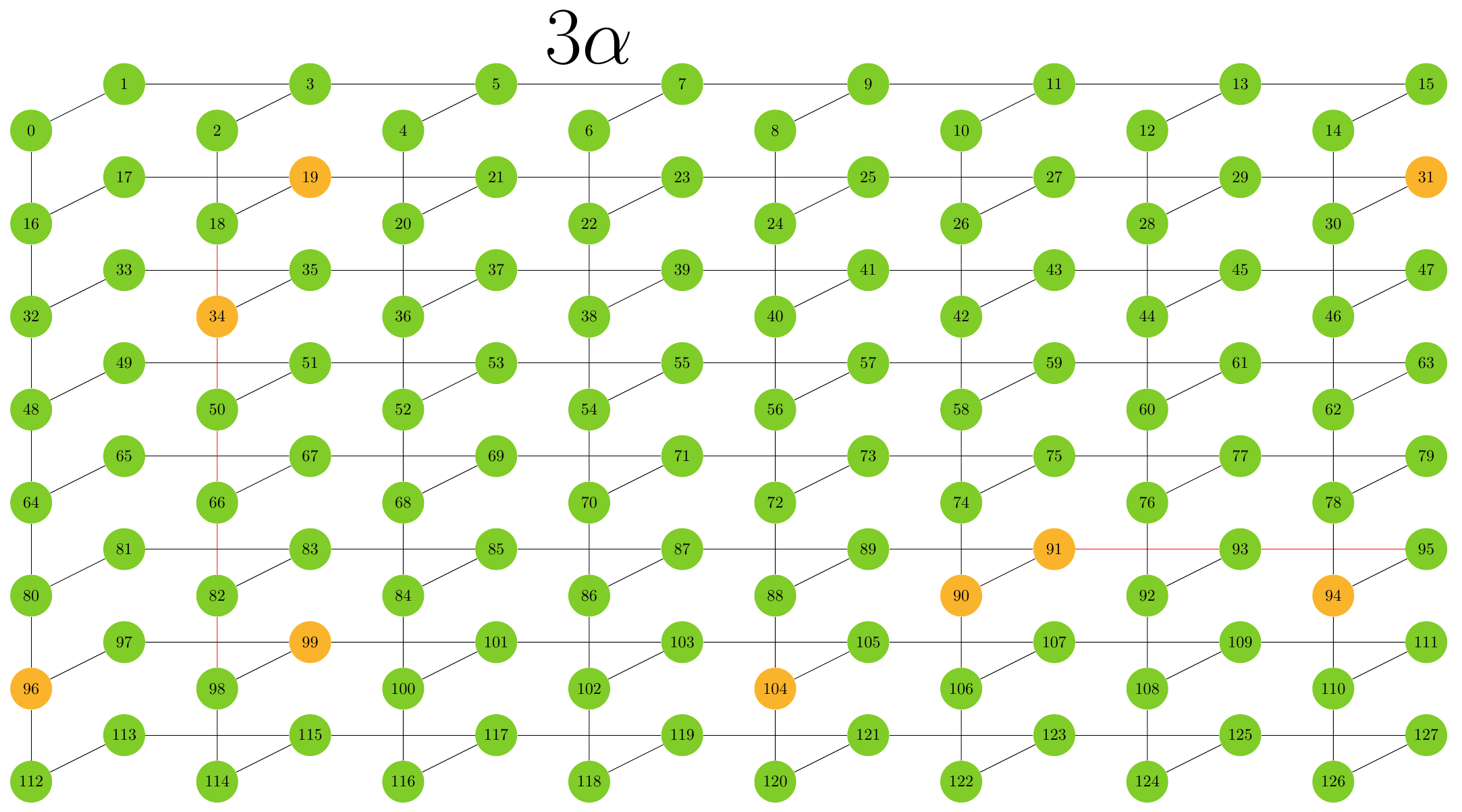}\label{fig:1c}} 
\end{center}
\caption{\small \textbf{Unit cell and encoded graph.} (a) Schematic of one of the $64$ {unit cells} of the DW2 processor (see Appendix~\ref{app:Chimera}). Unit cells are arranged in an $8\times 8$ array forming a ``Chimera" graph between qubits. Each circle represents a physical qubit, and each line a programmable Ising coupling $\sigma^z_i\sigma^z_j$. Lines on the right (left) couple to the corresponding qubits in the neighboring unit cell to the right (above). 
(b) Two ``logical qubits" ($i$, red and $j$, blue) embedded within a single unit cell. Qubits labeled 1-3 are the ``problem qubits", the opposing qubit of the same color labeled $P$ is the ``penalty qubit". Problem qubits couple via the black lines with tunable strength $\alpha$ both inter- and intra-unit cell. Light blue lines of magnitude $\beta$ are ferromagnetic couplings between the problem qubits and their penalty qubit. (c) Encoded processor graph obtained from the Chimera graph by replacing each logical qubit by a circle. This is a non-planar graph (see Appendix~\ref{app:non-planarity} for a proof) 
with couplings of strength $3\alpha$.
Green circles represent  complete logical qubits. Orange circles represent logical qubits lacking their penalty qubit (see Fig.~\ref{fig:Chimera}). Red lines are {groups of} couplers that cannot {all} be simultaneously activated. 
}
\label{fig:1}
\end{figure}

Numerous experiments have demonstrated the utility of quantum error correction in gate-model quantum computing with up to $9$ qubits using, e.g., NMR \cite{PhysRevLett.81.2152,PhysRevLett.109.100503}, trapped ions \cite{Chiaverini:04,Schindler27052011}, and optical systems \cite{Lu12082008,Aoki:2009:541}, and superconducting circuits \cite{2012Natur.482..382R}. However, such demonstrations  require far more control than is available in PQA. 
Likewise, methods developed for adiabatic quantum computing \cite{jordan2006error,PhysRevLett.100.160506,PhysRevA.86.042333,Young:13},
require operations which are not included in the PQA repertoire. Here we show how PQA can nevertheless be error corrected. We provide an experimental demonstration using up to $344$ superconducting flux qubits in D-Wave processors \cite{Dwave,harris_experimental_2010_1} which have recently been shown to physically implement PQA \cite{DWave-16q,q-sig,QA108}.
The qubit connectivity graph of these processors is depicted in Fig.~\ref{fig:1a}.


\begin{figure*}[th]
\begin{center}
\subfigure[\, $\alpha=1$]{\includegraphics[width=0.48\textwidth]{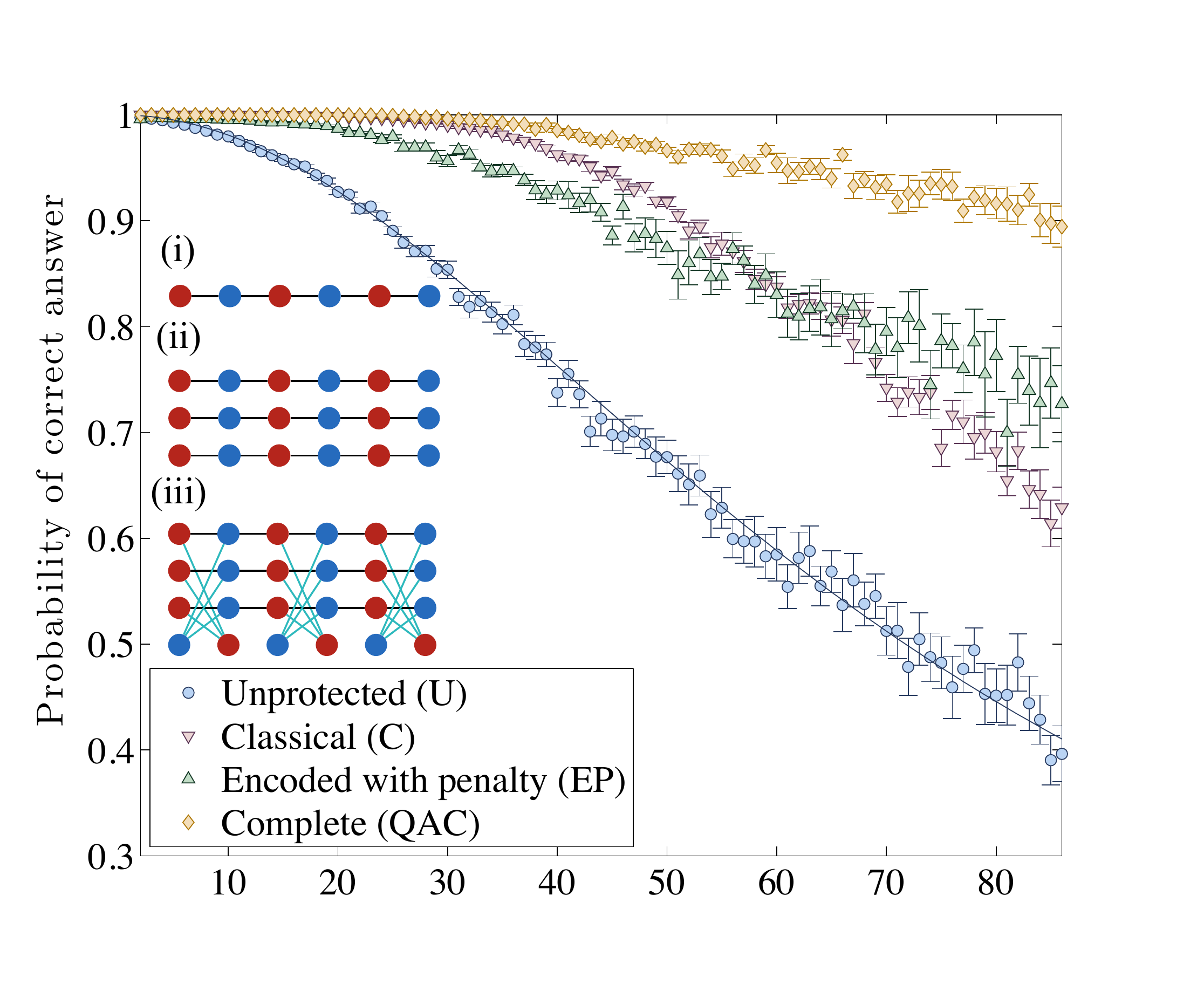}\label{fig:2a}} 
\subfigure[\, $\alpha=0.6$]{\includegraphics[width=0.48\textwidth]{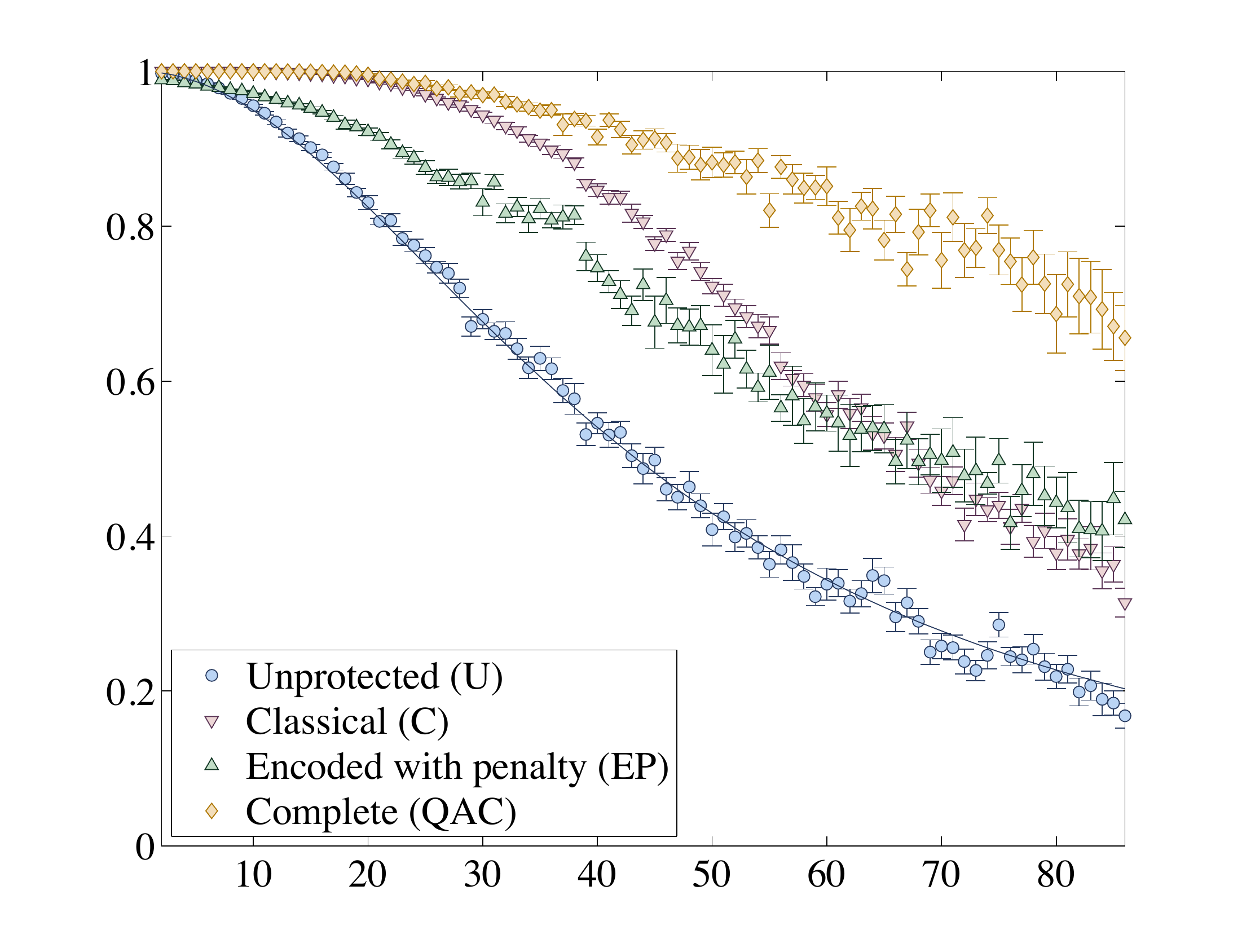}\label{fig:2b}} 
\subfigure[\, $\alpha=0.3$]{\includegraphics[width=0.48\textwidth]{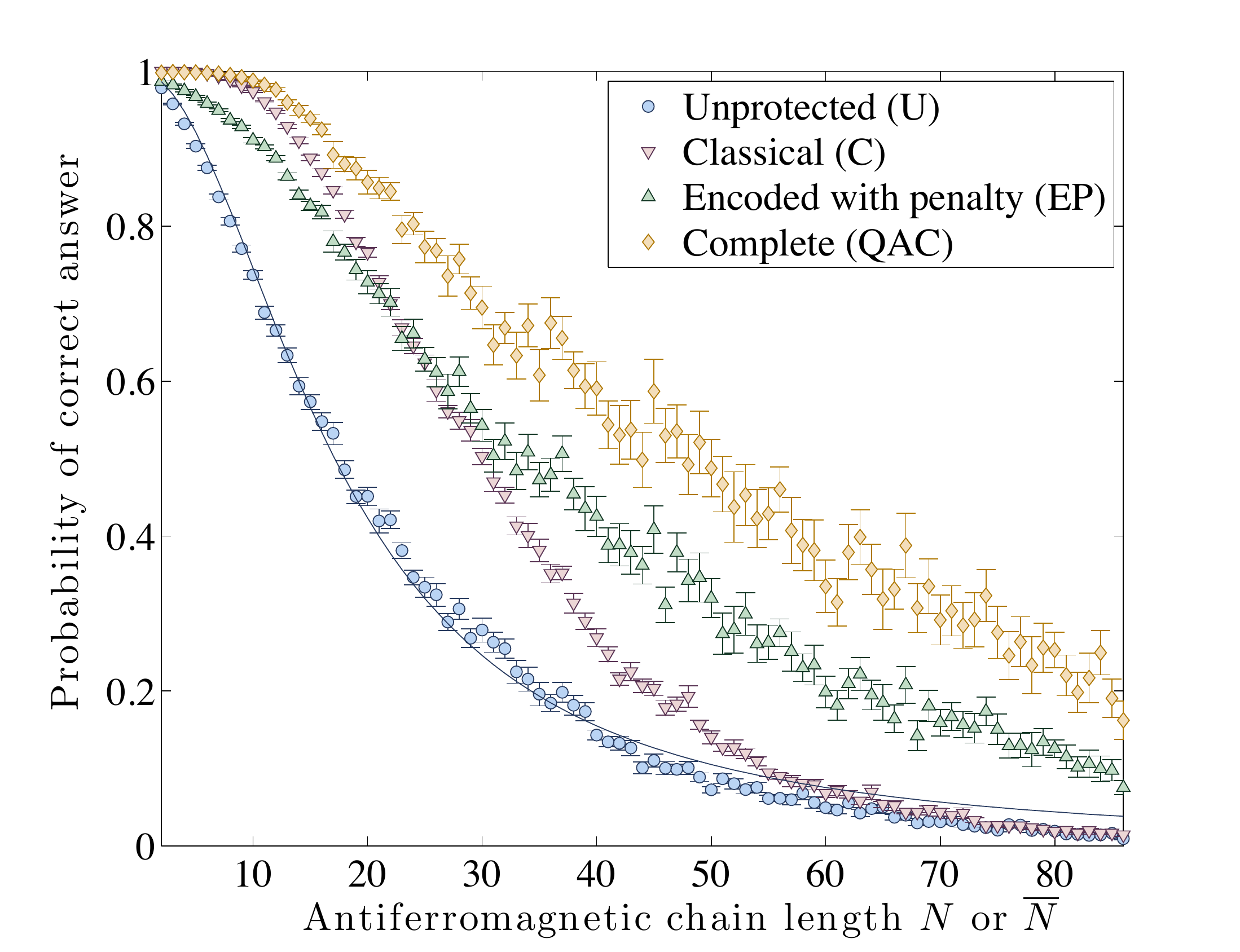} \label{fig:2c}}
\subfigure[\, U and QAC \textit{vs} $\alpha$, $N=\bar{N}=86$]{\includegraphics[width=0.48\textwidth]{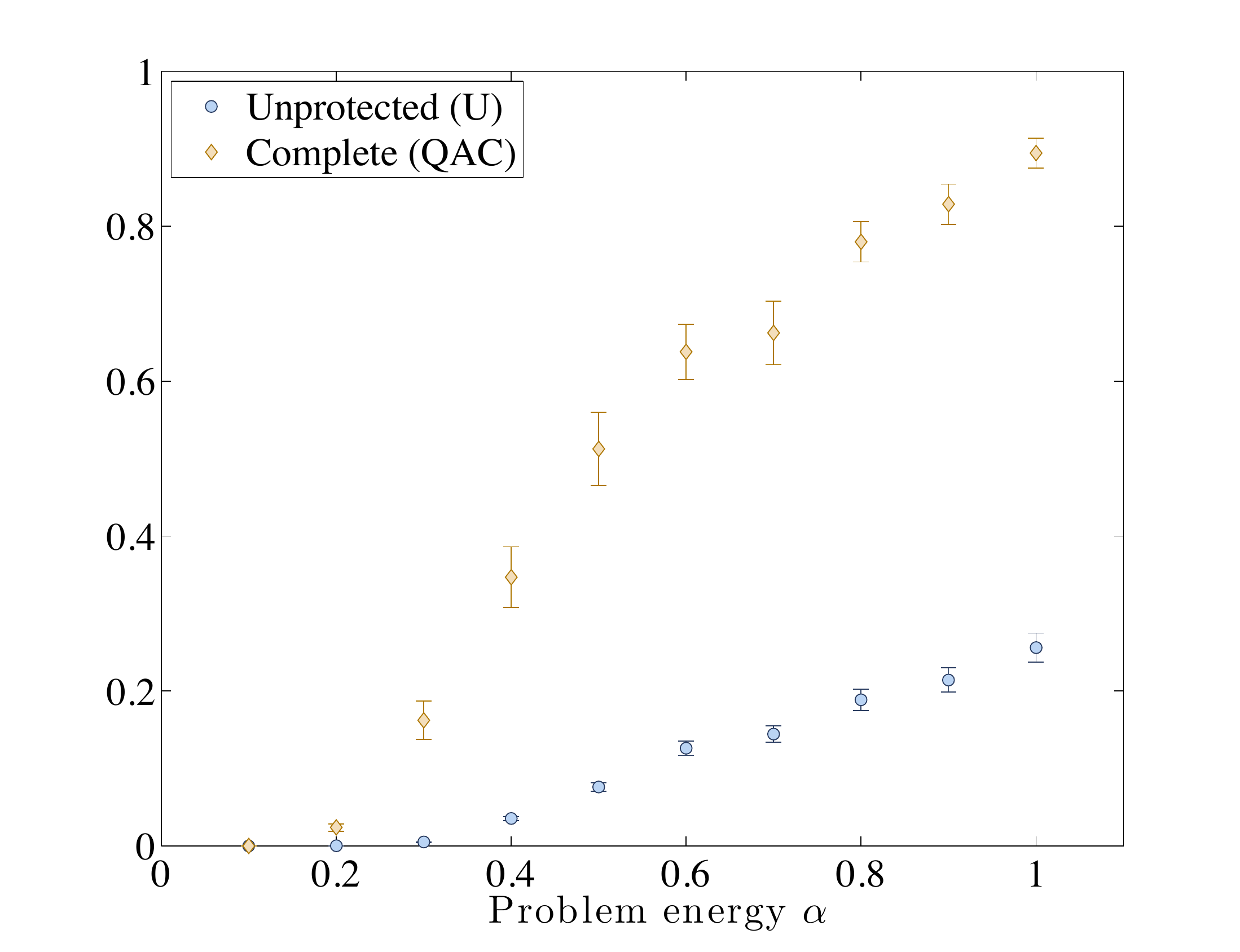} \label{fig:2d}}
\end{center}
\caption{\small 
\textbf{Success probabilities of the different strategies}. Panels (a)-(c) show the results for antiferromagnetic chains as a function of chain length. The solid blue lines in the U case are best fits to  $1/(1+ p N^2)$ (Lorentzian), yielding $p=1.94 \times 10^{-4}, 5.31 \times 10^{-4}, 3.41 \times 10^{-3}$ for $\alpha = 1,0.6,0.3$ respectively.  Panel (d) compares the U and QAC strategies at $N=\bar{N}=86$ and $\alpha\in\{0.1,0.2,\dots,1.0\}$. 
Chains shown in Panel (a) respectively depict the U (i), C (ii), and EP and QAC (iii) cases. In (ii) and (iii), respectively, vertically aligned and coupled physical qubits of the same color form a logical qubit.  Error bars in all our plots were calculated over the set of embeddings and express the standard error of the mean $\sigma/\sqrt{S}$, where $\sigma^2 = \frac{1}{S}\sum_{i=1}^S(x_i-\overline{x})^2$ is the sample variance and $S$ is the number of samples. Additional details, including from experiments on a previous generation of the processor (the D-Wave One (DW1) ``Rainier") are given in Appendix~\ref{app:more-fidelity-results}.
}
\label{fig:2}
\end{figure*}


 \section{Quantum annealing and computational errors} 
Many hard and important optimization problems can be encoded into the lowest energy configuration (ground state) of an Ising Hamiltonian
\beq
H_{\mathrm{Ising}}
= \sum_{i=1}^N h_i \sigma_i^z + \sum_{i<j}^N J_{ij} \sigma_i^z \sigma_j^z \ ,
\eeq
where each $\sigma_i^z = \pm 1$ is a classical binary variable, and the dimensionless local fields $h:=\{h_i\}$ and couplings $J:=\{J_{ij}\}$ are the ``program parameters" that specify the problem \cite{Barahona1982,2013arXiv1302.5843L}. In PQA the solution of the optimization problem is found by replacing the classical variables by $N$ quantum binary variables (qubits), that evolve subject to the time-dependent Hamiltonian 
\beq \label{eqt:QA}
H(t) = A(t) H_X + B(t) H_{\mathrm{Ising}} \ ,\quad t\in [0,t_f]\ .
\eeq
Here $H_X =  \sum_{i=1}^N \sigma_i^x$ is a transverse field Hamiltonian, $\sigma^z_i$ and $\sigma^x_i$ denote the spin-$1/2$ Pauli operators whose eigenstates are, respectively, $\ket{0},\ket{1}$ and $\ket{\pm} = (\ket{0}\pm\ket{1})/\sqrt{2}$, with eigenvalues $\pm 1$.
$A(t)$ and $B(t)$ are time-dependent functions (with dimensions of energy) satisfying $A(t_f)=B(0)=0$, and $t_f$ is the annealing time.     
A physical PQA always operates in the presence of a thermal environment at temperature $T$. Provided $A(0) \gg k_B T$, the PQA is initialized in the ground state of $H_X$, namely the uniform superposition state $(\ket{0\cdots0}+\cdots+\ket{1\cdots1})/\sqrt{2^N}$.  Provided $B(t_f) \gg k_B T$, the final state at the end of the annealing process is stable against thermal excitations when it is measured. If the evolution is adiabatic, i.e., if $H(t)$ is a smooth function of time and if the gap $\Delta:=\min_{t\in[0,t_f]}\epsilon_1(t)-\epsilon_0(t)$ between the first excited state energy $\epsilon_1(t)$ and the ground state energy $\epsilon_0(t)$ is sufficiently large compared to both $1/t_f$ and $T$, then the adiabatic approximation for open systems \cite{PhysRevA.65.012322,SarandyLidar:04,ABLZ:12,Qiang:13} guarantees that the desired  ground state of $H_{\mathrm{Ising}} $ will be reached with high fidelity at $t_f$. However, hard problems are characterized by gaps that close polynomially or even exponentially with increasing problem size \cite{Farhi20042001,PhysRevLett.101.170503}.
If the gap is too small, then both non-adiabatic transitions and thermal excitations
can result in computational errors, manifested in the appearance of excited states at $t_f$. While the non-adiabatic transition rate can in principle be suppressed to an arbitrarily high degree by enforcing a smoothness condition on the annealing functions $A(t)$ and $B(t)$ \cite{lidar:102106}, thermal excitations will cause errors at any non-zero temperature.
Additionally, even if $\Delta$ is large enough, inaccuracies in the implementation of $H_{\mathrm{Ising}}$ may result in the evolution ending up in the ``wrong" ground state. Overcoming such errors requires error correction.


\section{Quantum annealing correction}
We devise a strategy we call ``quantum annealing correction" (QAC), comprising the introduction of an energy penalty along with encoding and error correction. Our main tool is the ability to independently control pairwise Ising interactions, which can be viewed as the generators of the bit-flip stabilizer code \cite{Gaitan:book}. We first encode $H_{\mathrm{Ising}}$, replacing each  $\sigma^z_i$ term by its encoded counterpart $\overline{\sigma^z_i}=\sum_{\ell=1}^n  \sigma^z_{i_{\ell}}$ and each $\sigma_i^z \sigma_j^z$ by
$\overline{\sigma_i^z \sigma_j^z} =  \sum_{\ell=1}^n\sigma^z_{i_\ell}\sigma^z_{j_\ell}$, where the subindices $\ell$ refer to the problem qubits as depicted in Fig.~\ref{fig:1b}. After these replacements we obtain an encoded Ising Hamiltonian
\beq 
\label{eqt:Encoded_Ising}
\overline{H}_{{\mathrm{Ising}}}
= \sum_{i=1}^{\overline{N}} h_i\overline{\sigma_i^z} + \sum_{i<j}^{\overline{N}}  J_{ij} \overline{ \sigma_i^z \sigma_j^z}  \ ,
\eeq
where ${\overline{N}}$ is the number of encoded qubits. 
The ``code states" $\ket{\overline{0}_i} = \ket{0_{i_1}\cdots 0_{i_n}}$ and $\ket{\overline{1}_i} = \ket{1_{i_1}\cdots 1_{i_n}}$ are eigenstates of $\overline{\sigma^z_i}$ with eigenvalues $n$ and $-n$ respectively. ``Non-code states" are the remaining $2^n-2$ eigenstates, having at least one bit-flip error. The states $\ket{\overline{0}_i}\ket{\overline{0}_j},\ket{\overline{1}_i}\ket{\overline{1}_j}$ and $\ket{\overline{0}_i}\ket{\overline{1}_j},\ket{\overline{1}_i}\ket{\overline{0}_j}$ are eigenstates of $\overline{\sigma_i^z \sigma_j^z}$, also with eigenvalues $n$ and $-n$ respectively. 
Therefore the ground state of $\overline{H}_{{\mathrm{Ising}}}$ is identical, in terms of the code states, to that of the original unencoded Ising Hamiltonian, with $N=\overline{N}$. 

\begin{figure*}[t]
\begin{center}
\subfigure[\, ]{\includegraphics[width=0.48\textwidth]{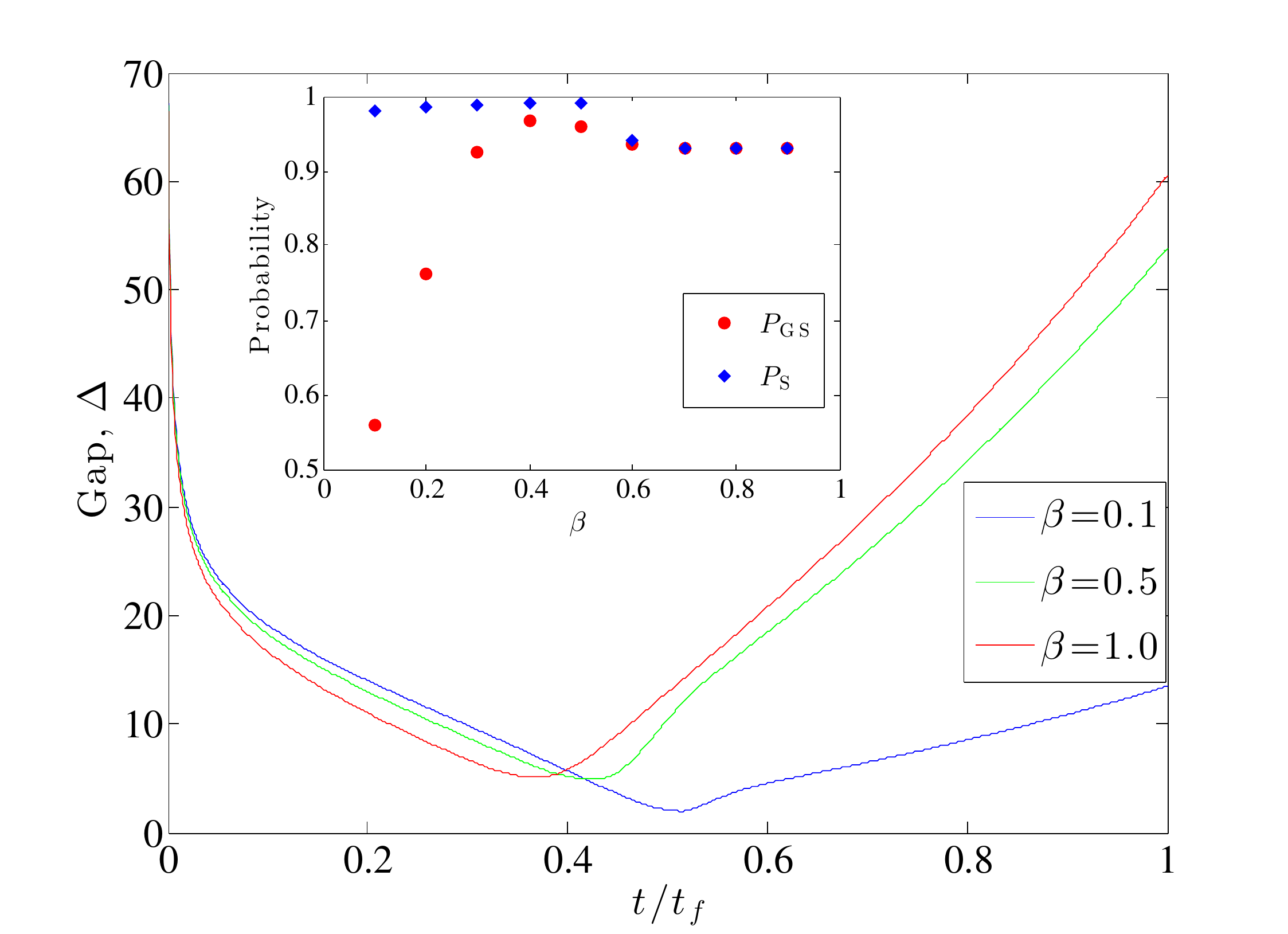}\label{fig:3a}} 
\subfigure[\, ]{\includegraphics[width=0.48\textwidth]{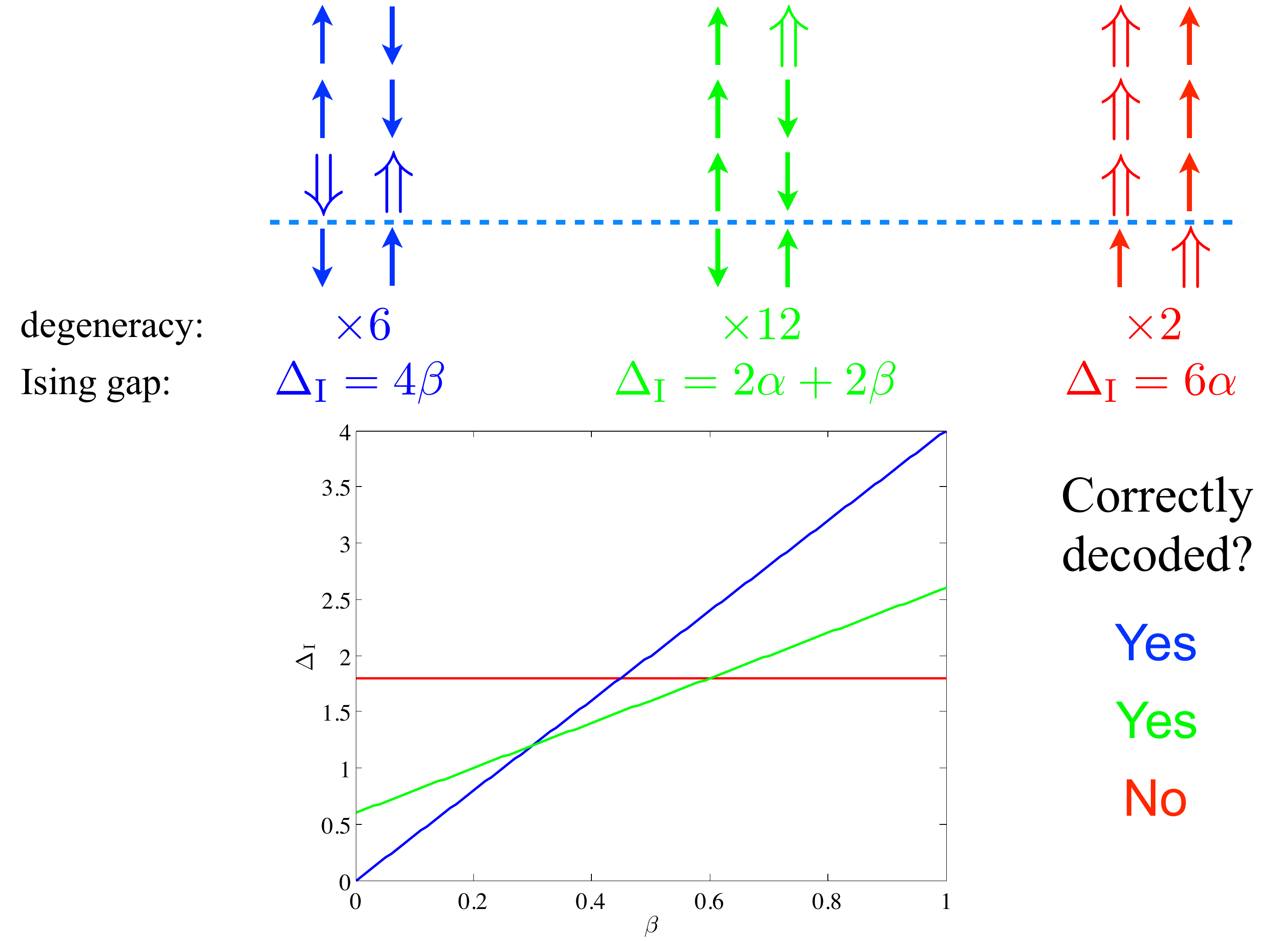}\label{fig:3b}} 
\end{center}
\caption{\small \textbf{Effect of varying the penalty strength $\beta$}. Panel (a) shows the numerically calculated gap to the lowest relevant  excited state for 
two antiferromagnetically coupled logical qubits for $\alpha=0.3$ and different values of $\beta$. Inset: undecoded (decoded) ground state probability $P_\textrm{GS}$ ($P_\textrm{S}$).
Panel (b) (top) shows three configurations of two antiferromagnetically coupled logical qubits. Physical qubits denoted by heavy arrows point in the wrong direction. 
In the left configuration both logical qubits have a bit-flip error, in the middle configuration only one logical qubit has a single bit-flip error, and in the right configuration one logical qubit is completely flipped. 
The corresponding degeneracies and gaps ($\Delta_{\textrm I}$) from the final ground state are indicated, and the gaps plotted (bottom).
}
\label{fig:3ab}
\end{figure*}

This encoding allows for protection against bit-flip errors in two ways.  First, the overall problem energy scale is increased by a factor of $n$, where $n=3$ in our implementation on the D-Wave processors. 
Note that since we cannot also encode $H_X$ (this would require $n$-body interactions), it does not directly follow that the gap energy scale also increases; we later present numerical evidence that this is the case (see Fig.~\ref{fig:3a}), so 
that thermal excitations will be suppressed.
Second, the excited state spectrum has been labeled in a manner which can be decoded by performing a post-readout majority-vote on each set of $n$ problem qubits, thereby error-correcting non-code states into code states in order to recover some of the excited state population. The $(n,1)$ repetition code has minimum Hamming distance $n$, i.e., 
a non-code state with more than $\lfloor n/2 \rfloor$ bit-flip errors will be incorrectly decoded; we call such states ``undecodable'', while ``decodable states''  are those excited states that are decoded via majority-vote to the correct code state. 

To generate additional protection we next introduce a ferromagnetic penalty term 
\beq
H_P = - \sum_{i=1}^{\overline{N}} \left(\sigma^z_{i_1}  + \cdots + \sigma^z_{i_n}\right)  \sigma^z_{i_{P}}  \ ,
\eeq
the sum of stabilizer generators of the $n+1$-qubit repetition code, which together detect and energetically penalize  \cite{jordan2006error} all bit-flip errors except the full logical qubit flip. The role of $H_P $ can also be understood as to lock the problem qubits into agreement with the penalty qubit, reducing the probability of excitations from the code space into non-code states; 
see Fig.~\ref{fig:1b} for the D-Wave processor implementation of this penalty. 
The encoded graph thus obtained in our experimental implementation is depicted in Fig.~\ref{fig:1c}.

Including the penalty term, the total encoded Hamiltonian we implement is 
\beq
\overline{H}(t) 
= A(t) H_X + B(t) \overline{H}_{\mathrm{Ising},P}(\alpha,\beta) \ ,
\label{eq:HencP}
\eeq
where $\overline{H}_{\mathrm{Ising},P} (\alpha,\beta):= \alpha \overline{H}_{\mathrm{Ising}} + \beta H_P$, and the two controllable parameters $\alpha$ and $\beta$ are the ``problem scale" and ``penalty scale", respectively, which we can tune between $0$ and $1$ in our experiments and optimize. 
Note that our scheme, as embodied in Eq.~\eqref{eq:HencP}, implements
\emph{quantum} annealing correction: $H_P$ energetically penalizes every error $E$ it does not commute with, e.g., every single-qubit error $E\in U(2)$ such that $E\not\propto \sigma^z_i$.\\ 


\begin{figure*}[t]
\begin{center}
\subfigure[\, EP, $\alpha=0.3$]{\includegraphics[width=0.32\textwidth]{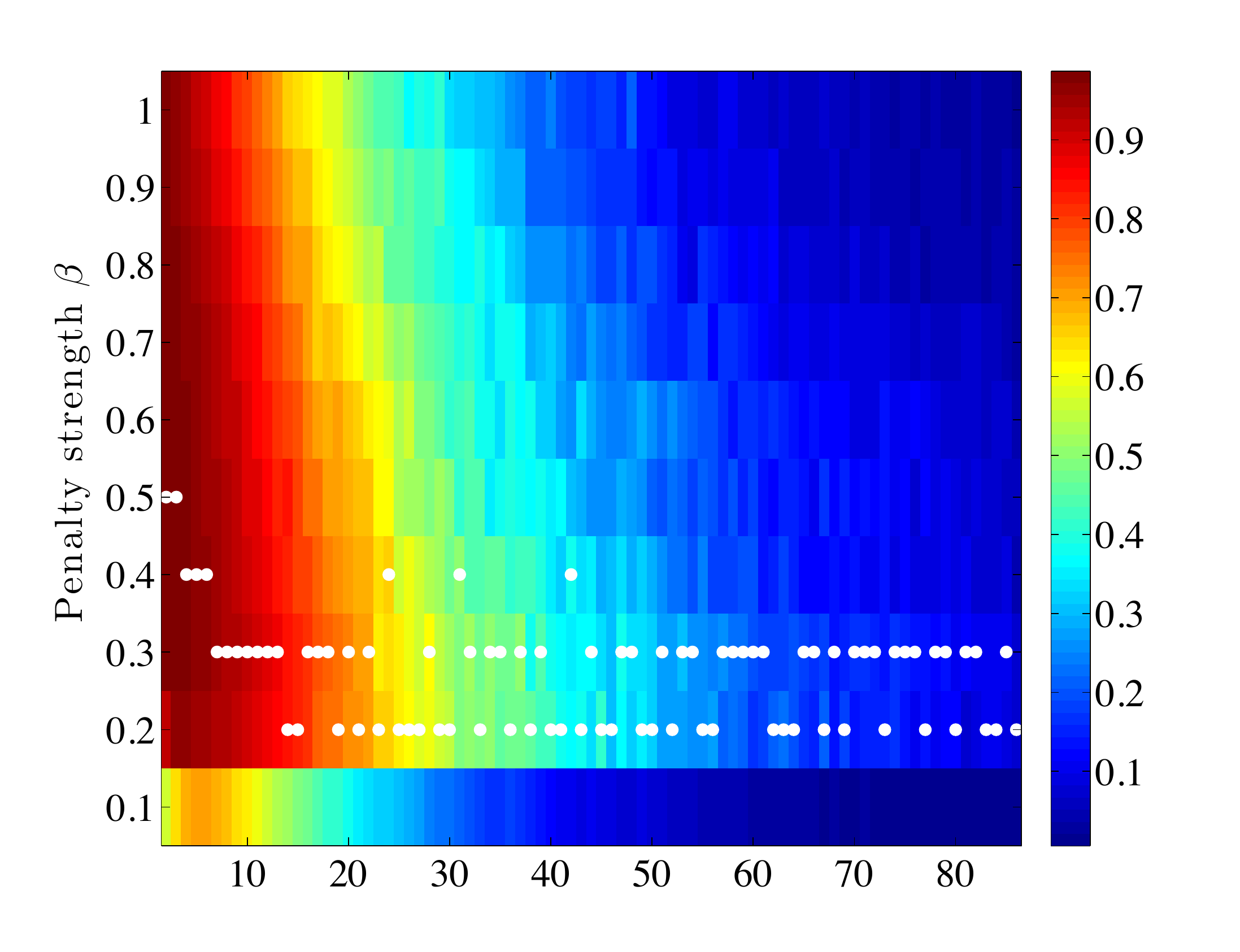} \label{fig:beta-EP-alpha=0.3}}
\subfigure[\, EP, $\alpha=0.6$]{\includegraphics[width=0.32\textwidth]{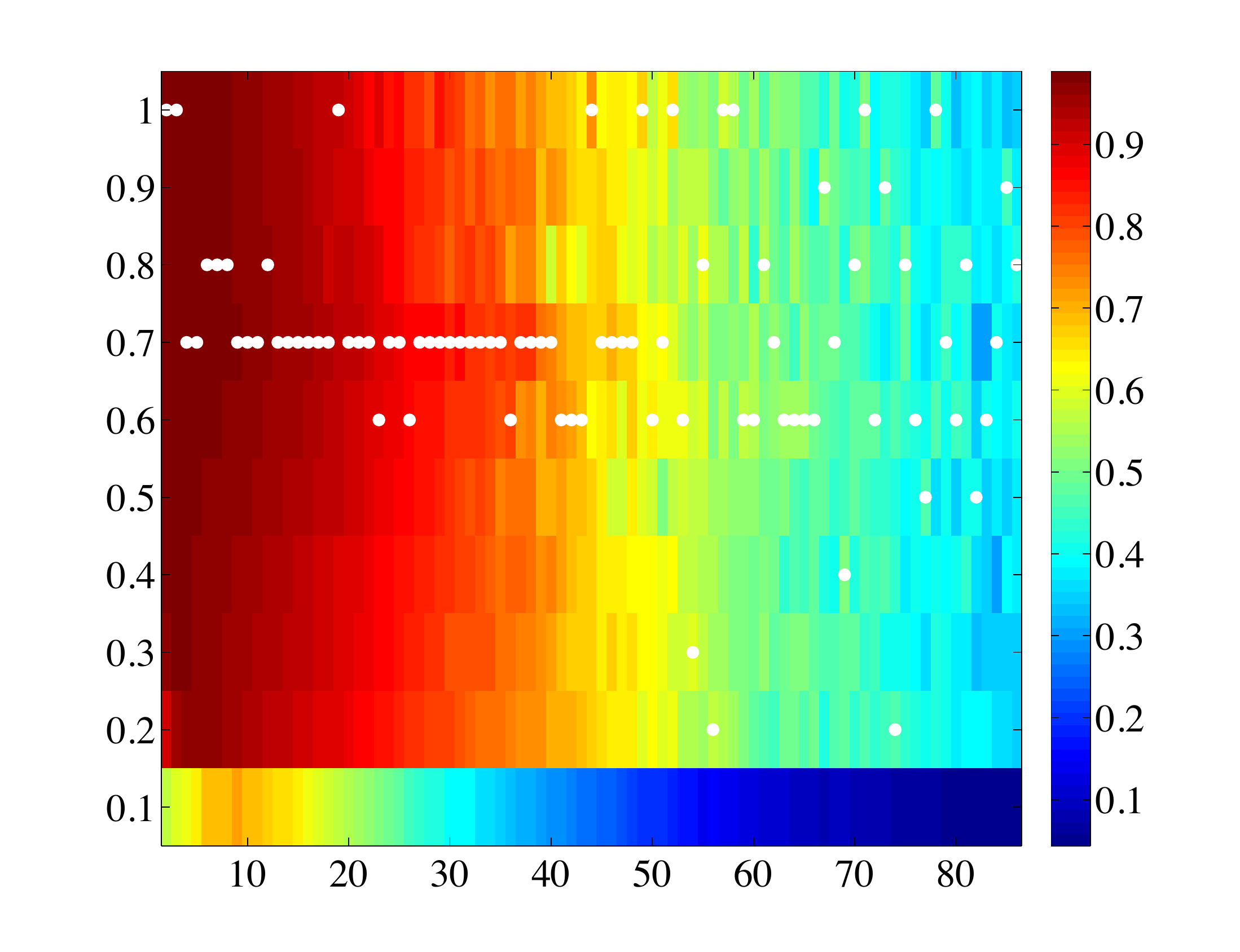}\label{fig:beta-EP-alpha=0.6}}
\subfigure[\, EP, $\alpha=1$]{\includegraphics[width=0.32\textwidth]{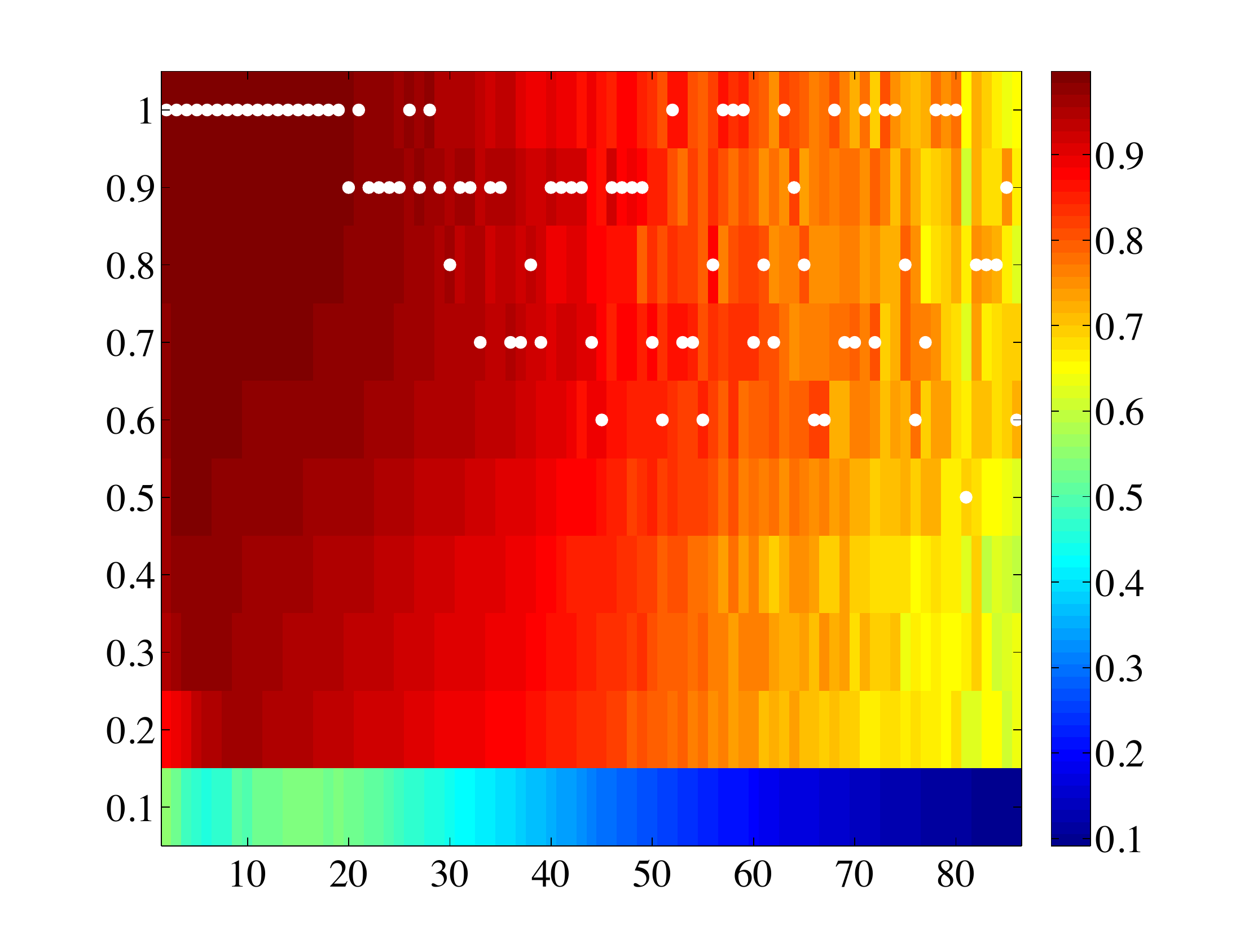}\label{fig:beta-EP-alpha=1}}
\subfigure[\, QAC, $\alpha=0.3$]{\includegraphics[width=0.32\textwidth]{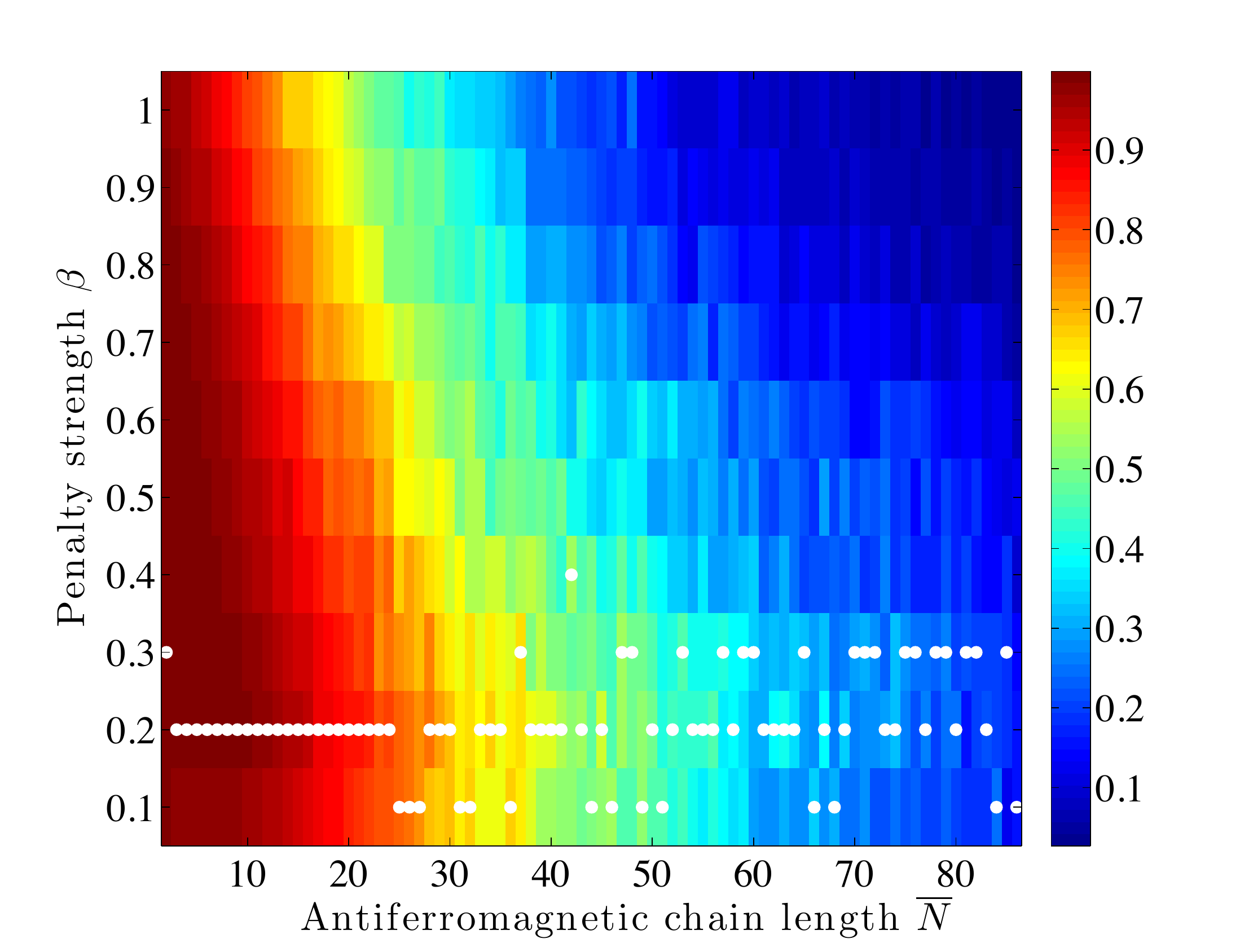}\label{fig:3d}} 
\subfigure[\, QAC, $\alpha=0.6$]{\includegraphics[width=0.32\textwidth]{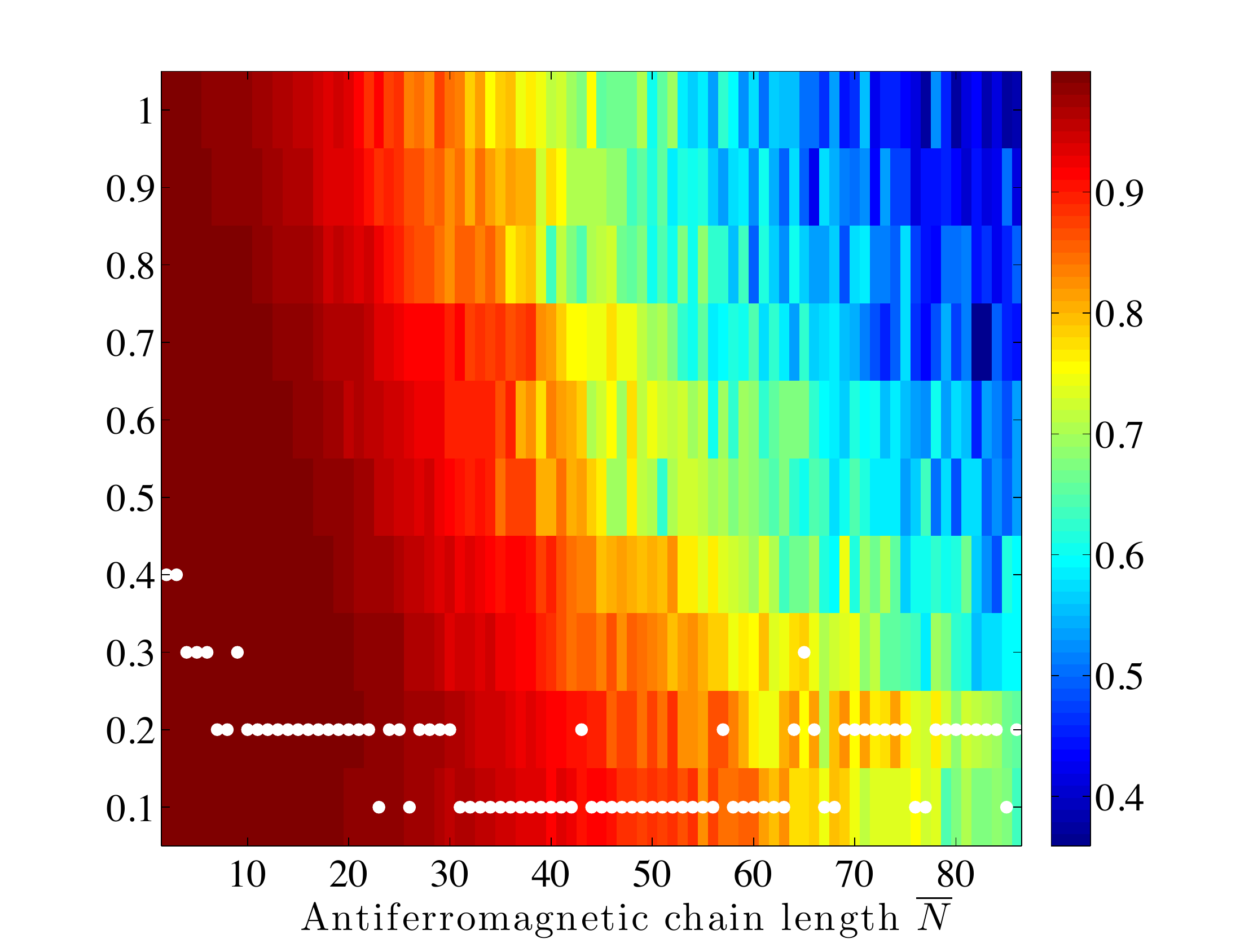}\label{fig:3dd}} 
\subfigure[\, QAC, $\alpha=1$]{\includegraphics[width=0.32\textwidth]{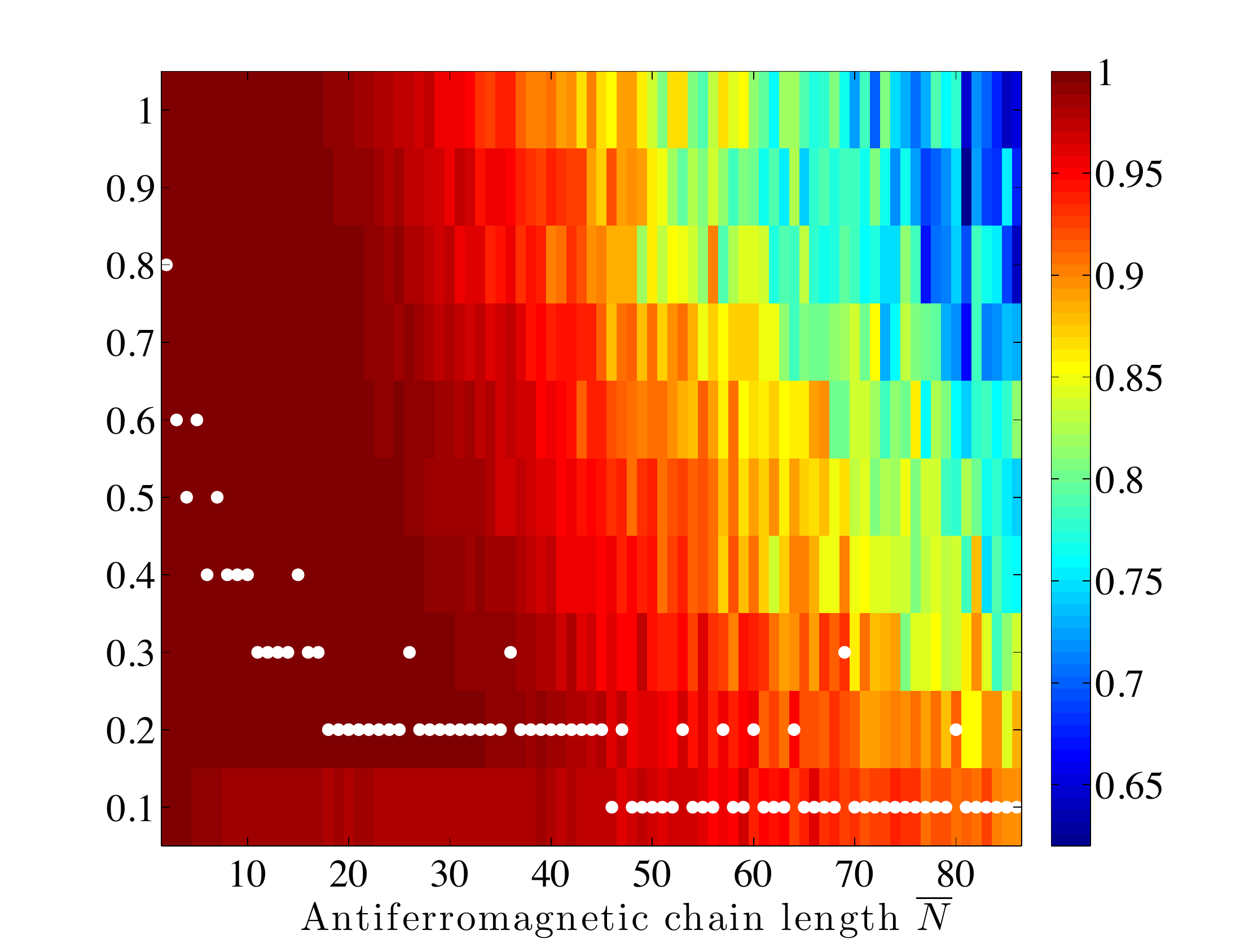}\label{fig:3c}} 
\end{center}
\caption{\small \textbf{Experimental optimization of the penalty strength $\beta$.} The top (bottom) row shows color density plots of the experimental success probability of the EP (QAC) strategy as a function of $\beta \in \{0.1,0.2,\dots,1.0\}$ and $\bar{N}\in\{2,3,\dots,86\}$, at $\alpha=0.3$ (left), $0.6$ (middle) and $1$ (right). The optimal $\beta$ values are indicated by the white dots.
}
\label{fig:3cd}
\end{figure*}


\section{Benchmarking using antiferromagnetic chains} 
Having specified the general scheme, which in particular is applicable to any problem that is embeddable on the encoded graph shown in Fig.~\ref{fig:1c}, we now focus on antiferromagnetic chains.  In this case, the classical ground states at $t=t_f$ are the trivial doubly degenerate states of nearest-neighbor spins pointing in opposite directions.  This allows us to benchmark our QAC strategy while focusing on the role of the controllable parameters, instead of the complications associated with the ground states of frustrated Ising models \cite{QA108,MAX2SAT}.  Moreover, chains are dominated by domain wall errors \cite{Dwave}, which as we explain below are a particularly challenging scenario for our QAC strategy.

As a reference problem we implemented an $N$-qubit antiferromagnetic chain with ${H}_{\mathrm{Ising}} (\alpha) = \alpha \sum_{i=1}^{N-1} \sigma_i^z \sigma_{i+1}^z$.
We call this problem ``unprotected" (U) since it involves no encoding or penalty. As a second reference problem we implemented three unpenalized parallel $N$-qubit chains:
$\overline{H}_{\mathrm{Ising}}(\alpha)   = \alpha\sum_{j=1}^3 \sum_{i=1}^{N-1} \sigma_{i_j}^z \sigma_{i_j+1}^z$. We call this problem ``classical" (C) since this results in a purely classical repetition code, whereby each triple of bits $\{i_j\}_{j=1}^3$ forms a logical bit $i$, decoded via majority-vote. 
As a third reference problem we implemented a chain of $\overline{N}$ encoded qubits with an energy penalty (EP):
$\overline{H}_{\mathrm{Ising},P} (\alpha,\beta)  = \alpha \sum_{i=1}^{\overline{N}-1} \overline{\sigma_{i}^z \sigma_{i+1}^z}  + \beta H_{P} $.
When we add majority-vote decoding to the EP strategy we have our complete QAC strategy. Comparing the probability of finding the ground state in the U, C, EP and QAC cases allows us to isolate the effects of the various components of the error correction strategy. 
Because antiferromagnetic chains have two degenerate ground states, below we consider the ground state for any given experimentally measured state to be that with which the majority of the decoded qubits align.

\section{Key experimental results -- success probabilities}
The performance of the different strategies are shown in Fig.~\ref{fig:2}.
Our key finding is the high success probability of the complete QAC strategy for $\alpha=1$ (Fig.~\ref{fig:2a}), improving significantly over the three other strategies, and resulting in a fidelity $>90\%$ for all chain lengths. The relative improvement is highest for low values of $\alpha$, as seen in Fig.~\ref{fig:2d}.
The C strategy is competitive with QAC for relatively short chains, but drops eventually. The EP probability is initially intermediate between the U and C cases, but always catches up with the C data for sufficiently long chains. This shows that the energy penalty strategy by itself is insufficient, and must be supplemented by decoding as in the complete QAC strategy.

Since $\alpha$ sets the overall problem energy scale, it is inversely related to the effective noise strength. This is clearly visible in Fig.~\ref{fig:2a}-\ref{fig:2d} (see also Fig.~\ref{fig:3cd}), where the overall success probability improves significantly over a range of $\alpha$ values.
The unprotected chains are reasonably well fit by a Lorentzian, whereas a classical model of independent errors (see Appendix~\ref{app:indep-err}) fails to describe the data as it predicts an exponential dependence on $N$. 
We turn next to an analysis and explanation of our results.


\begin{figure*}[t]
\begin{center}
\subfigure[\, ]{\includegraphics[width=0.48\textwidth]{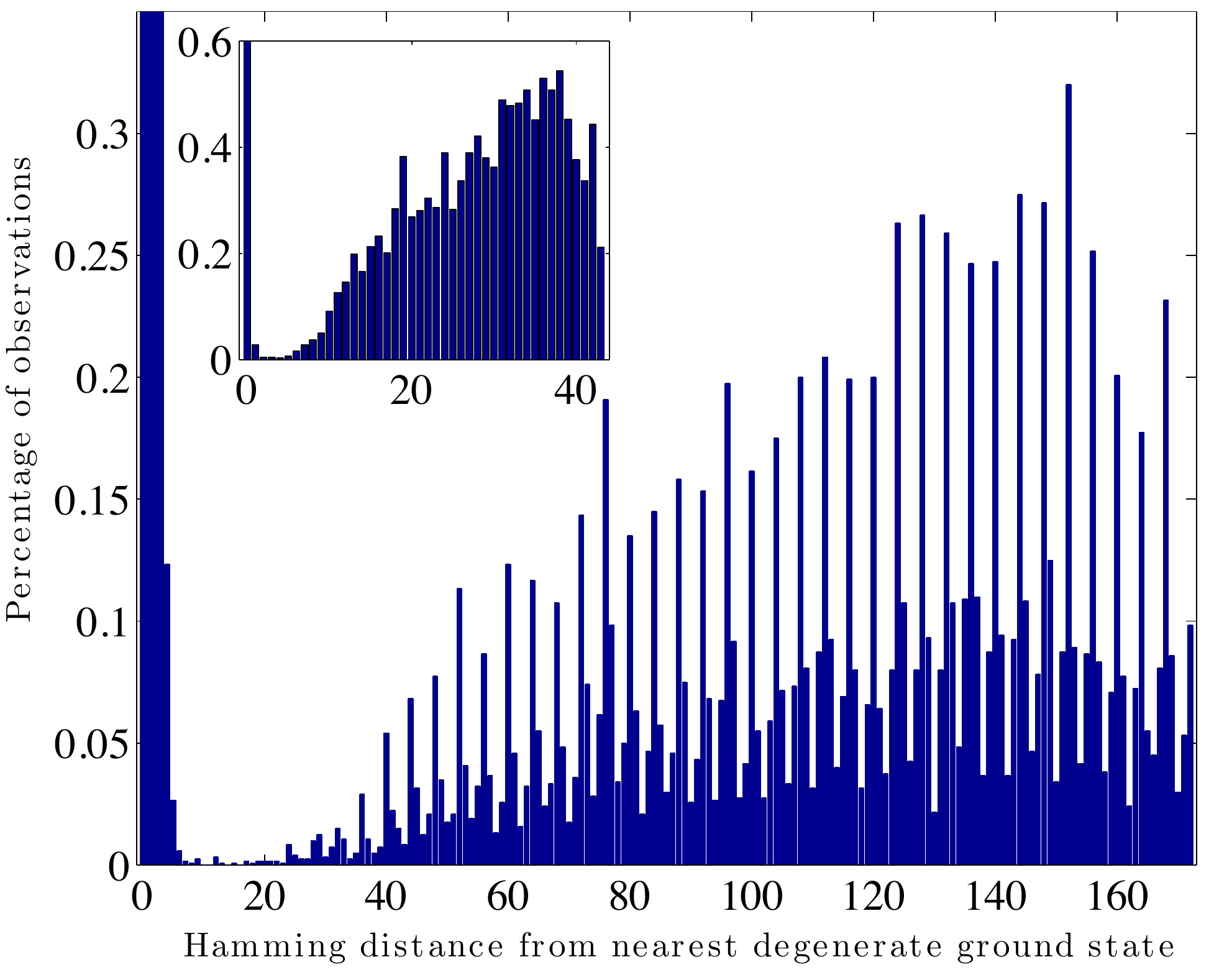}\label{fig:4a}} 
\subfigure[\, ]{\includegraphics[width=0.48\textwidth]{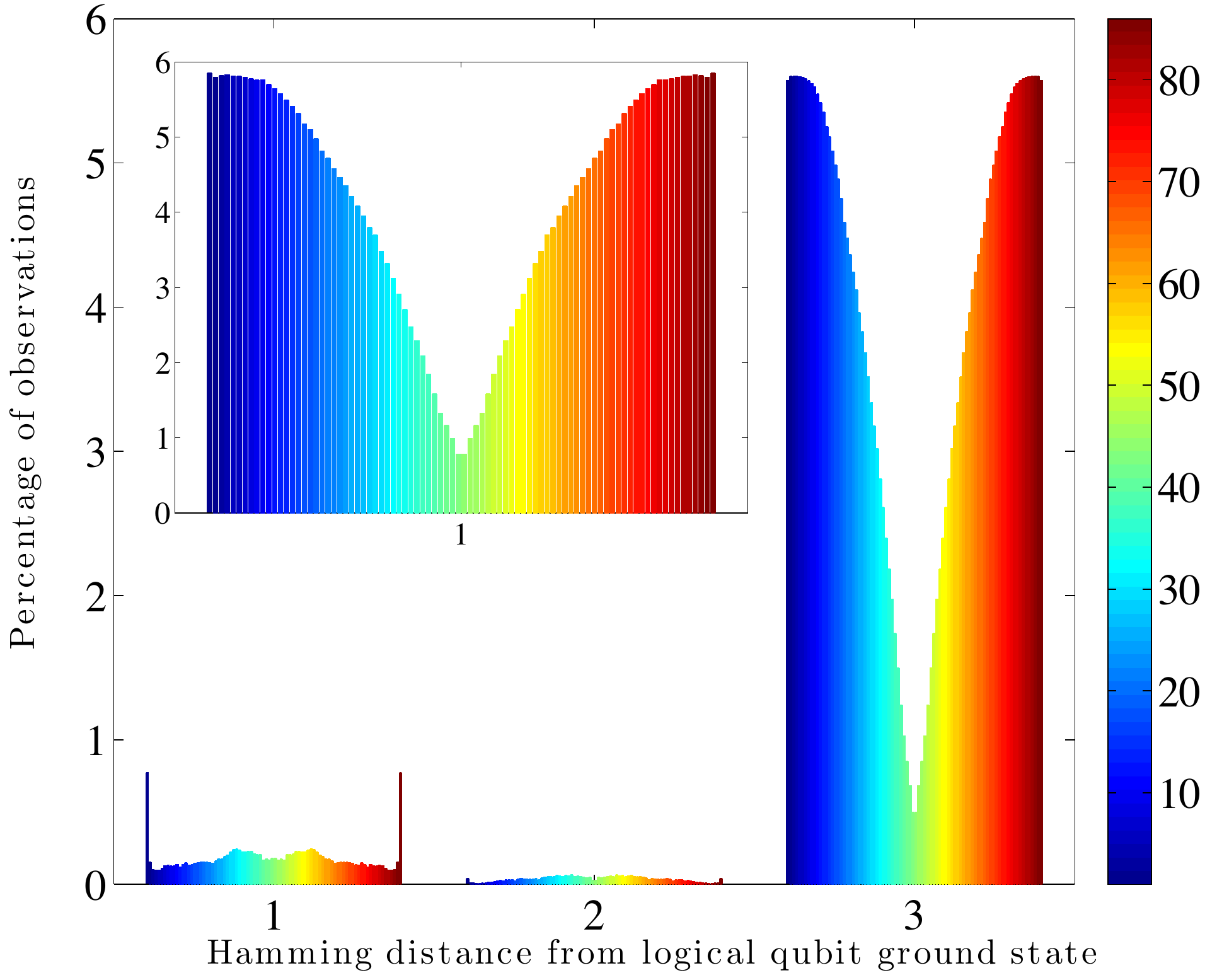}\label{fig:4b}} 
\end{center}
\caption{\small \textbf{Hamming distance histograms.} Observed errors 
in encoded $86$ qubit antiferromagnetic chains, at $\alpha=1$ and the near-optimal $\beta=0.2$. Panel (a) is a histogram of Hamming distances from the nearest of the two degenerate ground states, measured in terms of physical qubits. Inset: in terms of encoded qubits. The peaks at Hamming distance zero are cut off and extend to 63.6\% (88.3\%) for the physical (encoded) case. 
Panel (b) is a histogram of the errors as a function of logical qubit position (color scale) within the chain. Errors on encoded problem qubits are at Hamming distance $1$, $2$ or $3$. Flipped penalty qubits are shown in the inset. 
The mirror symmetry is due to averaging over the two equivalent chain directions. 
}
\label{fig:4}
\end{figure*}


\section{Optimizing the penalty scale $\beta$} 
To obtain the performance of the EP and QAC strategies shown in Fig.~\ref{fig:2} we optimized $\beta$ separately for each strategy and for each setting of $\alpha$ and $\bar{N}$. In order to understand the role of $\beta$ consider first how  increasing $\beta$ affects the size and position of the gap $\Delta$. The  excitations relevant to our error correction procedure are to the second excited state and above, since the ground state becomes degenerate at $t_f$.
In Fig.~\ref{fig:3a} we show that the relevant gap grows with increasing $\beta$, as desired. The gap position also shifts to the left, which is advantageous since it leaves less time for thermal excitations to act while the transverse field dominates. However, the role of $\beta$ is more subtle than would be suggested by considering only the gap. When $\beta \ll \alpha$  the penalty has no effect, and when $\beta \gg \alpha$ the penalty dominates the problem scale and the chains effectively comprise decoupled logical qubits. Thus there should be an optimal $\beta$ for each $(\bar{N},\alpha$) pair, which we denote $\bo$. Without decoding we expect $\bo\sim\alpha$ based on the argument above, which is confirmed in Fig.~\ref{fig:3cd} (top row). Note that when $\beta=0.1$ the penalty is too small to be beneficial, hence the poor performance for that value in the EP case.

In the QAC case another effect occurs: the spectrum is reordered so that undecodable states become lower in energy than decodable states. This is explained in Fig.~\ref{fig:3b}. Consider the three configurations shown. While the left and middle configurations are decodable, the right-side configuration is not. For sufficiently large $\alpha$ the undecodable state is always the highest of the three indicated excited states. 
The graph at the bottom of panel (b) shows the Ising gap as a function of $\beta$ for $\alpha=0.3$. 
While for sufficiently small $\beta$ such that $4 \beta , 2 \alpha + 2 \beta < 6 \alpha$ both decodable states are lower in energy than the undecodable state, the undecodable state becomes the first excited state for sufficiently large $\beta$. 
This adversely affects the success probability after decoding, as is verified numerically in the inset of panel (a), which shows the results of an adiabatic master equation \cite{ABLZ:12} calculation for the same problem, yielding the undecoded ground state probability $P_\textrm{GS}$ and the decoded ground state probability $P_\textrm{S}$ (for model details and parameters see Appendix~\ref{app:ME}). 
While for $\beta < 0.6$ decoding helps, this is no longer true when for $\beta >0.6$ the undecodable state becomes the first relevant excited state. 
Consequently we again expect there to be an optimal value of $\beta$ for the QAC strategy that differs from $\bo$ for EP. These expectations are borne out in our experiments: Fig.~\ref{fig:3cd} (bottom row) shows that $\bo$ is significantly lower than in the EP case, which differs only via the absence of the decoding step. The decrease in $\bo$ with increasing $\alpha$ and chain length can be understood in terms of domain wall errors (see below), which tend to flip entire logical qubits, thus resulting in a growing number of undecodable errors.  
For additional insight into the roles of the penalty qubits, $\alpha$ and $\beta$ see Appendices~\ref{app:penalty-role} and \ref{app:ab-role}.

\section{Error mechanisms}
Solving for the ground state of an antiferromagnetic Ising chain is an ``easy'' problem, so why do we observe decreasing success probabilities?  As alluded to earlier, domain walls are the dominant form of errors for antiferromagnetic chains, and we show next how they account for the shrinking success probability.  We analyze the errors on the problem \textit{vs} the penalty qubits and their distribution along the chain. 
Figure~\ref{fig:4a} is a histogram of the observed decoded states at a given Hamming distance $d$ from the ground state of the $ \bar{N}=86$ chains. The large peak near $d=0$ shows that most states are either correctly decoded or have just a few flipped bits. The quasi-periodic structure seen emerging at $d\geq 20$ can be understood in terms of domain walls. The period is four, the number of physical qubits per logical qubit, so this periodicity reflects the flipping of an integer multiple of logical qubits, as in Fig.~\ref{fig:3b}. Once an entire logical qubit has flipped and violates the antiferromagnetic coupling to, say, its left (thus creating a kink), it becomes energetically preferable for the nearest neighbor logical qubit to its right to flip as well, setting off a cascade of logical qubit flips all the way to the end of the chain. 
The inset is the logical Hamming distance histogram, which looks like a condensed version of the physical Hamming distance histogram because it is dominated by these domain wall dynamics. 

Rather than considering the entire final state, Fig.~\ref{fig:4b} integrates the data in Fig.~\ref{fig:4a} and displays the observed occurrence rates of the various classes of errors 
per logical qubit in $\bar{N}=86$ chains. 
The histograms for one, two, and three problem qubits flipping in each location are shown separately. 
Flipped penalty qubits are shown in the inset and are essentially perfectly correlated with $d=3$ errors, meaning that a penalty qubit flip will nearly always occur in conjunction with all problem qubits flipping as well. Thus the penalty qubits function to lock the problem qubits into agreement, as they should (further analysis of the role of the penalty qubit in error suppression is presented in Appendix~\ref{app:penalty-role}).
The overwhelming majority of errors are one or more domain walls between logical qubits. The domains occur with higher probability the closer they are to the ends of the chain, since kink creation costs half the energy at the chain boundaries. The same low barrier to flipping a qubit at the chain ends also explains the large peaks at $d=1$. 


\begin{figure}[t]
\includegraphics[width=0.5\textwidth]{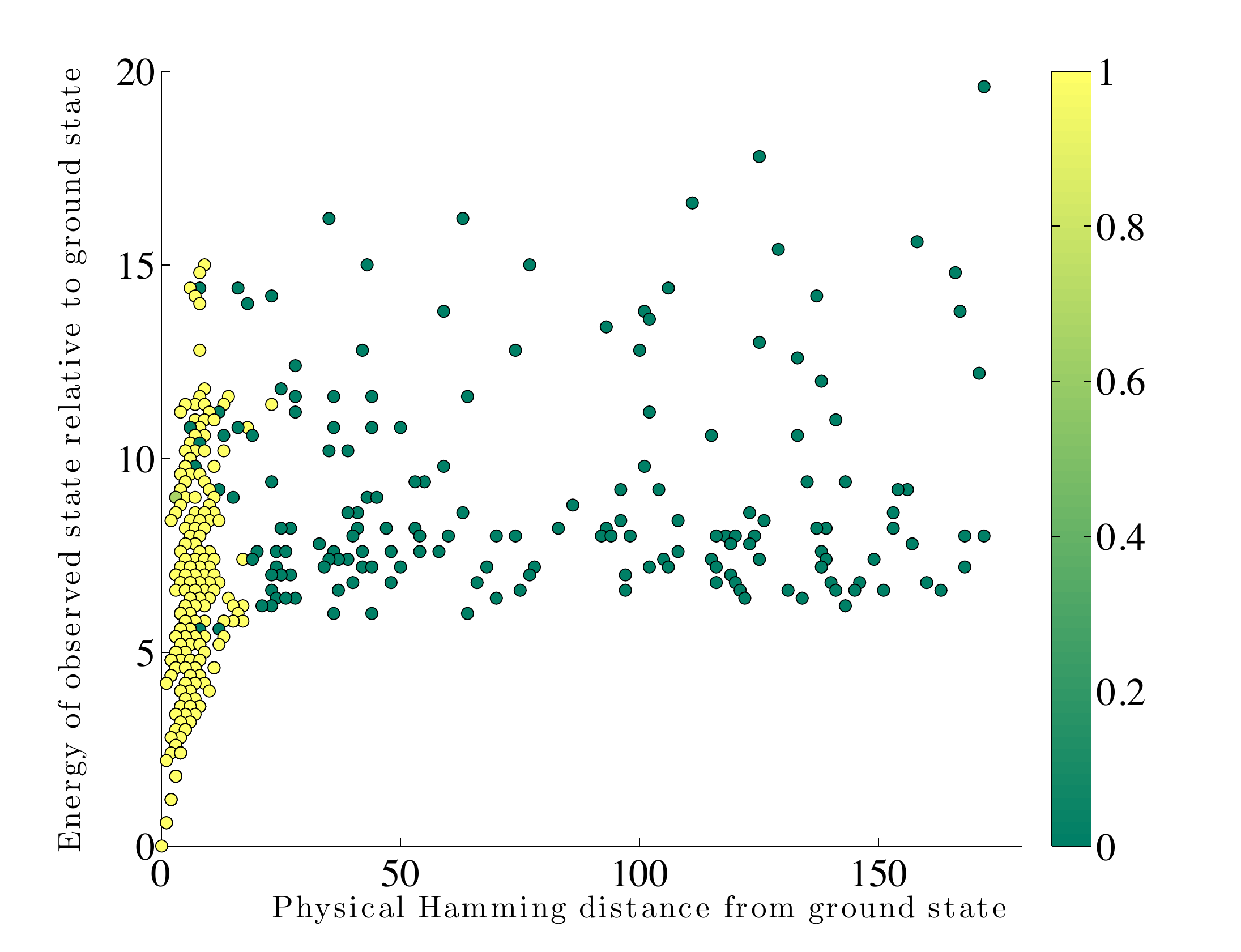}
\caption{\small \textbf{Decodability analysis.} The fraction of decodable states out of all states (color scale) observed at a given Hamming distance from the nearest degenerate ground state (measured in physical qubits), and given energy above the ground state (in units of $J_{ij}=1$), for $\bar{N}=86$ and $\alpha=1$.}
\label{fig:5}
\end{figure}


Our majority-vote decoding strategy correctly decodes errors with $d=1$, incorrectly decodes the much less frequent $d=2$ errors, and is oblivious to the dominant $d=3$ domain wall errors, which present as logical errors. Therefore the preponderance of domain wall errors at large $\bar{N}$ is largely responsible for the drop seen in the QAC data in Fig.~\ref{fig:2}. The two-qubit problem analyzed in Fig.~\ref{fig:3a} and \ref{fig:3b} suggests that logical errors can dominate the low energy spectrum. We observe this phenomenon in Fig.~\ref{fig:5}, which shows that decodable and undecodable states separate cleanly by Hamming distance but not by energy, with many low energy states being undecodable states. In this sense the problem of chains we are studying here is in fact unfavorable for our QAC scheme, and we can expect better performance for computationally hard problems involving frustration.

Figure~\ref{fig:5} lends itself to another interesting interpretation. Quantum annealing is normally understood as an optimization scheme that succeeds by evolving in the ground state, but how much does the energy of the final state matter when we implement error correction? Figure~\ref{fig:5} shows that a small Hamming distance is much more strongly correlated with decodability than the final state energy: the latter can be quite high while the state remain decodable. Thus the decoding strategy tolerates relatively high energy final states.\\

\section{Conclusion and outlook}
This work demonstrates that QAC  can significantly  improve the performance of programmable quantum annealing even for the relatively unfavorable problem of antiferromagnetic chains, which are dominated by logical qubit errors manifested as domain walls. We have shown that (i) increasing the problem energy scale via encoding into logical qubits, (ii) introducing an optimum penalty strength $\beta$ to penalize errors that do not commute with the penalty term, and (iii) decoding the excited states, reduces the overall error rate relative to any strategy that does less than these three steps which comprise the complete QAC strategy. 

The next step is to extend QAC to problems where the correct solution is not known in advance, and is in fact the object of running the quantum annealer. Optimization of the decoding scheme would then be desirable. For example, detected errors could be corrected by solving a local optimization problem, whereby the values of a small cluster of logical qubits that were flagged as erroneous and their neighbors are used to find the lowest energy solution possible. Other decoding schemes could be devised as needed, drawing, e.g., on recent developments in optimal decoding of surface codes \cite{Fowler:2012vn}. Another important venue for future studies is the development of more efficient QAC-compatible codes capable of handling larger weight errors. Ultimately, the scalability of quantum annealing depends on the incorporation of fault-tolerant error correction techniques, which we hope this work will help to inspire.

\acknowledgments
We thank Gerardo Paz for useful discussions. This research was supported by the Lockheed Martin Corporation, by ARO-MURI grant W911NF-11-1-0268, by ARO-QA grant number W911NF-12-1-0523, and and by NSF grant numbers PHY-969969 and PHY-803304.

\clearpage

\appendix

\section{Hardware parameters of the DW1 and DW2}
\label{app:Chimera}

The experiments described in the main text were performed on the D-Wave Two (DW2) ``Vesuvius" processor at the Information Sciences Institute of the University of Southern California. The device has been described in detail elsewhere \cite{harris_experimental_2010_1,Harris2010,berkley2010scalable}. The annealing functions $A(t)$ and $B(t)$ are specified below. All our results were averaged over $24$ embeddings of the chains on the processor (except U in Fig.~\ref{fig:2d}, which used $188$ embeddings), where an embedding assigns a specific set of physical qubits to a given chain. After programming the couplings, the device was cooled for $10\, \textrm{ms}$, and then $5000$ annealing runs per embedding were performed using an annealing time of $t_f=20\,\mu \textrm{s}$ for every problem size $N,\overline{N}\in\{2,3,\dots,86\}$, $\alpha\in\{0.1,0.2,...,1.0\}$ and $\beta\in\{0.1,0.2,...,1.0\}$. Annealing was performed at a temperature of $17\, \textrm{mK}$ ($\approx 2.2\, \textrm{GHz}$), with an initial transverse field starting at $A(0)\approx 33.8\, \textrm{GHz}$, going to zero during the annealing, while the couplings are ramped up from near zero to $B(t_f)\approx 20.5\, \textrm{GHz}$. 

\begin{figure*}[t]
\begin{center}
\subfigure[\, DW1]{\includegraphics[width=0.49\textwidth]{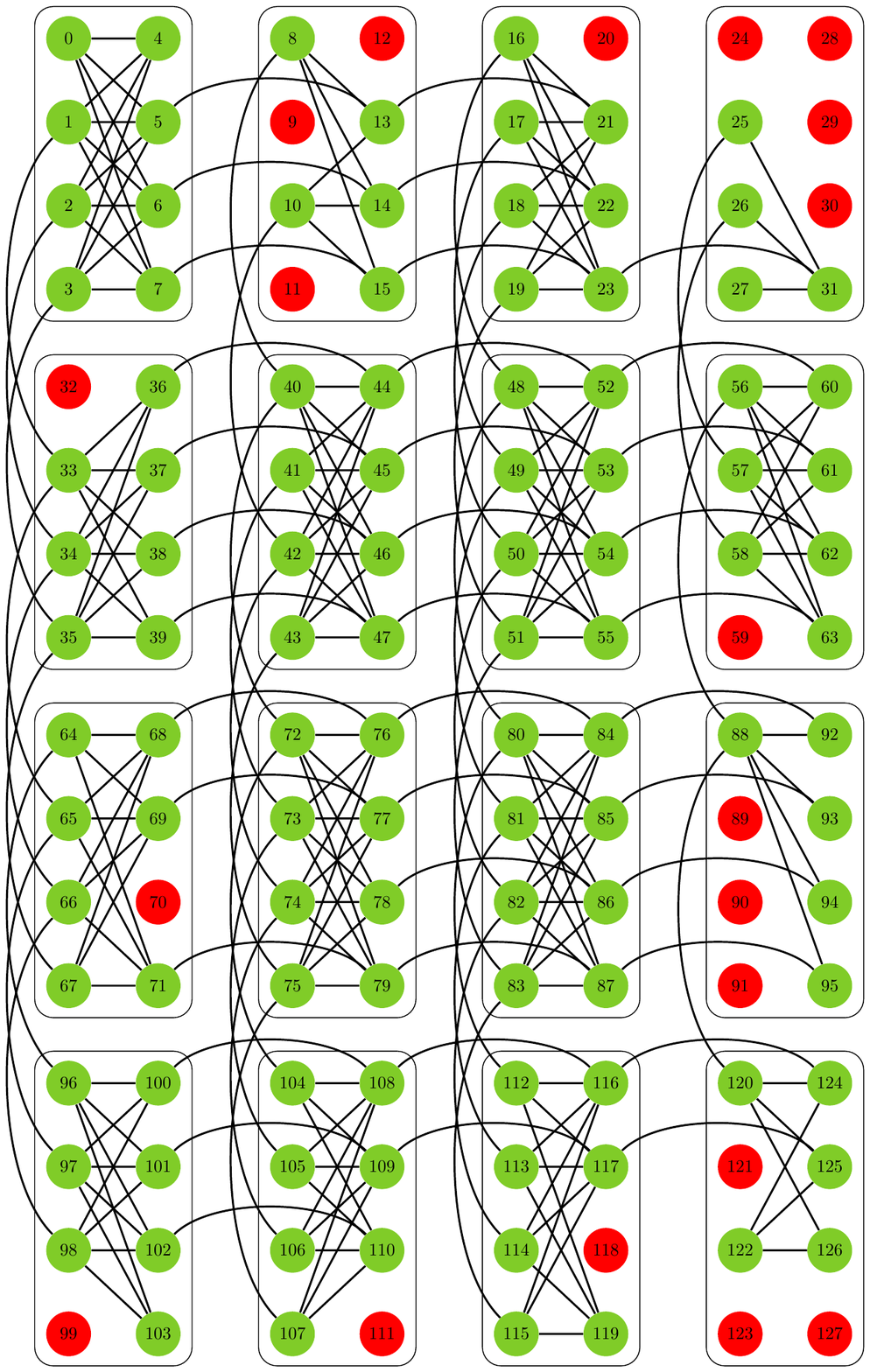}\label{fig:DW1}} 
\subfigure[\, DW2]{\includegraphics[width=0.49\textwidth]{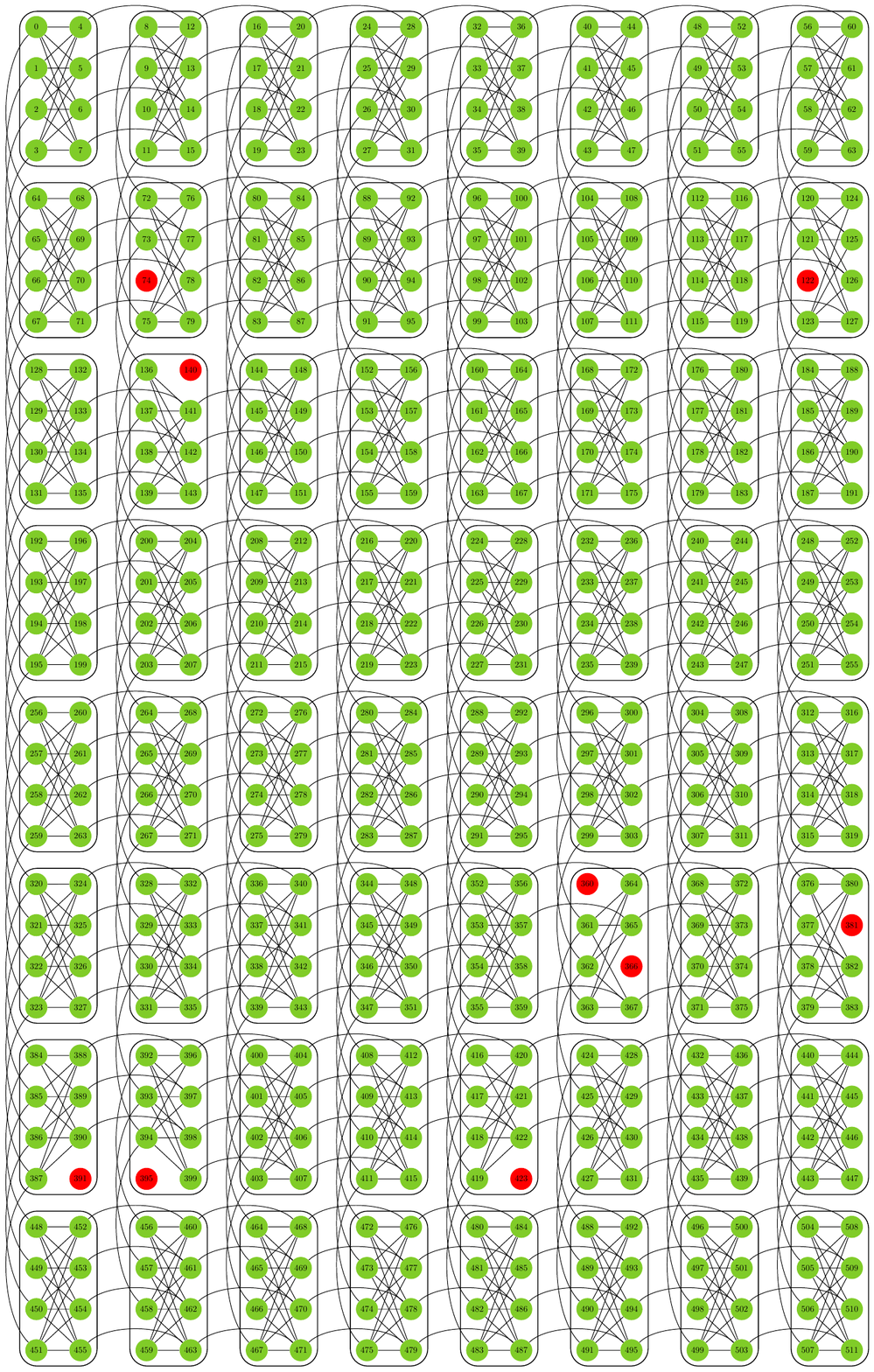}\label{fig:DW2}} 
\end{center}
\caption{The connectivity graph of the D-Wave One (DW1) ``Rainier" processor shown on the left consists of $4\times4$ unit cells of eight qubits (denoted by circles), connected by programmable inductive couplers (lines). The $108$ green (red) circles denote functional (inactive) qubits. Most qubits connect to six other qubits. The D-Wave Two (DW2) Vesuvius processor shown on the right consists of $8\times8$ unit cells. The $503$ green (red) circles denote functional (inactive) qubits. In the ideal case, where all qubits are functional and all couplers are present, one obtains the non-planar ``Chimera" connectivity graph.}
\label{fig:Chimera}
\end{figure*}

\begin{figure}[t]
\begin{center}
\includegraphics[width=.48\textwidth]{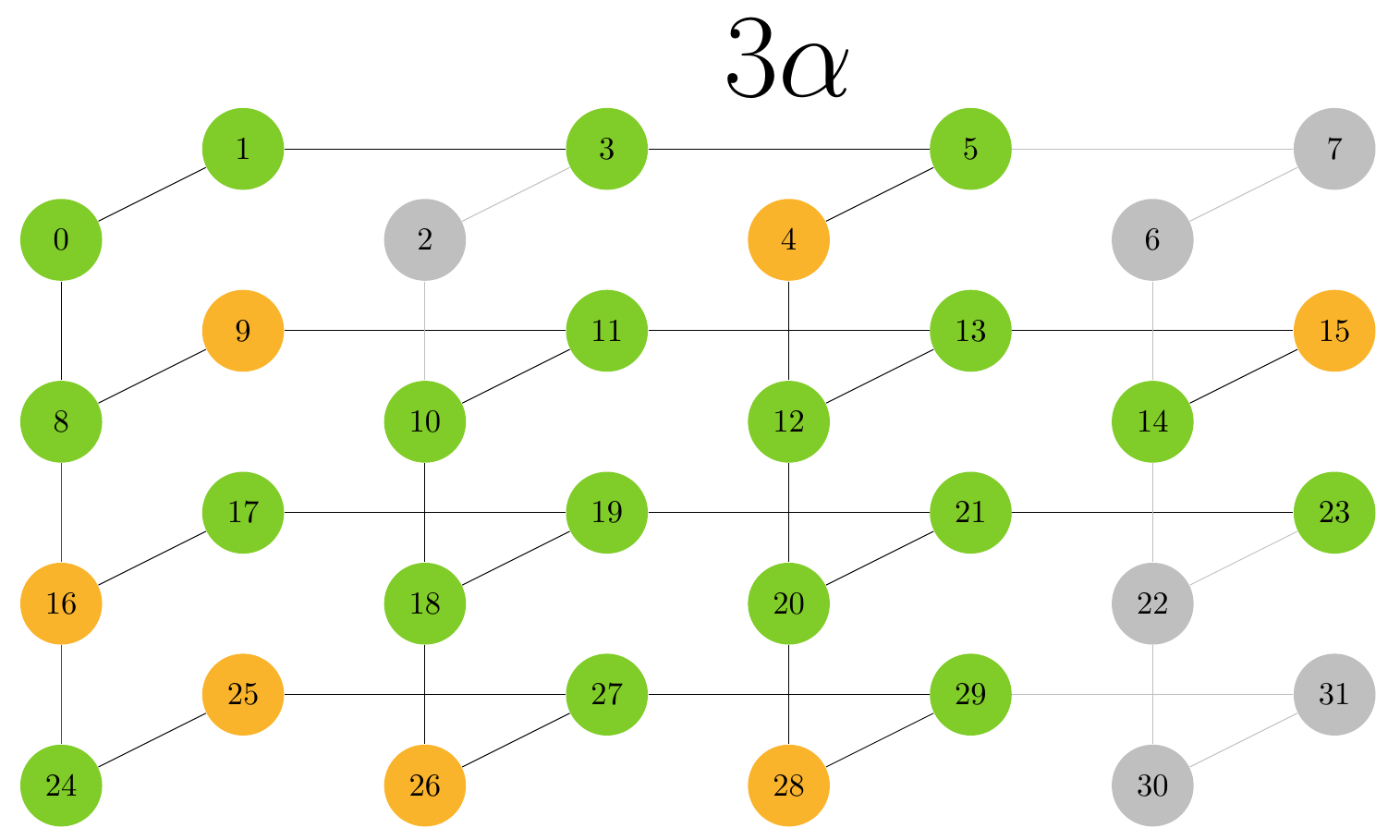}
\end{center}
\caption{The encoded DW1 graph. Symbols are the same as Fig.~\ref{fig:1}(c) in the main text for the encoded DW2 graph. In addition, grey circles represent logical qubits with defective data qubits that were not used in our experiments; the corresponding grey lines represent couplings that were not used. }
\label{fig:enc-Rainier}
\end{figure}

\begin{figure}[t]
\begin{center}
\subfigure[\, DW1]{\includegraphics[width=0.49\textwidth]{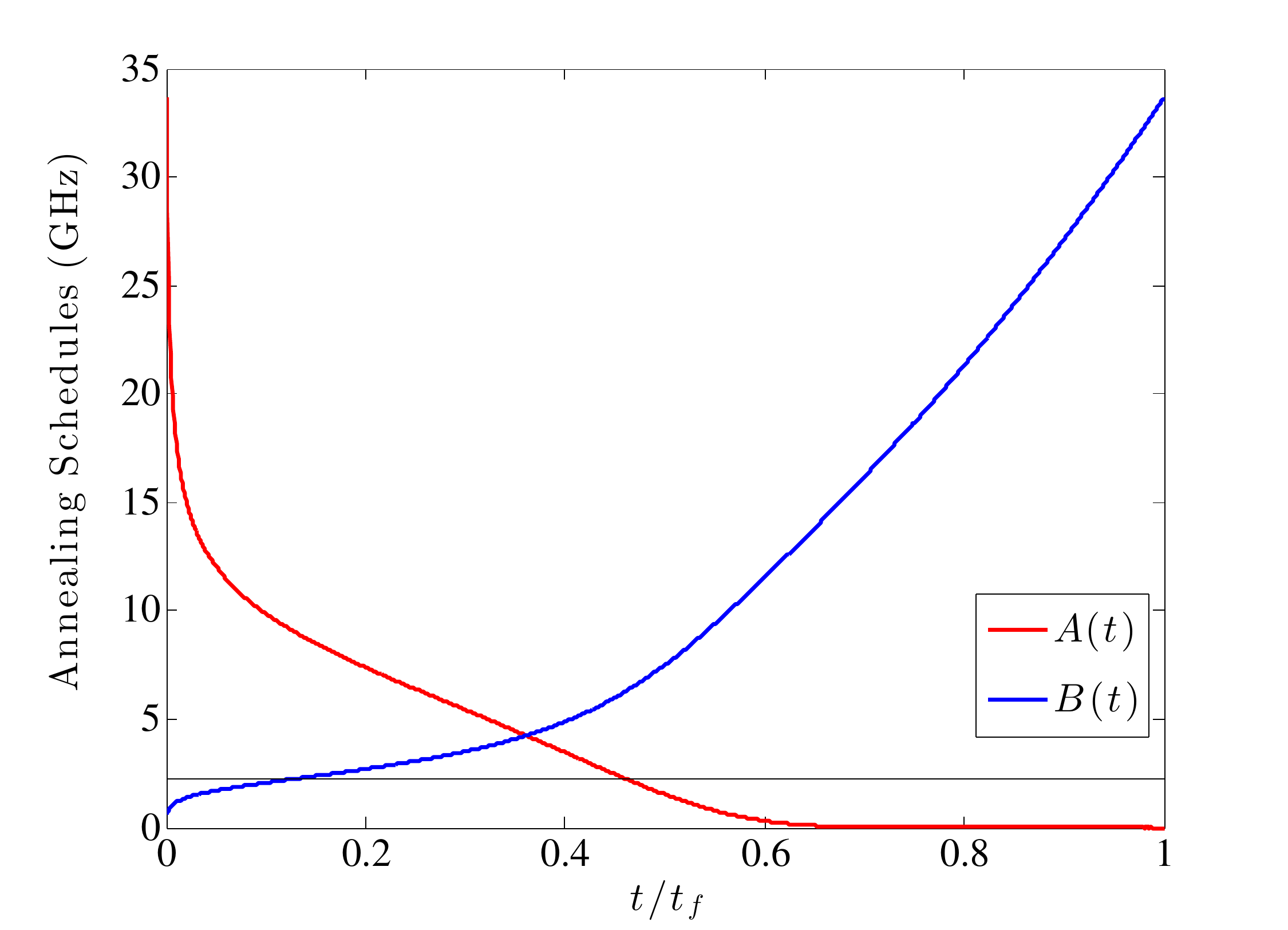}\label{fig:DW1-schedule}}
\subfigure[\, DW1]{\includegraphics[width=0.49\textwidth]{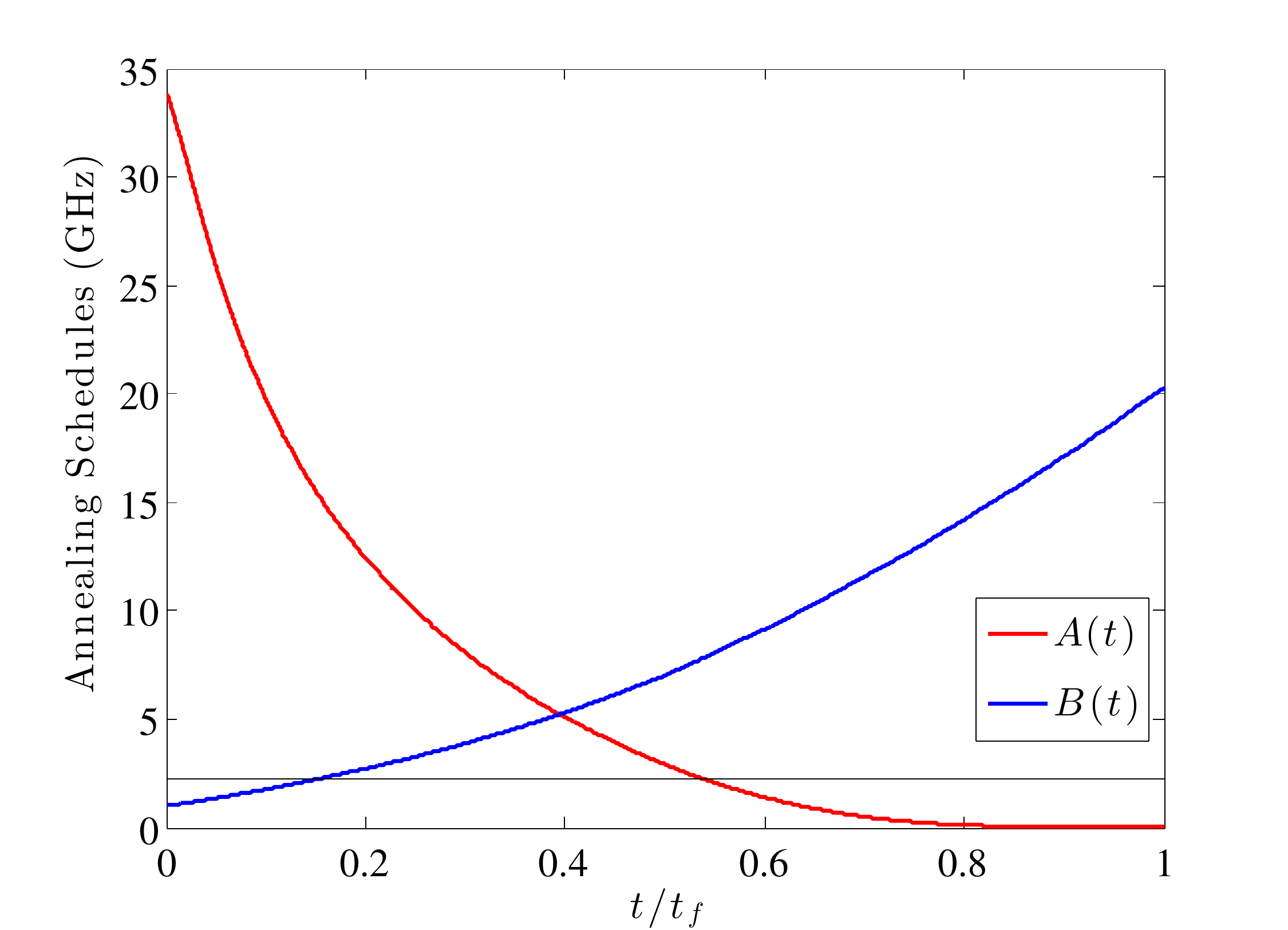}\label{fig:DW2-schedule}}
\end{center}
\caption{The DW1 (a) and DW2 (b) annealing schedules. The functions $A$ and $B$ are the ones appearing in Eqs.~\eqref{eqt:QA} and \eqref{eq:HencP}.  The solid horizontal black line is the operating temperature energy.}
\label{fig:anneal-schedule}
\end{figure}

The D-Wave processors are organized into unit cells consisting of eight qubits arranged in a complete, balanced bipartite graph, with each side of the graph connecting to a neighboring unit cell, as seen in Fig.~\ref{fig:Chimera}, known as the ``Chimera" graph \cite{Choi1,Choi2}. The D-Wave One (DW1) ``Rainier" processor is the predecessor of the DW2, and was used in our early experiments. The minimum DW1 annealing time is $5\, \mu\textrm{s}$, compared to $20\, \mu\textrm{s}$ for the DW2. The analog part of the DW2 circuitry is very similar to the DW1 processors other than improved qubit parameters (lower inductance and capacitance, shorter qubit length, higher critical current). The DAC (digital to analog) and readout technologies were also improved (non-dissipative readout scheme instead of dc SQUIDs). The DW2 has an XYZ addressing scheme (to eliminate static power dissipation when programming), and a smaller spread on coupling strength of key on-chip control transformers allowing better synchronization between the $h$ and $J$ parameters \cite{Trevor}. The reduction in control noise sources is noticeable in our experimental data, as shown in Sec.~\ref{app:more-fidelity-results}. Figure~\ref{fig:enc-Rainier} shows the encoded hardware graph used in our DW1 experiments. The annealing schedules for the DW1 and DW2 are shown in Fig.~\ref{fig:anneal-schedule}.

Details for the DW1 experiments are as follows (for the analogous DW2 details see Methods in the main text). After programming the couplings, the device was cooled for $1.5\, \textrm{s}$
and then $20,000$ annealing runs per embedding were performed using an annealing time of $t_f=5\, \mu \textrm{s}$ 
for every problem size $N,\overline{N}\in\{2,3,\dots,16\}$. Only a single embedding was used. Error bars in all our DW1 plots were calculated over the set of $S=20,000$ annealing runs and express the standard error of the mean $\sigma/\sqrt{S}$, where $\sigma^2 = \frac{1}{S}\sum_{i=1}^S(x_i-\overline{x})^2$ is the sample variance.

\begin{figure*}
\begin{center}
\subfigure[\, ]{\includegraphics[width=0.2\textwidth]{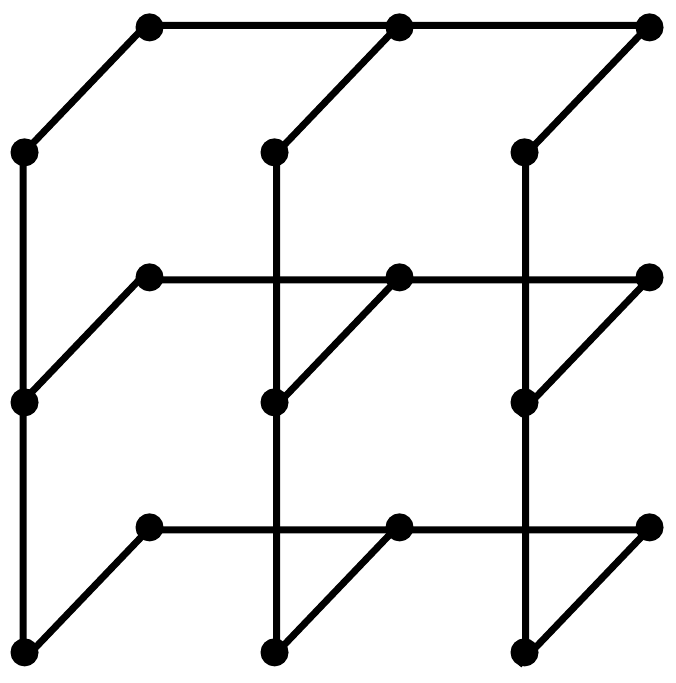}
		\label{fig:planarity1}}
\subfigure[\, ]{		\includegraphics[width=0.2\textwidth]{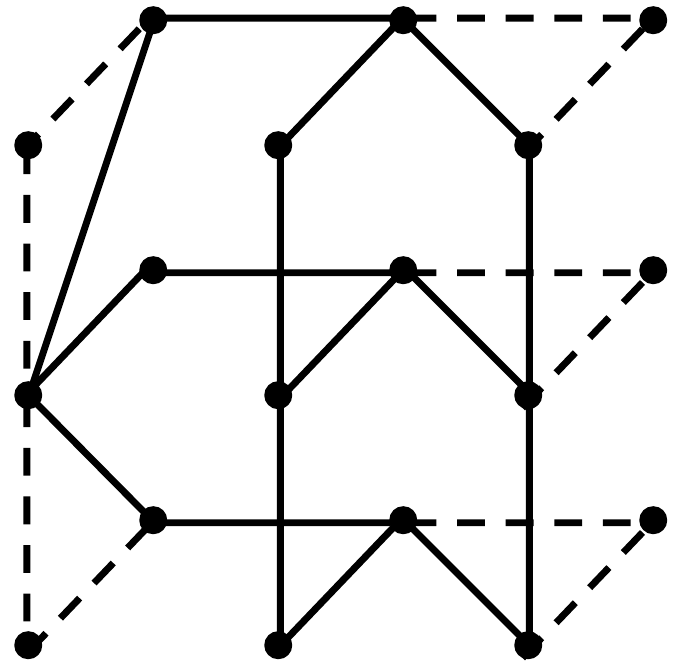}
		\label{fig:planarity2}}
\subfigure[\, ]{		\includegraphics[width=0.2\textwidth]{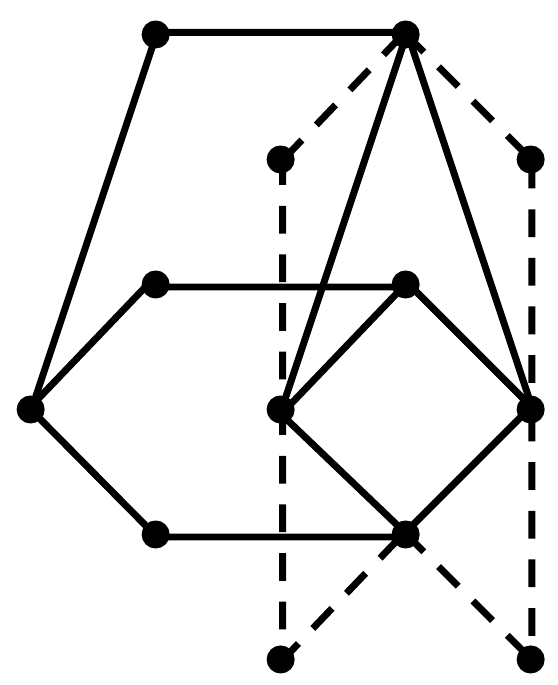}
		\label{fig:planarity3}}
\subfigure[\, ]{		\includegraphics[width=0.2\textwidth]{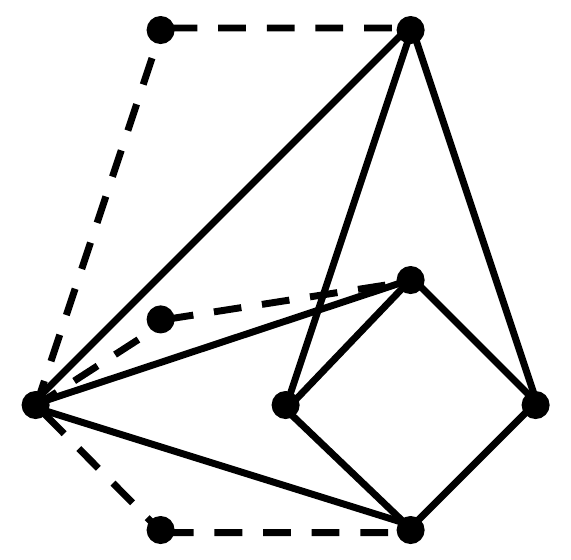}
		\label{fig:planarity4}}
\end{center}
\caption{(a) A portion of the encoded graph over logical qubits. (b)-(d) Contraction of paths in the original graph into edges. Paths consisting of two edges and a vertex are selected (represented in the figure as dotted lines), then contracted into a single edge connecting the ends of the chosen path (shown as a new solid line).}
\label{fig:planarity_moves}
\end{figure*}

\begin{figure}
\subfigure[\, ]{		\includegraphics[width=0.2\textwidth]{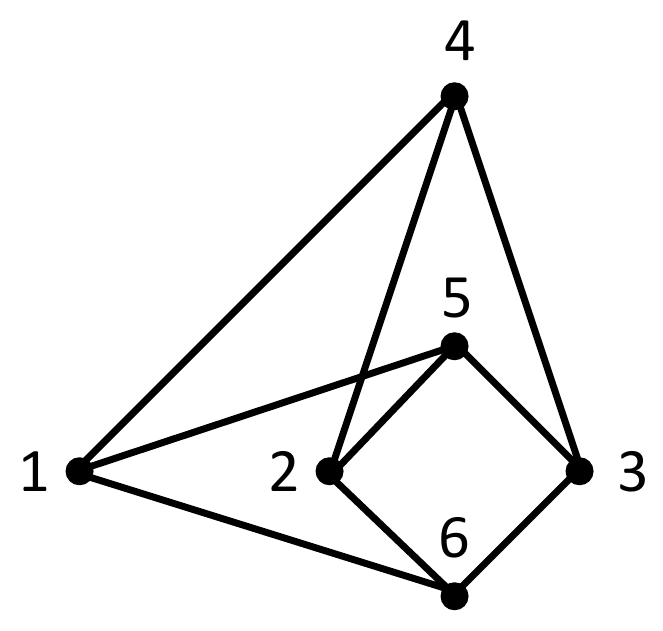}
		\label{fig:planarity5}}
\subfigure[\, ]{		\includegraphics[width=0.1\textwidth]{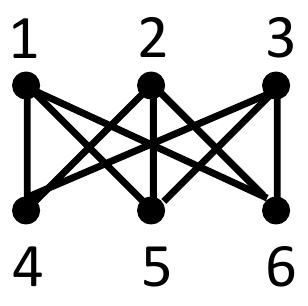}
		\label{fig:planarity6}}
\caption{The condensed graph (a) is isomorphic to the standard representation of the $K_{3,3}$ bipartite graph (b).}
\label{fig:bipartite_labeling}
\end{figure}

\section{Proof that the encoded graph is non-planar}
\label{app:non-planarity}

The solution of the Ising model over the encoded graph over the processor, shown in Fig.~\ref{fig:1}  in the main text, is an NP-hard problem, just as the same problem over the original hardware graph is NP-hard. The key lies in the three-dimensional nature of both graphs; the ground state of Ising spin glasses over non-planar lattices is an NP-hard problem \cite{Barahona1982}.

We provide a graphical proof of non-planarity for the encoded graph here. The existence of a subgraph homeomorphic to the $K_{3,3}$ complete bipartite graph with three vertices on each side is sufficient to prove that a given graph is non-planar \cite{Boyer}. This subgraph may take as its edges paths within the graph being studied. We take a section of the encoded graph and, by performing a series of allowed moves of condensing paths to edges, show that the section is indeed homeomorphic to $K_{3,3}$.

We begin with an 18-qubit section of the regular encoded graph, shown in Fig.~\ref{fig:planarity_moves}(a).
This encoded graph is then condensed along its paths by repeatedly removing two edges and a vertex and replacing them with a single edge representing the path. A clear sequence of these moves is shown in Fig.~\ref{fig:planarity_moves}(b)-(d).
The studied subgraph has now been condensed into the form of the desired $K_{3,3}$ graph, as is made clear by labeling and rearranging the vertices as in Fig.~\ref{fig:bipartite_labeling}. The encoded graph is thereby proved non-planar.

\section{A classical independent errors model}
\label{app:indep-err}

It is tempting to explain the decay of success probability seen in Fig.~\ref{fig:2} in the main text in terms of a classical model of uncorrelated errors. To this end, consider an antiferromagnetic chain of $N$ spins, with Hamiltonian $H_{\textrm{Ising}} = \alpha \sum_{i=1}^{N-1} \sigma_i^z \sigma_{i+1}^z$. Its excitations are kinks, resulting in domain walls. A kink is a single disagreement between nearest-neighbor spins. There are $N-1$ possible kink locations since there are $N-1$ nearest-neighbor spin pairs. Each kink costs an energy of $2\alpha$. The ground state energy is $-\alpha (N-1)$ and so the energy of a state with $k$ kinks is $E_N(k) = \alpha  (-N+1+2k)$. The degeneracy $d_N(k)$ of a state with $k$ kinks is the number of ways of placing $k$ kinks in $N-1$ slots, i.e.,  $d_N(k) = \binom{N-1}{k}$. Assuming a classical system in thermal equilibrium at temperature $k_B T$, the probability of a state with $k$ kinks is
\beq
P_N(k) = \frac{1}{Z_N} d_N(k) e^{-E_N(k)/k_B T}\ ,
\eeq
where $Z_N = \sum_{k=0}^{N-1}P_N(k)$ is the normalization factor. We have
\bes
\begin{align}
Z_N &=  \sum_{k=0}^{N-1} \binom{N-1}{k} e^{- \alpha  (-N+1+2k)/k_B T} \\
&= e^{\alpha (N-1)/k_B T}\sum_{k=0}^{N-1} \binom{N-1}{k} \left( e^{-2\alpha/k_B T}\right)^k \\
& = (2\cosh(\alpha/k_B T))^{N-1} .
\end{align}
\ees
Thus the probability of no kinks, i.e., no errors in the chain, is
\bes
\begin{align}
P_N(0) &= \frac{e^{-\alpha (N-1)/k_B T}}{(2\cosh(\alpha/k_B T))^{N-1}}  = \left(\frac{1}{1+e^{2\alpha/k_B T}}\right)^{N-1} \\
& \equiv (1-p(\alpha))^{N-1}\ ,
\end{align}
\ees
whence we identify the bit-flip probability from the main text as
\beq
p(\alpha) = \frac{1}{1+e^{-2\alpha/k_B T}}\ .
\label{eq:p-alpha}
\eeq

This purely classical thermal error model predicts an exponential fidelity decay with $N$, that does not agree well with our data. Indeed, the success probability depends quadratically on $N$ for small chain lengths, whence the Lorentzian fits shown in Fig.~\ref{fig:2} in the main text.  This suggests that a different error mechanism is at work, which we can capture using an adiabatic master equation presented in Sec.~\ref{app:ME}. 

\section{Comparison between the DW1 and DW2}

\label{app:more-fidelity-results}
\begin{figure}[t]
\begin{center}
\includegraphics[width=.48\textwidth]{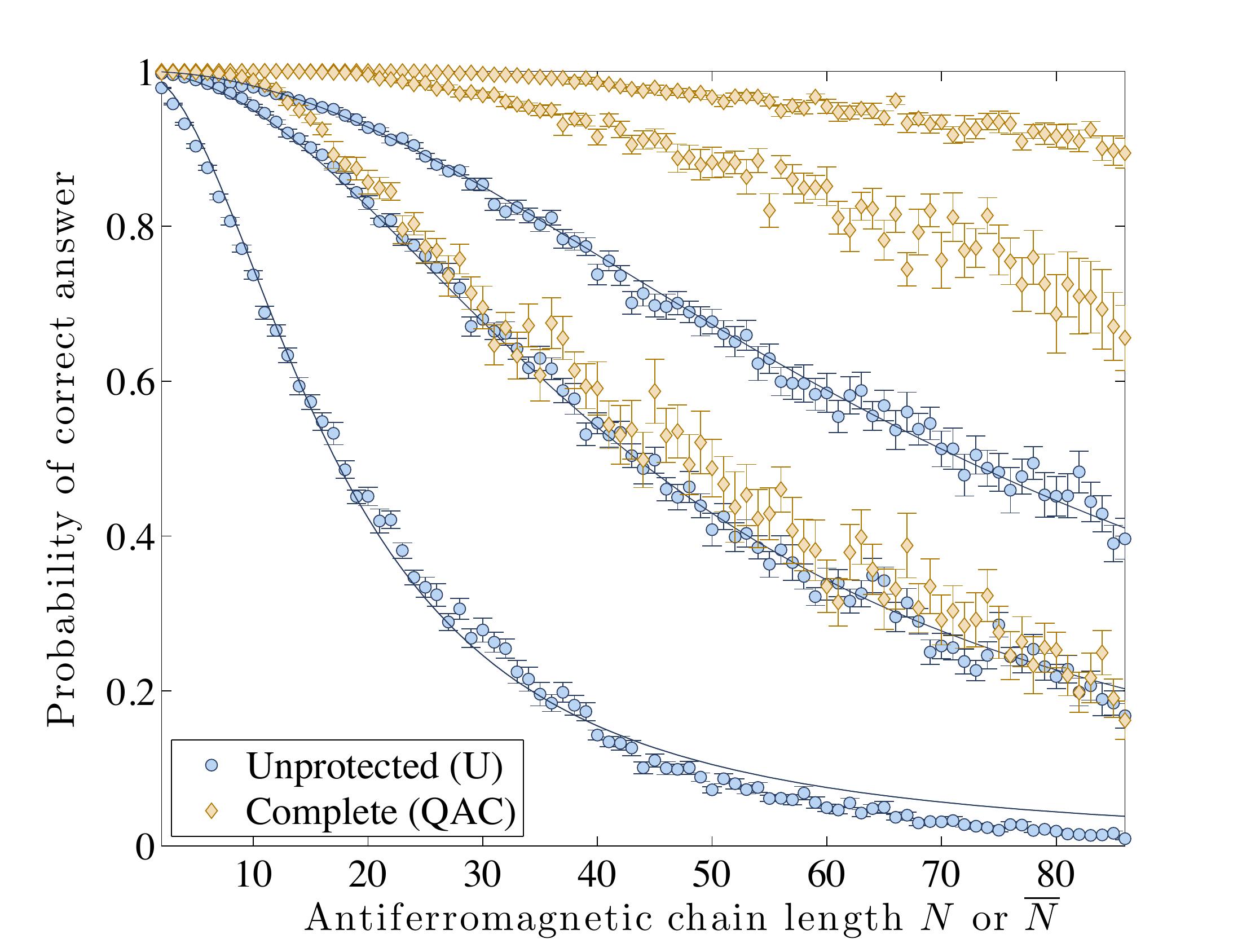}
\end{center}
\caption{Recap of the results shown in Fig.~\ref{fig:2} in the main text, for the U and QAC cases, for $\alpha = 1$ (top) $\alpha = 0.6$ (middle), and $\alpha = 0.3$ (bottom).}
\label{fig:ves_results_comparison}
\end{figure}

In the main text we provided DW2 results at various values of $\alpha$. Figure~\ref{fig:ves_results_comparison} collects these results for the U and QAC cases, at three different values of $\alpha$, to simplify the comparison. It clearly demonstrates the advantage of QAC over the U case at every value of the problem scale $\alpha$, and shows how increasing $\alpha$ increases the success probability.

\begin{figure*}[ht]
\begin{center}
\subfigure[\, DW1, $\alpha=0.3$.]{\includegraphics[width=3.0in]{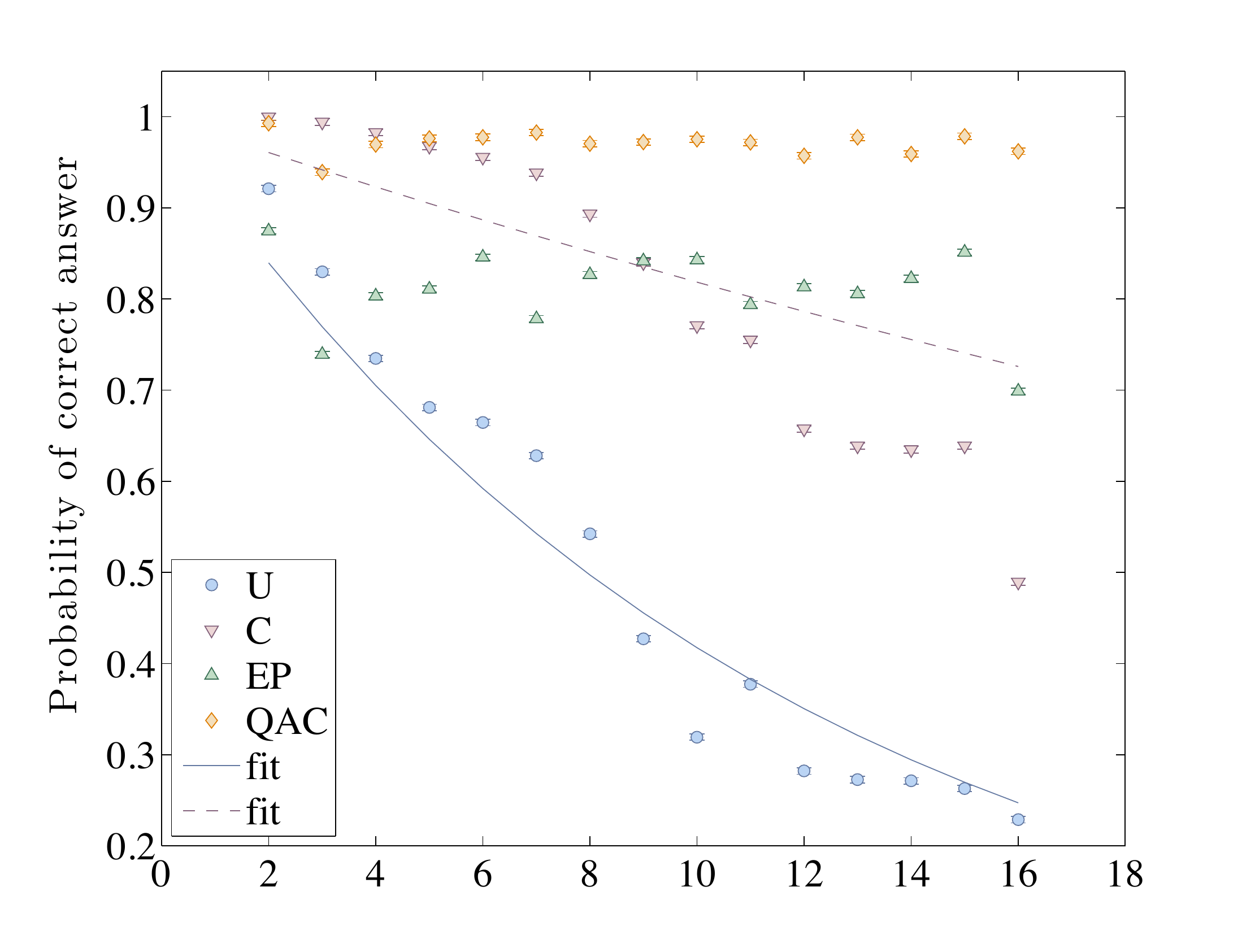}  \label{fig:energy_03}} \hspace{0.5cm}
\subfigure[\, DW1, $\alpha=0.6$.]{\includegraphics[width=3.0in]{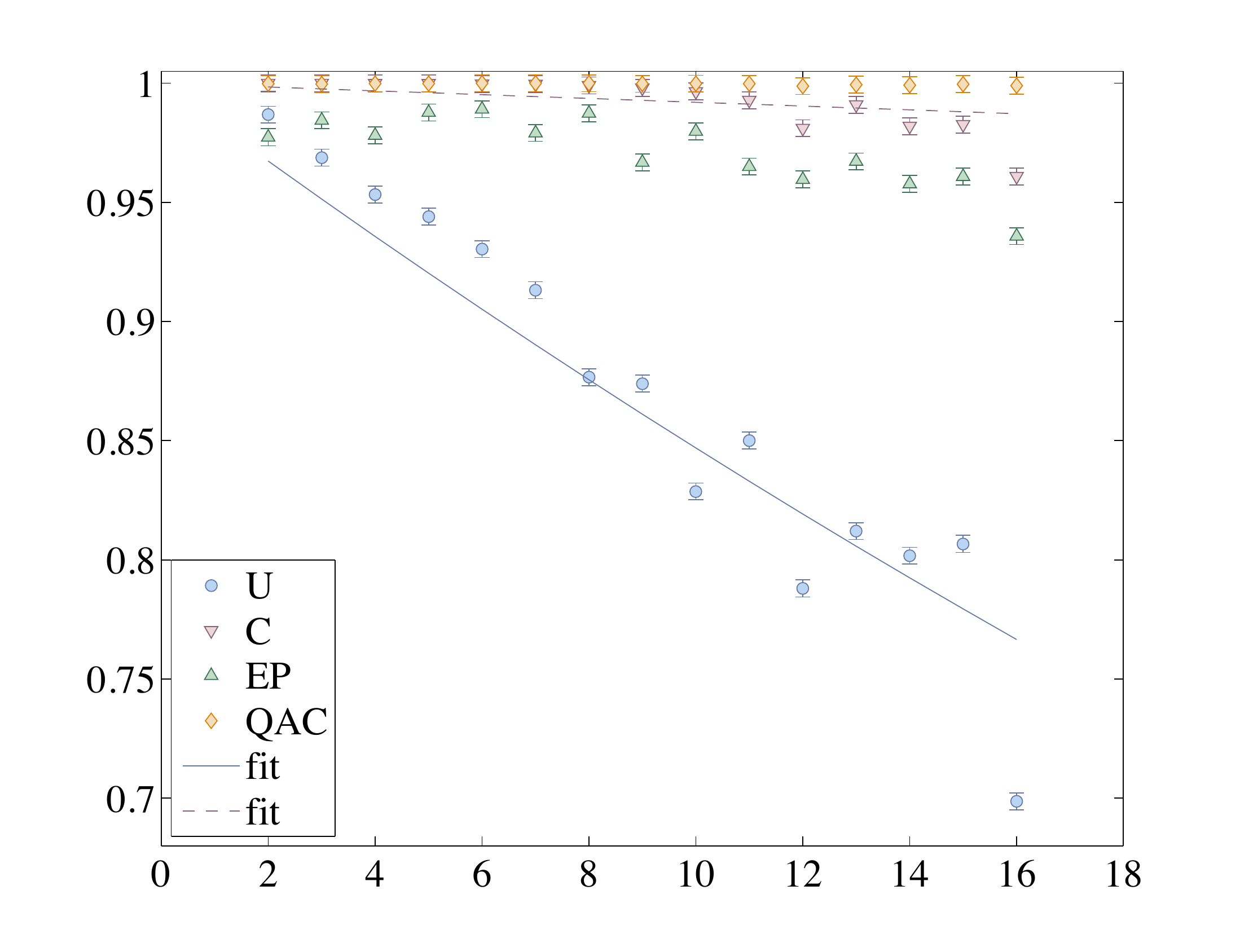}  \label{fig:energy_06}} \hspace{0.5cm}
\subfigure[\, DW1, $\alpha=0.9$.]{\includegraphics[width=3.0in]{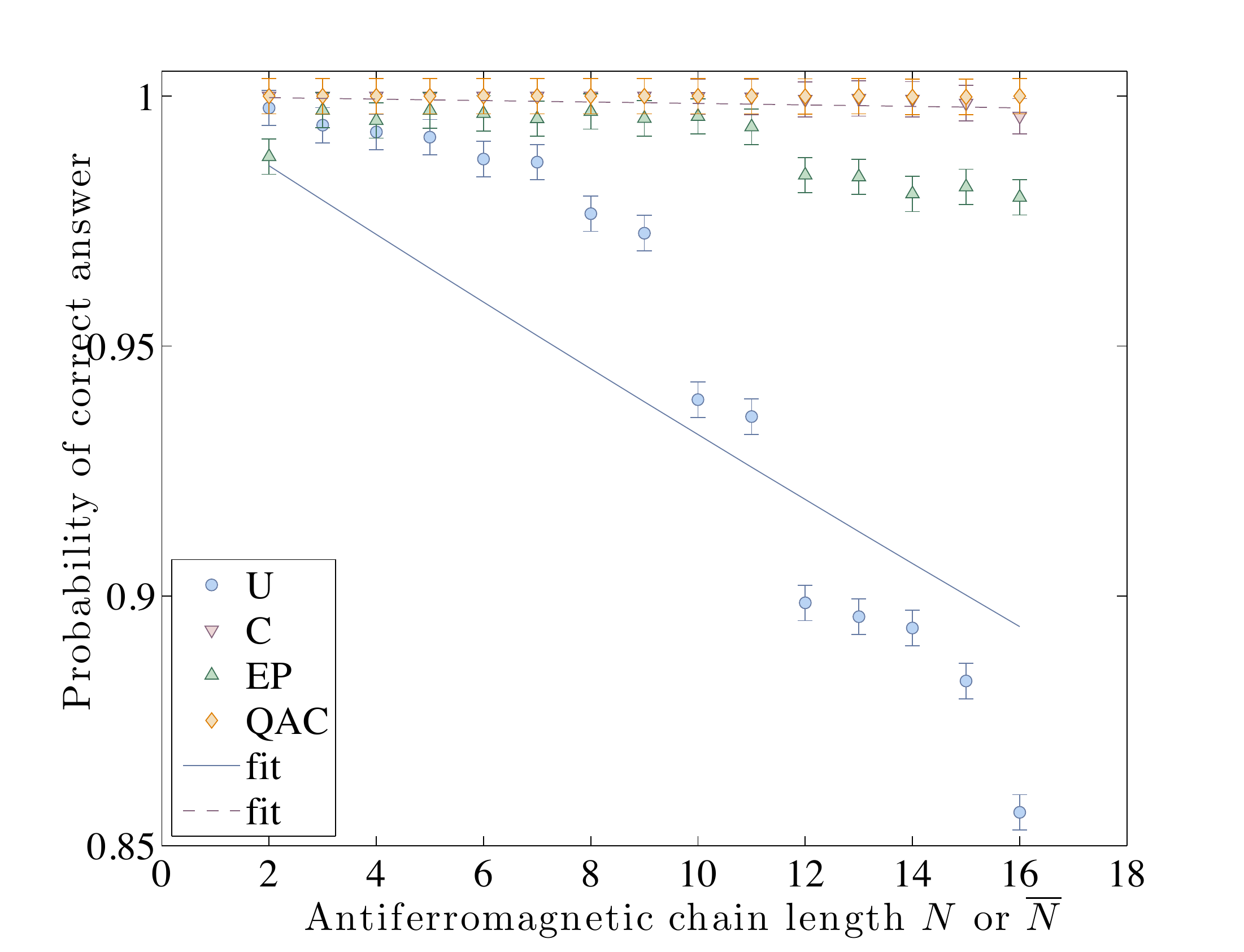} \label{fig:energy_09}}   
\subfigure[\, DW2, $\alpha=0.9$.]{\includegraphics[width=3.0in]{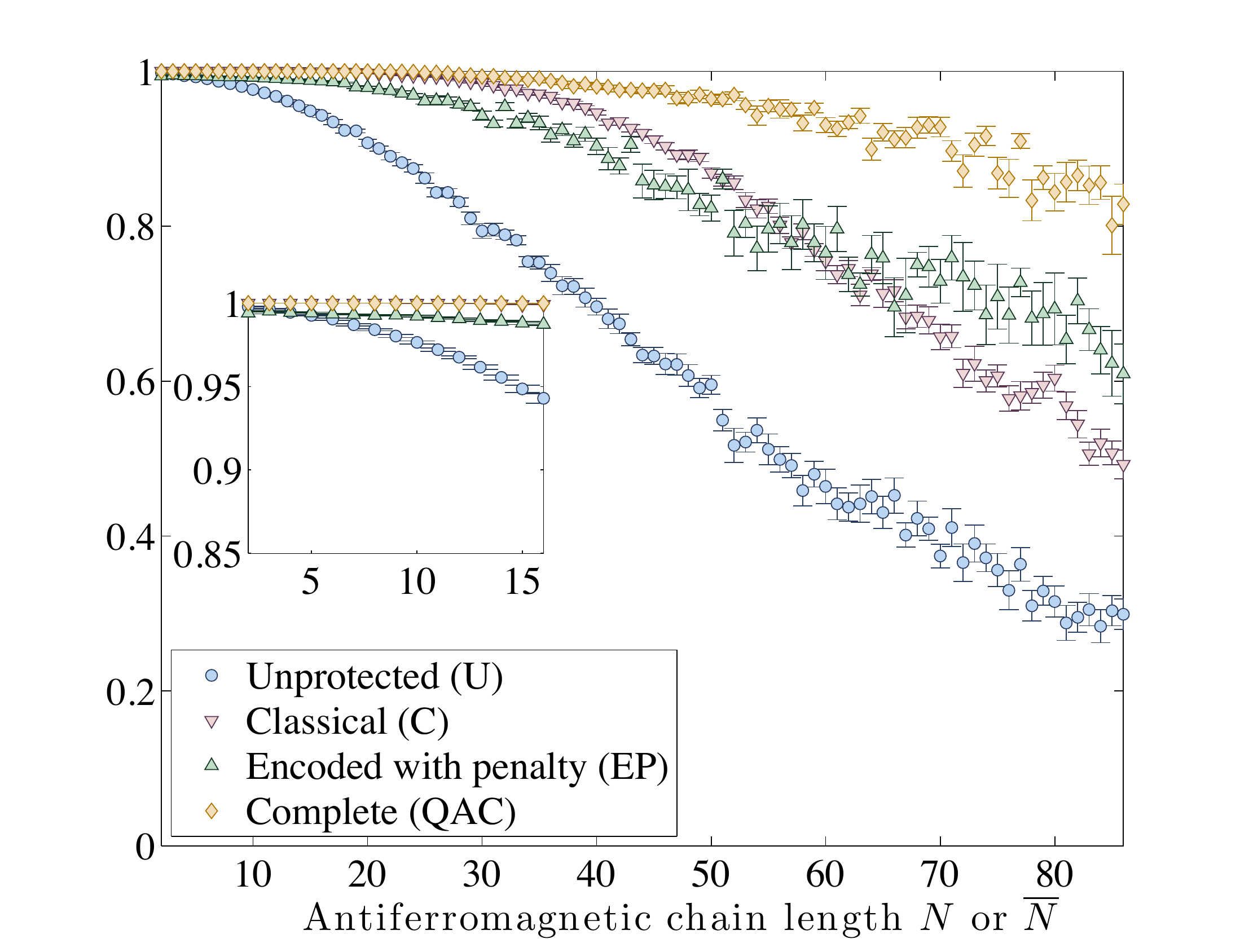} \label{fig:energy_09-DW2}}   
\end{center}
\caption{\small Panels (a)-(c) are DW1 results, panel (d) is DW2 results. Shown are success probabilities for the various cases discussed in the main text as a function of chain length $N,\bar{N}\in{2,...,16}$, complementing the results shown in Fig.~\ref{fig:2}. Blue and purple lines are best fits to an independent errors model as described in Sec.~\ref{app:indep-err}. Panel (d) shows an additional DW2 data set, at $\alpha=0.9$. The inset shows the same data on the same scale as in panel (c), to emphasize the improved performance of the DW2 over the DW1. The penalty scale $\beta$ was chosen as the experimental optimum for each $\alpha$ and $\bar{N}$ in all cases.}
\label{fig:large_a_enc_success}
\end{figure*}

Figure~\ref{fig:large_a_enc_success} complements the DW2 data with results from the DW1, and an additional DW2 data set at $\alpha=0.9$, alongside the corresponding DW1 data. The DW1 results are seen to agree overall with those from the DW2, though the absence of penalty qubits in up to three logical qubits in the DW1 (as explained in the caption of Fig.~\ref{fig:6}), and larger control errors, resulted in lower success probabilities overall.  Moreover, the DW1 accommodated at most $16$ logical qubits, so some of the larger scale trends observed in the DW2 case, such as the eventual decline of the QAC success probability as seen in Fig.~\ref{fig:2}, are not visible in the DW1 data. 

\begin{figure*}[t]
\begin{center}
\subfigure[\, DW1]{\includegraphics[width=0.48\textwidth]{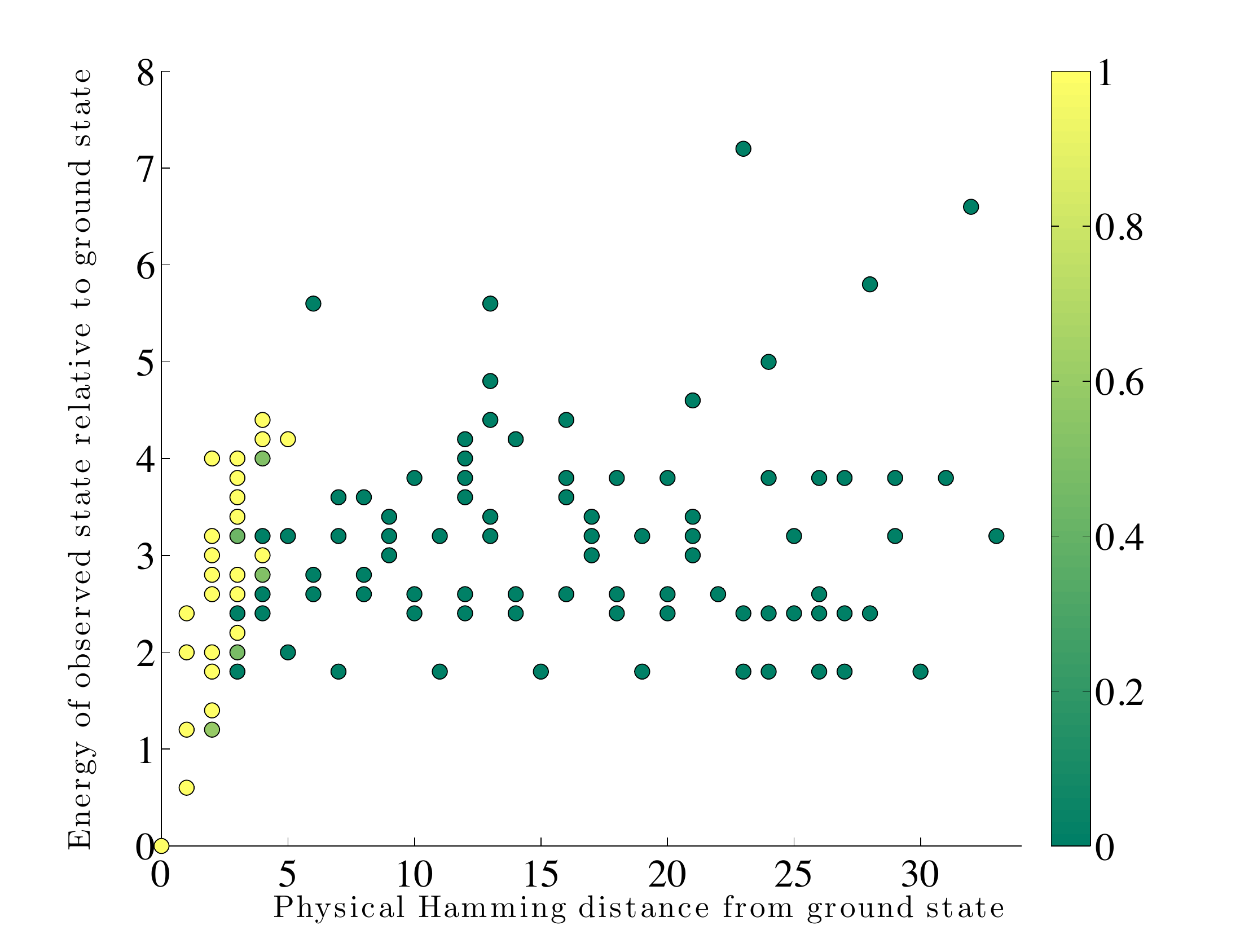}\label{fig:hamming-energy-decodable-DW1}} 
\subfigure[\, DW2]{\includegraphics[width=0.48\textwidth]{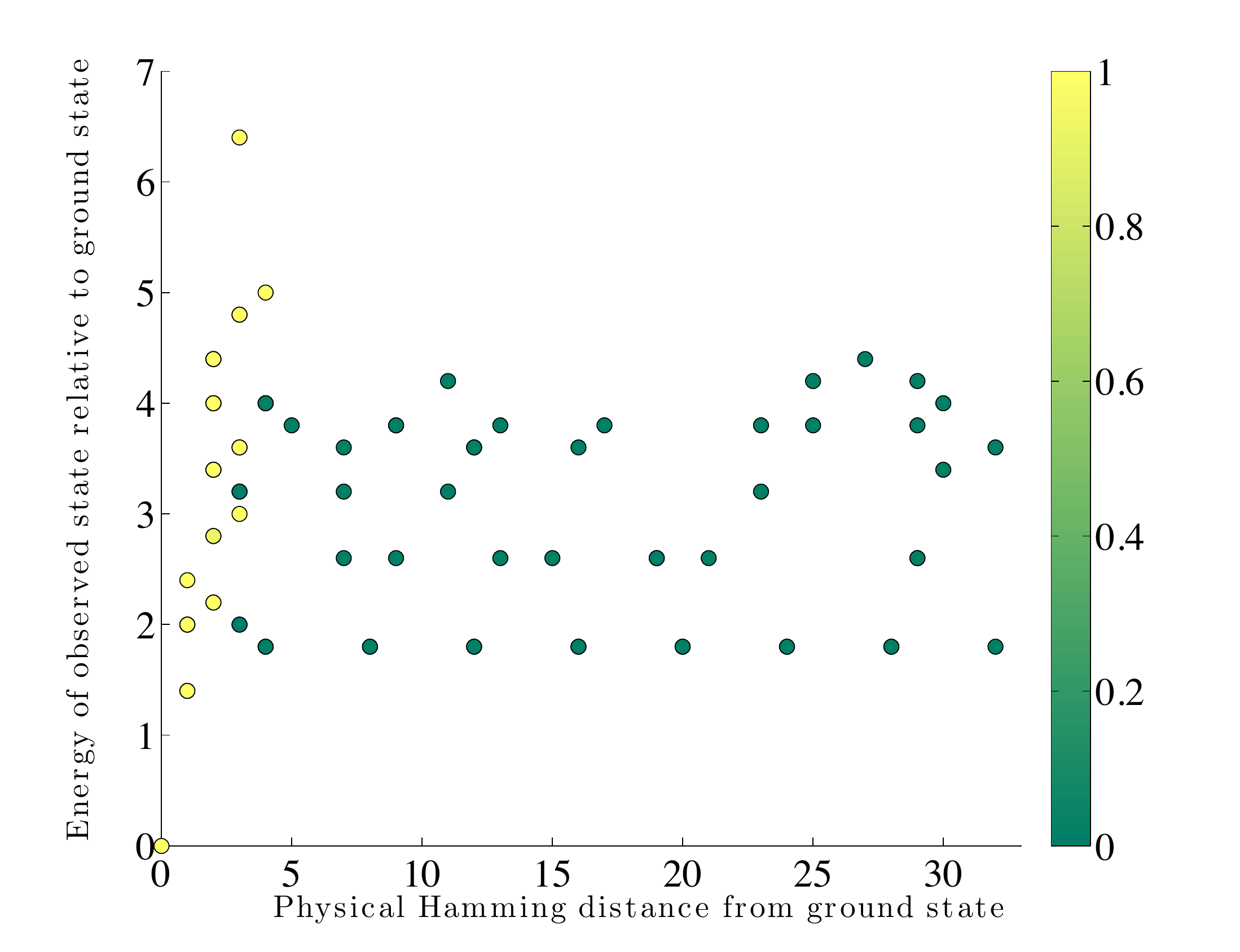} \label{fig:hamming-energy-decodable-DW2}}
\end{center}
\caption{\small The fraction of decodable states (color scale), for $\bar{N}=16$, $\alpha=0.3$ and $\beta=0.4$ (optimal for DW1), \textit{vs} the energy (in units of $J_{ij}=1$) of each observed state relative to the ground state and the Hamming distance from the nearest degenerate ground state, measured in physical qubits. 
Here we compare the DW1 (a) and the DW2 (b);  $\bo=0.2$ for DW2.
}
\label{fig:HD-energy}
\end{figure*}

In Fig.~\ref{fig:HD-energy} we show the correlation between Hamming distance, energy, and decodability, for the DW1 and the DW2 for chains of length $16$. This complements Fig.~\ref{fig:5} in the main text. In the DW2 plot, we see the points representing domain walls spaced at even four-qubit intervals because there are no missing penalty qubits. As can be seen from the color variation, the distribution for the DW1 is less bimodal than that of the DW2. In both cases states with physical Hamming distance less than $5$ from the ground state show decodability while states further away do not.

\section{Adiabatic master equation}
\label{app:ME}

In order to derive the adiabatic Markovian master equation used in performing the simulations, we consider a closed system with Hamiltonian
\beq
H(t) =H_S(t) + H_B +  g \sum_{\alpha} A_{\alpha} \otimes B_{\alpha} \ ,
\eeq
where $H_S(t)$ is the time-dependent system Hamiltonian (which in our case takes the form given in Eq.~\eqref{eqt:QA}), $H_B$ is the bath Hamiltonian, $\{A_{\alpha}\}$ are Hermitian system operators, $\{B_{\alpha}\}$ are Hermitian bath operators, and $g$ is the system-bath interaction strength (with dimensions of energy).  Under suitable approximations, a master equation can be derived from first principles \cite{ABLZ:12} describing the Markovian evolution of the system.  This equation takes the Lindblad form \cite{springerlink:10.1007/BF01608499}:
\begin{eqnarray}
\frac{d}{dt} \rho(t) \ &&= - \frac{i}{\hbar} \left[H_S(t) + H_{\textrm {LS}}(t), \rho(t) \right] + \frac{g^2}{\hbar^2} \sum_{\alpha,\beta} \sum_{\omega}  \gamma_{\alpha \beta}(\omega) \nonumber \\
&&  \hspace{-1.0cm}  \times  \left(L_{\beta,\omega}(t) \rho(t) L_{\alpha, \omega}^{\dagger}(t) - \frac{1}{2} \left\{ L_{\alpha,\omega}^{\dagger}(t) L_{\beta,\omega}(t) , \rho(t) \right\} \right) \ ,
\end{eqnarray}
where $H_{\textrm {LS}}$ is the Lamb shift term induced by the interaction with the thermal bath, $\omega$ is a frequency, $\gamma_{\alpha \beta}(\omega)$ is a positive matrix for all values of $\omega$, and $L_{\alpha, \omega}(t)$ are time-dependent Lindblad operators.  They are given by
\bes
\begin{align}
L_{\alpha, \omega}(t) & =  \sum_{\omega} \delta_{\hbar \omega,\Delta_{ba}(t)} \bra{\eps_a(t)} A_\alpha \ket{\eps_b(t)}\ketbra{\eps_a(t)}{\eps_b(t)}  \ , \\
\gamma_{\alpha \beta}(\omega) & =  \int_{-\infty}^{\infty} dt \ e^{i \omega t} \langle e^{i H_B t} B_{\alpha} e^{-i H_B t} B_{\beta} \rangle \ ,  \\
H_{\textrm {LS}}(t) & = \frac{g^2}{\hbar} \sum_{\alpha \beta} \sum_{\omega} S_{\alpha \beta} (\omega) L_{\alpha,\omega}^{\dagger}(t) L_{\beta,\omega}(t) \ ,  \\
S_{\alpha \beta}(\omega) & = \int_{-\infty}^{\infty} d \omega' \gamma_{\alpha \beta}(\omega') \mathcal{P} \left( \frac{1}{\omega - \omega'} \right) \ , 
\end{align}
\ees
where $\mathcal{P}$ is the Cauchy principal value, $\Delta_{ba}(t) \equiv \eps_b(t) - \eps_a(t)$, and the states $\ket{\eps_a(t)}$ are the instantaneous energy eigenstates of $H_S(t)$ with eigenvalues $\eps_a(t)$ satisfying
\beq
H_S(t) \ket{\eps_a(t)} = \eps_a(t) \ket{\eps_a(t)}\, ,
\eeq
For our simulations we considered independent dephasing harmonic oscillator baths (i.e., each qubit experiences its own thermal bath) such that:
\beq
\sum_{\alpha} A_{\alpha} \otimes B_{\alpha} = \sum_{\alpha} \sigma^z_{\alpha} \otimes \sum_k \left(b_{k,\alpha} + b_{k,\alpha}^{\dagger} \right)\, ,
\eeq
where $b_{k,\alpha}$ and $b_{k,\alpha}^{\dagger}$ are, respectively, lowering and raising operators for the $k$th oscillator of the bath associated with qubit $\alpha$ satisfying $\left[ b_{k,\alpha},b_{k',\alpha}^\dagger \right] = \delta_{k,k'}$.  Furthermore, we assume an Ohmic spectrum for each bath such that
\beq
g^2 \gamma_{\alpha \beta}(\omega) = \delta_{\alpha, \beta} \frac{2 \pi g^2 \eta \omega}{1 - e^{-\upbeta  \hbar \omega}} e^{-\omega/\omega_c} \ ,
\eeq
where $\upbeta$ is the inverse temperature, $\eta$ (with units of time squared) characterizes the Ohmic bath, and $\omega_c$ is a UV cut-off.  In our simulations, we fix $\omega_c = 8 \pi\,$GHz in order to satisfy the approximations made in deriving the master equation (see Ref.~\cite{ABLZ:12} for more details), and we fix $\upbeta^{-1}/\hbar \approx 2.2\,$GHz to match the operating temperature of $17\,$mK of the D-Wave device.  The only remaining free parameter is the effective system-bath coupling 
\beq
\kappa \equiv g^2 \eta / \hbar^2,
\eeq 
which we vary to find the best agreement with our experimental data. 

\begin{figure}[t]
\begin{center}
\includegraphics[width=.48\textwidth]{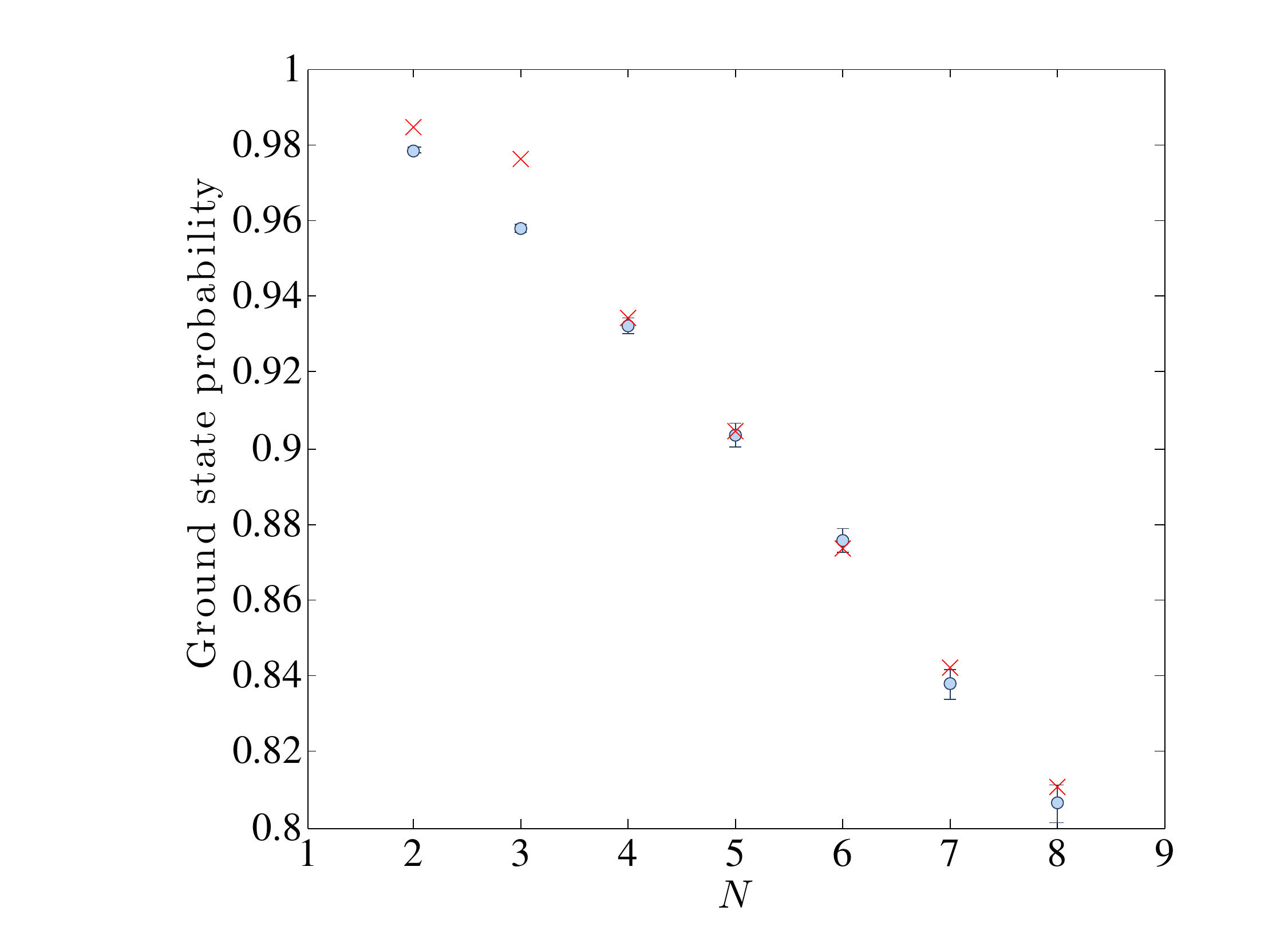}
\end{center}
\caption{{Experimental (circles) and simulation (crosses) results for an unencoded AF chain with $\alpha = 0.3$ and annealing time $t_f = 20\, \mu s$.  The effective system-bath coupling is $\kappa = 5.57 \times 10^{-6}$}.}
\label{fig:ME-vs-chip}
\end{figure}

As an example of the ability of our adiabatic master equation to capture the experimental data, we show in Fig.~\ref{fig:ME-vs-chip} the results of calculations for unprotected antiferromagnetic (AF) chains, along with the corresponding DW2 data. As can be seen the master equation correctly captures the initial quadratic decay as a function of chain length, which is missed by the classical kinks model presented in Sec.~\ref{app:indep-err}. Additional confirmation of the validity of the adiabatic master equation approach is given in Fig.~\ref{fig:6}(a), and earlier work \cite{q-sig,ALMZ:12}.

\section{The role of the penalty qubit}
\label{app:penalty-role}
\begin{figure*}[t]
\begin{center}
\subfigure[\, ]{\includegraphics[width=0.5\textwidth]{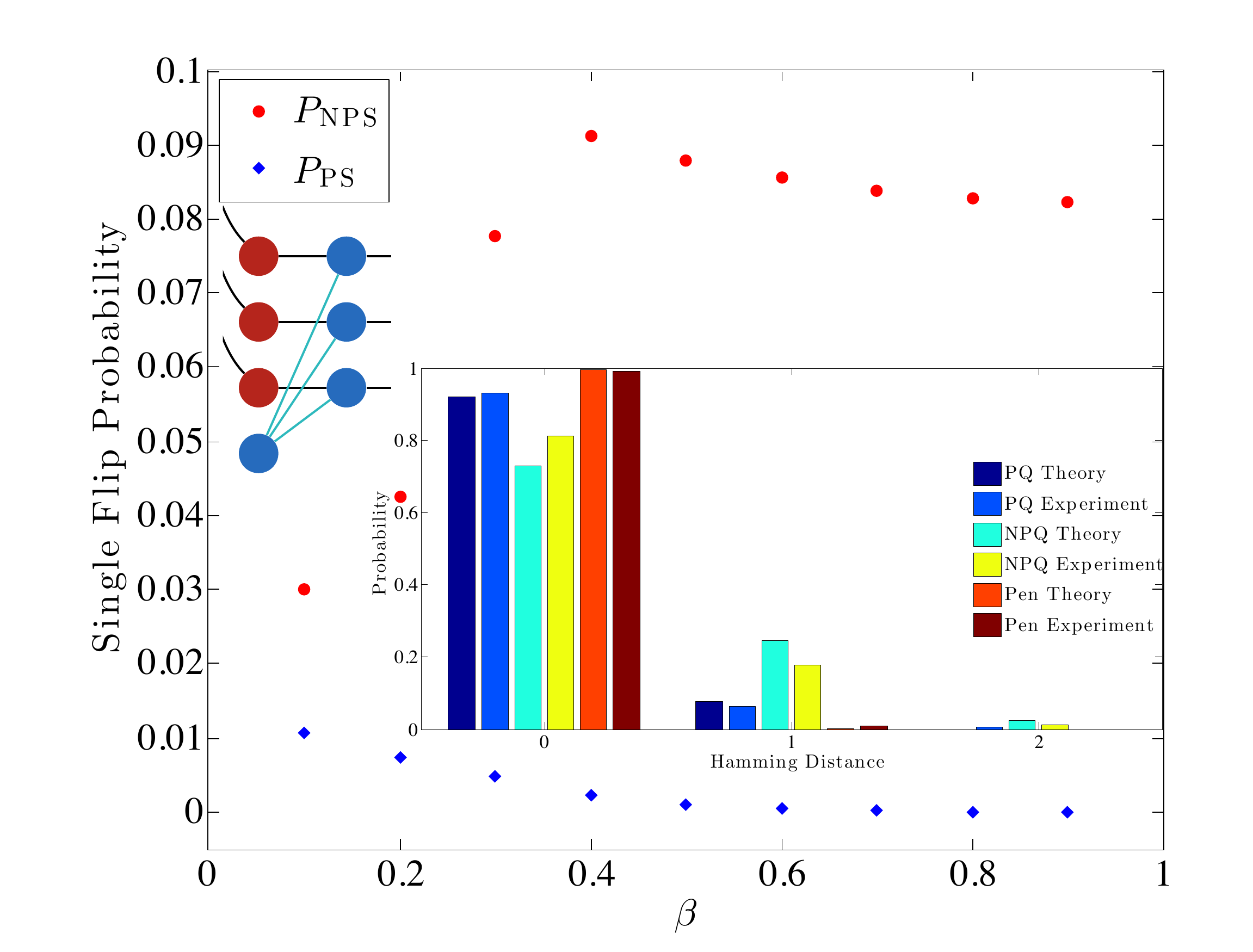}\label{fig:6a}}
\subfigure[\, ]{\includegraphics[width=0.48\textwidth]{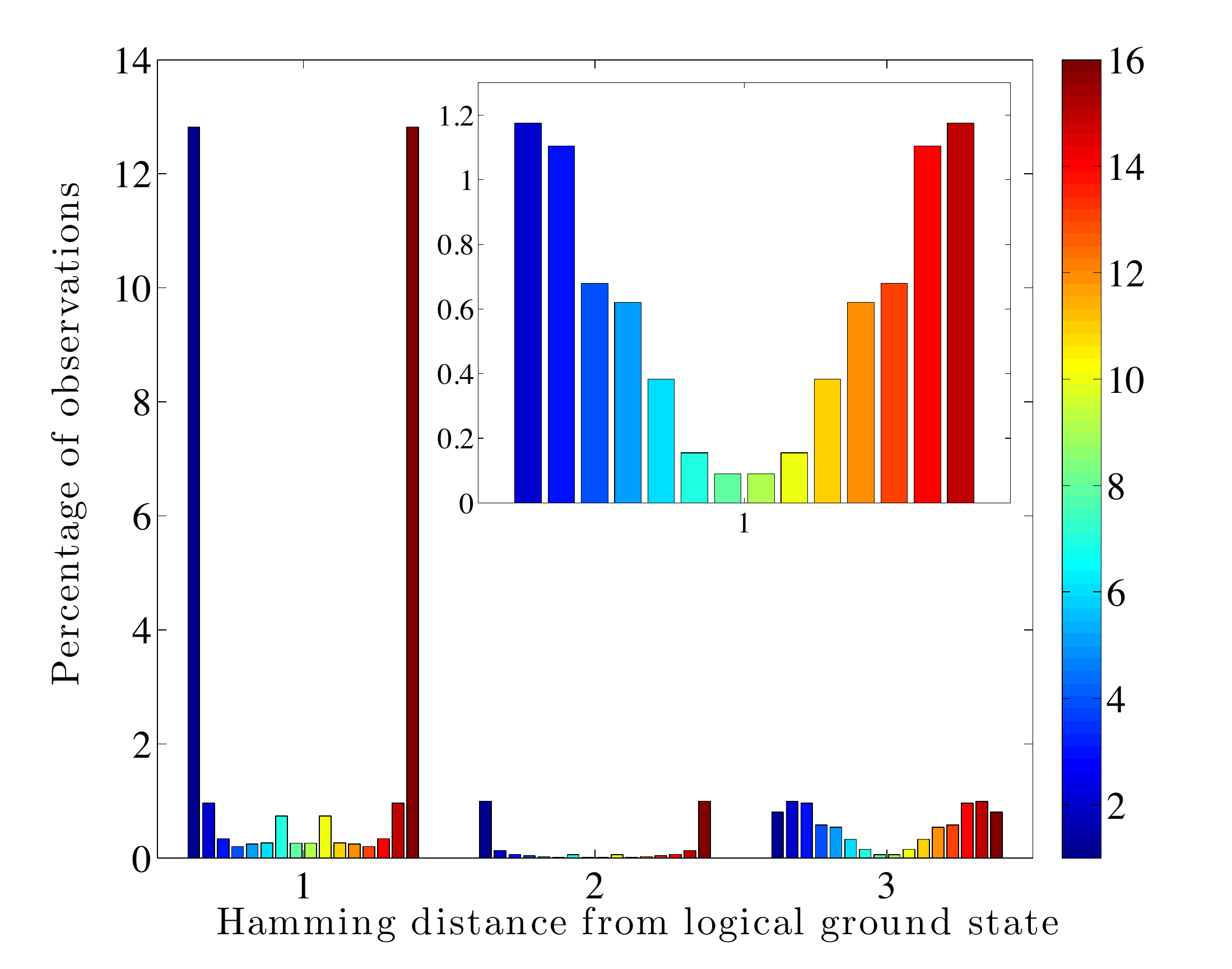}\label{fig:6b}} 
\end{center}
\caption{\small Effect of the penalty qubits. (a) Shown are adiabatic master equation simulation results {(with parameters $\kappa = 3.18 \times 10^{-4}$ and $t_f = 20\, \mu s$)} as a function of $\beta$, with $\alpha = 0.3$, for two antiferromagnetically coupled encoded qubits, one with a penalty qubit and one without (see diagram in upper left corner). Blue diamonds are the probability of a single flip in the problem qubits coupled to the penalty qubit (penalty side -- PS), red circles for the uncoupled case (no-penalty side -- NPS). 
Inset: Simulated and experimental probability of Hamming distance $d=1,2$ for the encoded qubit with (PQ) or without (NPQ)  the penalty qubit; ``Pen" denotes the penalty qubit (the probability at $d=3$ is negligible and is not shown). Good agreement is seen between the master equation and the experimental results. The NPQ case has substantially higher error probability, demonstrating the positive effect of the penalty qubit. (b) Similar to Fig.~\ref{fig:4a}, for an $\bar{N}=16$ qubit chain ($20,000$ samples collected on the DW1). The logical qubits numbered $1$, $7$, and $16$ did not have a penalty qubit, which is manifested by large peaks at these positions, at both $d=1$ and $d=2$.  The inset shows the percentage of errors on the penalty qubits (with numbers $1,7,16$ missing). The $d=3$ data illustrates that a flipped penalty qubit increases the probability of all problem qubits flipping even more than an absent penalty qubit.
}
\label{fig:6}
\end{figure*}

In this section we demonstrate that the penalty qubits are indeed responsible for error suppression. To do so we tested both experimentally and numerically the effect of removing one of the penalty qubits. Figure~\ref{fig:6}(a) clearly shows, using a simulation of two AF coupled logical qubits, that the physical problem qubits that are coupled to the penalty qubit have a smaller probability of flipping than those that are not. Figure~\ref{fig:6}(b) shows experimental data for an $\bar{N}=16$ chain with three missing penalty qubits; Hamming distance $1$ errors at the locations of the missing penalty qubits are greatly enhanced, reversing the dominance of $d=3$ errors seen in Fig.~\ref{fig:4b} in the main text.

\begin{figure}[ht]
\begin{center}
\subfigure[\, One penalty qubit]{\includegraphics[width=3.0in]{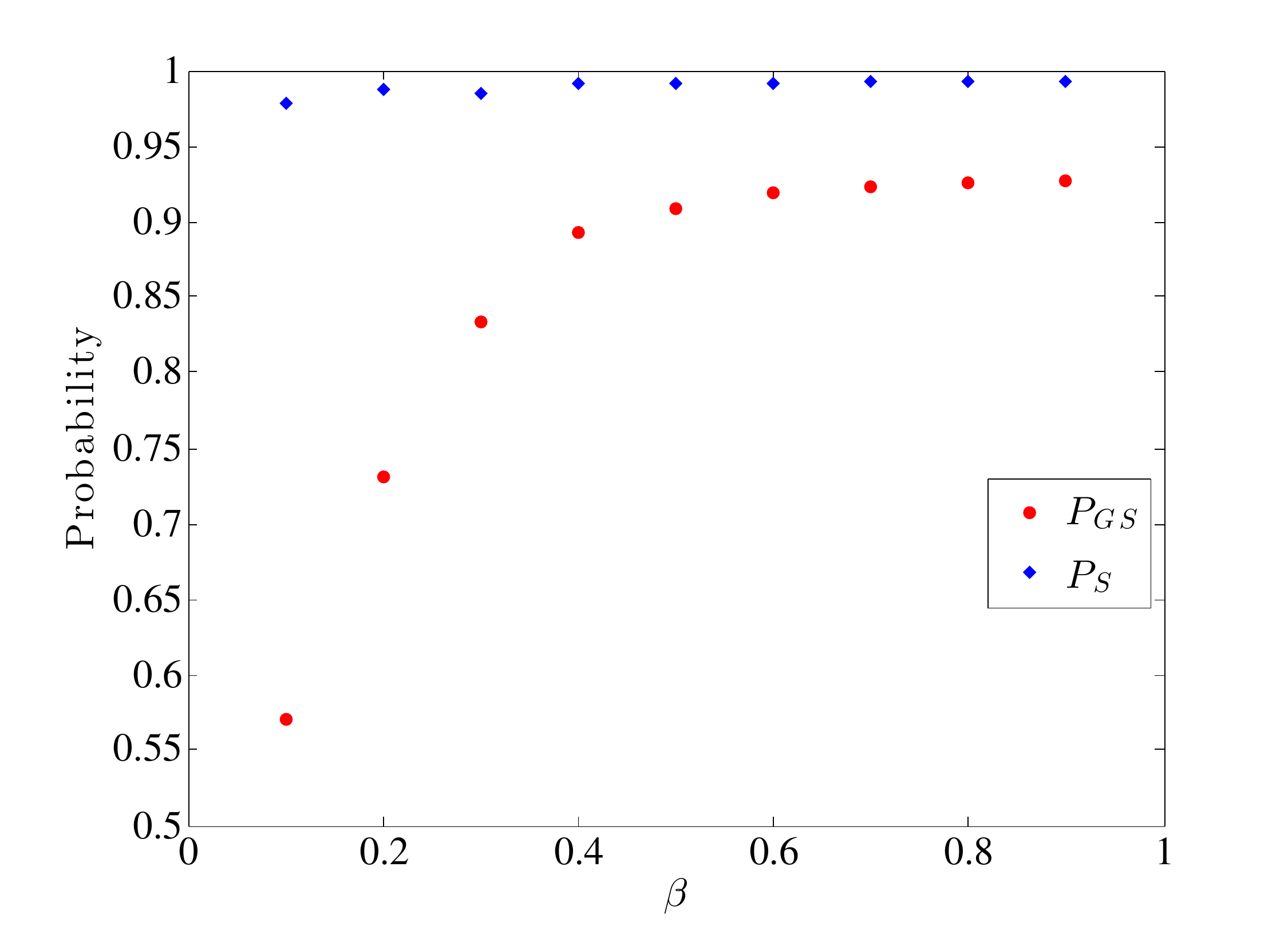} \label{fig:PGS_2encoded}} \hspace{0.5cm}
\subfigure[\, Two penalty qubits]{\includegraphics[width=3.0in]{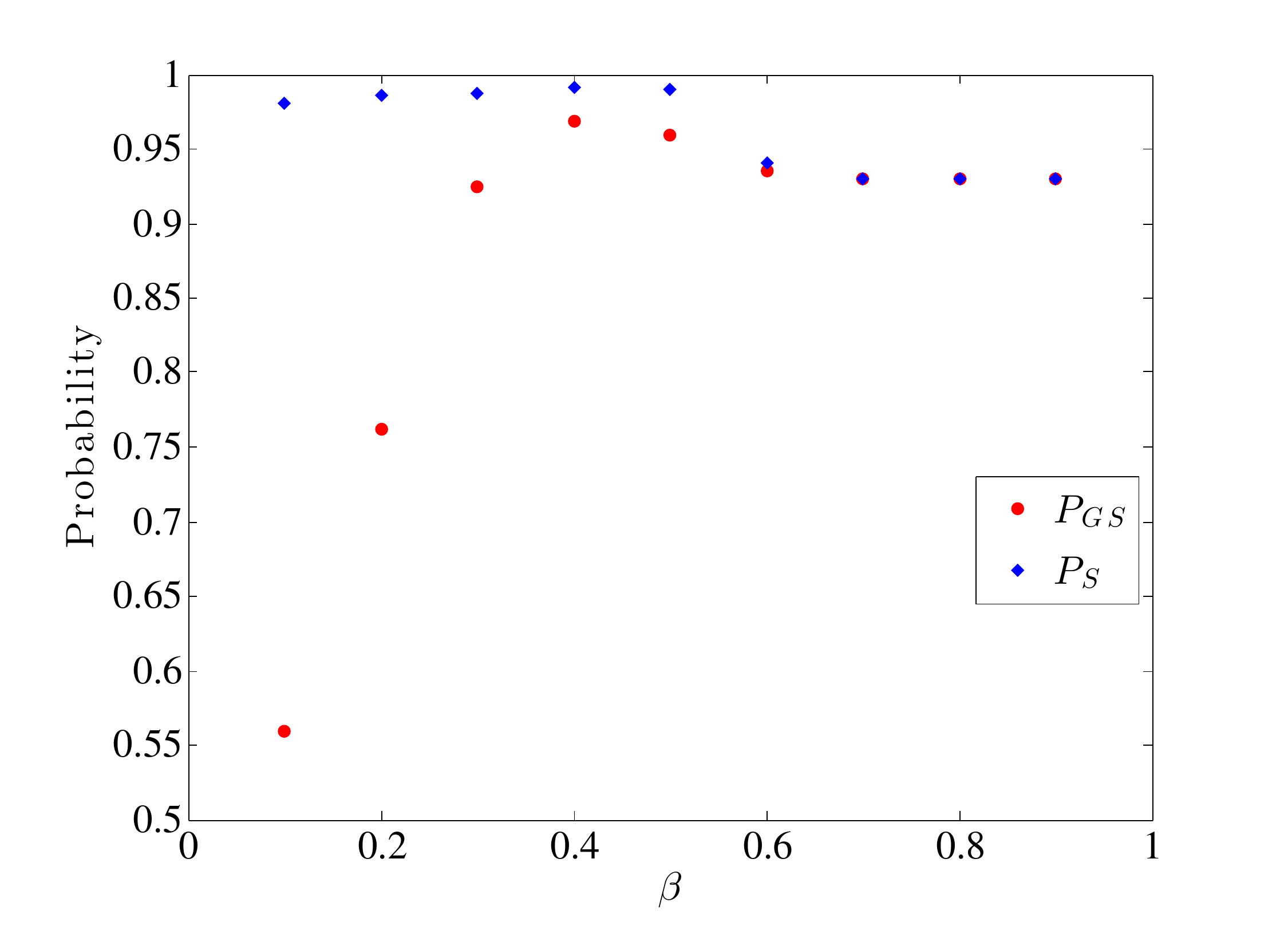} \label{fig:PS_2encoded}}   
\end{center}
\caption{\small Adiabatic master equation simulation results for the probability of finding the undecoded ground state $P_{\mathrm{GS}}$ and the decoded ground state $P_S$ for two logical qubits.  We set $\alpha=0.3$, $\kappa = 3.18 \times 10^{-4}$, and $t_f = 20 \mu \mathrm{s}$.}
\label{fig:2encoded}
\end{figure}

Let us now consider in some more detail the interplay between the optimal $\beta$ effect and the consequences of an absent penalty qubit. 
We saw the effects of changing the penalty strength in Figs.~\ref{fig:3a} and \ref{fig:3b} in the main text, in terms of a model of two coupled logical qubits. For a given problem, there exists an optimal $\beta$ at which the success probability is maximized. In Fig.~\ref{fig:6}(a) we saw that the absence of a penalty qubit increases the bit flip probability of the associated problem qubits. 
Fig.~\ref{fig:2encoded} shows the results of simulations using the adiabatic master equation of Sec.~\ref{app:ME}, for the probability of finding the ground state of the physical system versus finding the correct ground state after decoding.  For small $\beta$, the two penalty versus one penalty qubit instances exhibit similar behavior, with a ground state probability $P_\textrm{GS}$ that grows with increasing $\beta$, whereas the decoded success probability remains more or less constant.  This initial growth in $P_\textrm{GS}$ can be easily understood in terms of the behavior of the energy gap to the second excited state (recall that the final ground state is degenerate so any population in the first excited state joins the ground state by the end of the evolution).  As we showed in the main text, this energy gap grows and shifts to earlier times in the evolution.  This reduces the thermal excitation rate out of the ground state. The low lying excited states are decodable to the correct ground state, and therefore, any population excited to them is still recovered by decoding, which explains the constant decoded success probability.

As $\beta$ is further increased, we observe that the two-penalty problem exhibits a peak in $P_\textrm{GS}$ as a function of $\beta$, although no such peak is observed in $P_S$.  This means that there is significant probability loss to states that are still decoded to the correct ground state.  However, we notice that at $\beta =0.6$, there is a large drop in $P_S$ without an accompanying large drop in $P_{GS}$.  This means that there is a large probability lost to a state that is decoded incorrectly.  This indeed happens because for $\beta \geq 0.6$, the first excited state is actually the fully ferromagnetic state, which decodes to the incorrect ground state.  Notice however that this large drop does not happen for the single penalty case, because this important identity shift in the excited state spectrum does not happen over the range of $\beta$ explored.

\section{The role of the problem scale $\alpha$ and the penalty scale $\beta$}
\label{app:ab-role}
%
In order to understand the role of the problem scale $\alpha$ and the penalty strength $\beta$ on the quantum annealing evolution, we analyze the effect of each separately in terms of analytically tractable models.  

\subsection{The role of $\alpha$}
\begin{figure}[t]
\begin{center}
\includegraphics[width=0.48\textwidth]{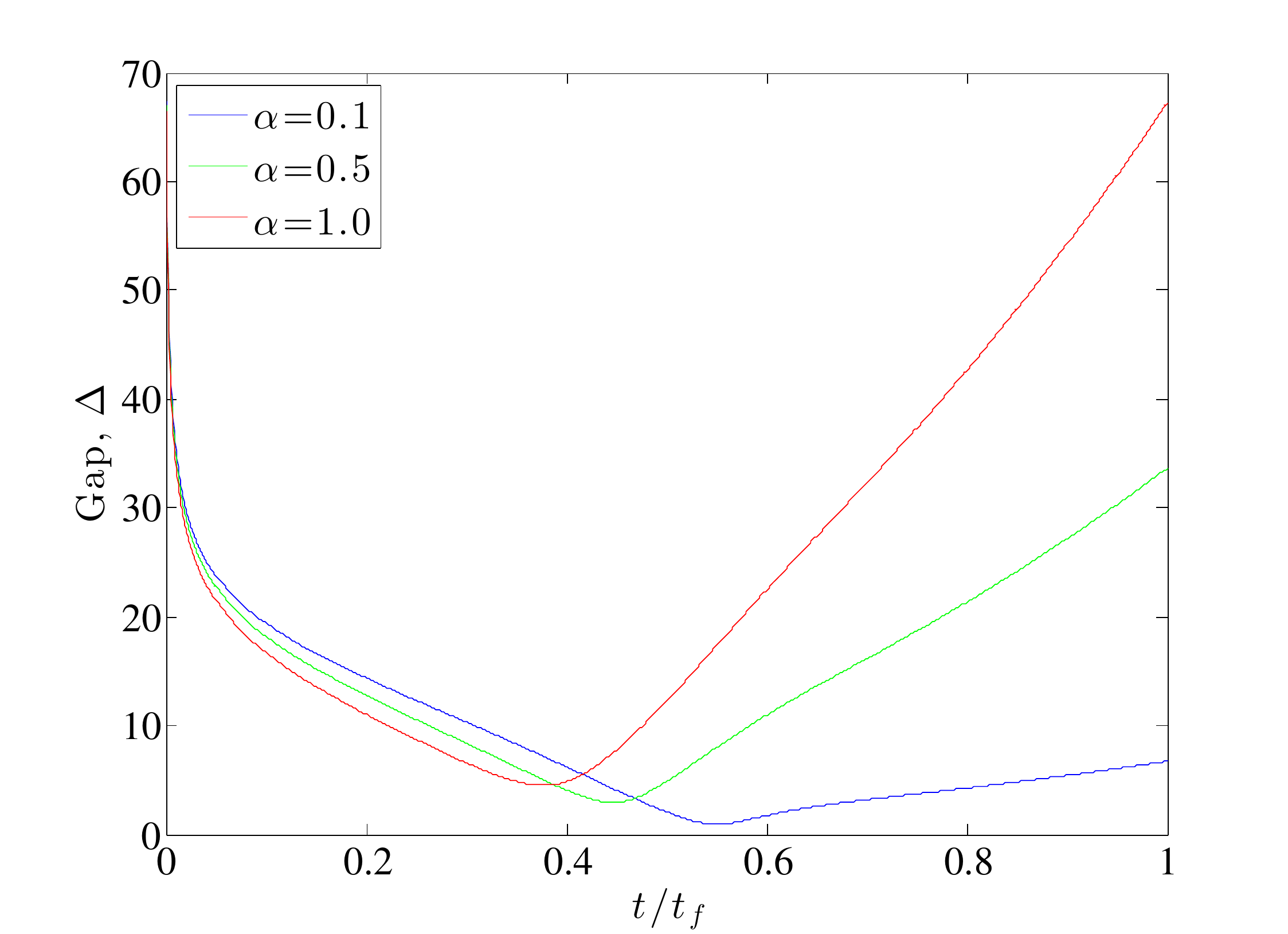} 
\end{center}
\caption{The gap to the lowest relevant (2nd and higher) excited state for 
an antiferromagnetically coupled $8$-qubit chain for $\beta=0$ and different values of $\alpha$. The gap increases and moves to the left as $\alpha$ is increased (see Fig.~\ref{fig:3a} in the main text for a similar plot with $\alpha$ fixed and varying $\beta$).}
\label{fig:5a-app}
\end{figure}

First, let us consider the case of $\beta = 0$, and consider a one dimensional antiferromagnetic transverse Ising model:
\beq
H = \alpha \left( \frac{h}{\alpha} \sum_{i=1}^N \sigma_i^x +  \sum_{i=1}^N \sigma_i^z \sigma_{i+1}^z \right) \ .
\eeq
It is well established that in the large $N$ limit, the system undergoes a quantum phase transition at $h_c = J$ \cite{Lieb1961407,Pfeuty197079}.  Let us denote the minimum gap for this case as $\Delta_{\mathrm{min}} = \alpha \Delta_0$.  Now let us consider the time-dependent Hamiltonian:
\beq
H(t) = A(t) \sum_i \sigma_i^x + \alpha B(t) \sum_i \sigma_i^z \sigma_{i+1}^z \ ,
\eeq
which is the case of interest in the main text.  The minimum gap would then occur at $A(t_{\mathrm{min}}) = \alpha B(t_{\mathrm{min}})$.  For monotonically decreasing $A(t)$ and monotonically increasing $B(t)$, increasing $\alpha$ means $t_{\mathrm{min}}$ decreases.  For example, consider linear interpolating functions:
\beq
A_L(s) = 2A_0 (1 - s) \ , \quad B_L(s) = 2A_0 s \ ,
\eeq
where $s = t/t_f$.  The minimum gap occurs at $s_{\mathrm{min}} = 1/ \left(1+ \alpha \right)$, such that the minimum gap is given by $\Delta_{\mathrm{min}} = 2A_0 \Delta_0 \alpha / \left( 1 + \alpha \right)$. Therefore, increasing $\alpha$ decreases $s_{\mathrm{min}}$, i.e., shifts the minimum gap to earlier in the evolution, and increases $\Delta_{\mathrm{min}}$. Figure~\ref{fig:5a-app} shows the results of numerical simulations for an an antiferromagnetically coupled chain, verifying both the increase in the gap and its shift to the left.

\subsection{The role of $\beta$}
In order to analytically study the effect of $\beta$ we resort to a much simpler model and perturbation theory.  Let us consider the annealing Hamiltonian for a single qubit with linear interpolating functions:
\beq \label{eqt:qubitH}
H(s) = A_0 (1-s)  \sigma^x + A_0 s \frac{1}{2} \omega \sigma^z \ .
\eeq
This has time-dependent eigenvalues given by
\beq 
A_0 \eps_{\pm}(s) = \pm \frac{A_0 }{2} \left( 4(1-s)^2 + s^2  \omega^2  \right)^{1/2} \equiv \pm \frac{A_0 }{2} \lambda\ ,
\eeq
with eigenvectors (in the computational basis):
\bes
\begin{align}
| \eps_- (s) \rangle = & \frac{1}{c_-(s)} \left( \frac{s \omega}{2 (1-s)} - \frac{\lambda}{2 (1-s)}, 1 \right) \ , \\
| \eps_+ (s) \rangle = & \frac{1}{c_+(s)} \left( \frac{s \omega}{2} + \frac{\lambda}{2 } , 1-s \right) \ ,
\end{align}
\ees
where $c_\pm(s)$ are positive normalization coefficients.  Let us now consider three decoupled qubits, representing the problem qubits.  The ground state of this three qubit system is simply a tensor product of the individual ground states:
\beq
| 0 \rangle = | \eps_- \rangle_1 \otimes | \eps_- \rangle_2 \otimes | \eps_- \rangle_3 \ ,
\eeq
with energy $\eps_0(s) = 3 \eps_-(s) $.  The first excited state is triply degenerate:
\bes
\begin{align}
| 1 \rangle = & | \eps_+ \rangle_1 \otimes | \eps_- \rangle_2 \otimes | \eps_- \rangle_3 \ , \\
| 2 \rangle = & | \eps_- \rangle_1 \otimes | \eps_+ \rangle_2 \otimes | \eps_- \rangle_3 \ , \\
| 3 \rangle = & | \eps_- \rangle_1 \otimes | \eps_- \rangle_2 \otimes | \eps_+ \rangle_3 \ ,
\end{align}
\ees
with energy $\eps_1(s) = 2 \eps_-(s) +  \eps_+(s) $. 

\subsubsection{A single logical qubit}

\begin{figure}[ht] 
   \centering
   \includegraphics[width=0.5\textwidth]{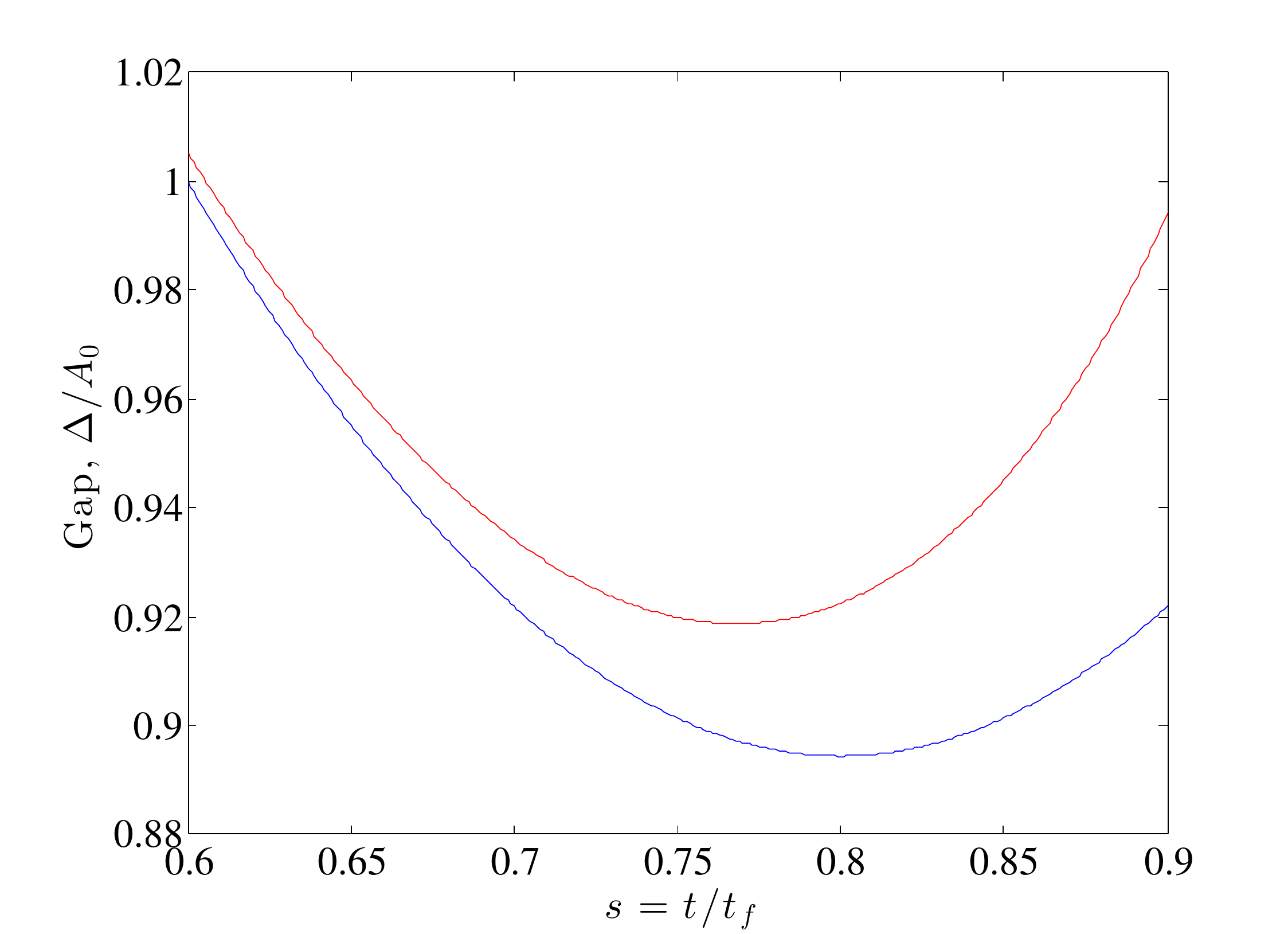} 
   \caption{The gap to the first relevant excited state for the unperturbed (blue curve) Hamiltonian and the perturbed (red curve) Hamiltonian with $\omega = 1$ and $\beta = \omega_0 = 0.1$.}
   \label{fig:SingleQubitBeta}
\end{figure}

To model a logical qubit which includes three problems qubits and a penalty qubit we now introduce a fourth qubit, with a Hamiltonian similar to Eq.~\eqref{eqt:qubitH} but with energy $\omega_0 \ll \omega$, i.e., 
\beq
H(s) = A_0 (1-s)  \sum_{i=1}^4 \sigma^x_i + A_0 s \frac{1}{2} (\sum_{i=1}^3\omega \sigma^z_i + \omega_0 \sigma^z_4) \ .
\eeq
This is chosen to ensure that the ground state and the degenerate first excited states all have the fourth qubit in its ground state $\ket{\epsilon_-}_4$.
Thus the ground state is
\beq
\ket{0} =  | \eps_- \rangle_1 \otimes | \eps_- \rangle_2 \otimes | \eps_- \rangle_3 \otimes | \epsilon_- \rangle_4 \ ,
\eeq 
and the excited states corresponding to the single bit-flip errors we are interested in are:
\bes
\begin{align}
| 1 \rangle = & | \eps_+ \rangle_1 \otimes | \eps_- \rangle_2 \otimes | \eps_- \rangle_3 \otimes | \epsilon_- \rangle_4 \ , \\
| 2 \rangle = & | \eps_- \rangle_1 \otimes | \eps_+ \rangle_2 \otimes | \eps_- \rangle_3 \otimes | \epsilon_- \rangle_4\ , \\
| 3 \rangle = & | \eps_- \rangle_1 \otimes | \eps_- \rangle_2 \otimes | \eps_+ \rangle_3\otimes | \epsilon_- \rangle_4 \ .
\end{align}
\ees
Next we introduce the ferromagnetic penalty term of the main text as a perturbation:
\beq
H_P = - A_0 s \beta \sum_{i=1}^3  \sigma^z_i \sigma^z_4 \ ,
\eeq
We can calculate the first order perturbation to the energy states, and from there obtain the gap.  This amounts to simply calculating the matrix element $\langle a | \sigma_i^z \sigma_4^z | a \rangle$, where $a = 0,1,2,3$. 
The $\sigma_4^z$ always gives a multiplicative contribution of $\langle \epsilon_- | \sigma_4^z | \epsilon_- \rangle$.  The remaining matrix elements are given by:
\bes
\begin{align}
\langle 0 | \sigma_i^z | 0 \rangle = & \frac{c_-^2 - 2}{c_-^2} \  , \quad i = 1,2,3  \ , \\
 \langle 1 | \sigma_2^z | 1 \rangle = & \langle 3 | \sigma_2^z | 3 \rangle = \langle 1 | \sigma_3^z | 1 \rangle = \langle 2 | \sigma_3^z | 2 \rangle =  \frac{c_-^2 - 2}{c_-^2}  \ ,\\
\langle 1 | \sigma_1^z | 1 \rangle = & \langle 2 |\sigma_2^z | 2 \rangle = \langle 3 | \sigma_3^z | 3 \rangle = \frac{c_+^2 - 2 (1-s)^2}{c_+^2} \ .
\end{align}
\ees
The remaining matrix elements vanish.  
The degeneracy of the excited states is not broken by this perturbation, and the perturbed energy spectrum is given by:
\bes
\begin{align}
E_{0} = &\  3 A_0 \eps_-(s) - A_0 \beta s \left( 3 \frac{2 - c_-^2 }{c_-^2} \frac{2 - \tilde{c}_-^2 }{\tilde{c}_-^2} \right)  \ , \\ 
E_{1} = & \  2 A_0 \eps_-(s) +A_0 \eps_+(s) \nonumber \\
& + A_0 \beta s \left( -  2 \frac{2 - c_-^2}{c_-^2} + \frac{c_+^2 - 2 (1-s)^2}{c_+^2} \right) \left( \frac{2 - \tilde{c}_-^2 }{\tilde{c}_-^2 }\right) \ .
\end{align}
\ees
where $\tilde{c}_-$ is the normalization for the ground state of the unperturbed fourth qubit.
Therefore the gap is
\begin{eqnarray}
\Delta &=& E_{1}- E_{0}  = A_0 \left[ \eps_+(s) - \eps_-(s)  \right. \\
&& \left. + \beta s \left( \frac{c_+^2 - 2 (1-s)^2}{c_+^2} + \frac{2 - c_-^2}{c_-^2} \right) \left( \frac{2 - \tilde{c}_-^2 }{\tilde{c}_-^2 }\right) \right]\, . \nonumber
\end{eqnarray}
The contribution to the gap from the perturbation is positive throughout the evolution for finite values of $\omega_0$.  Furthermore, as shown in Fig.~\ref{fig:SingleQubitBeta}, the minimum gap shifts to earlier in the evolution.  

\begin{figure}[ht] %
   \centering
   \includegraphics[width=0.5\textwidth]{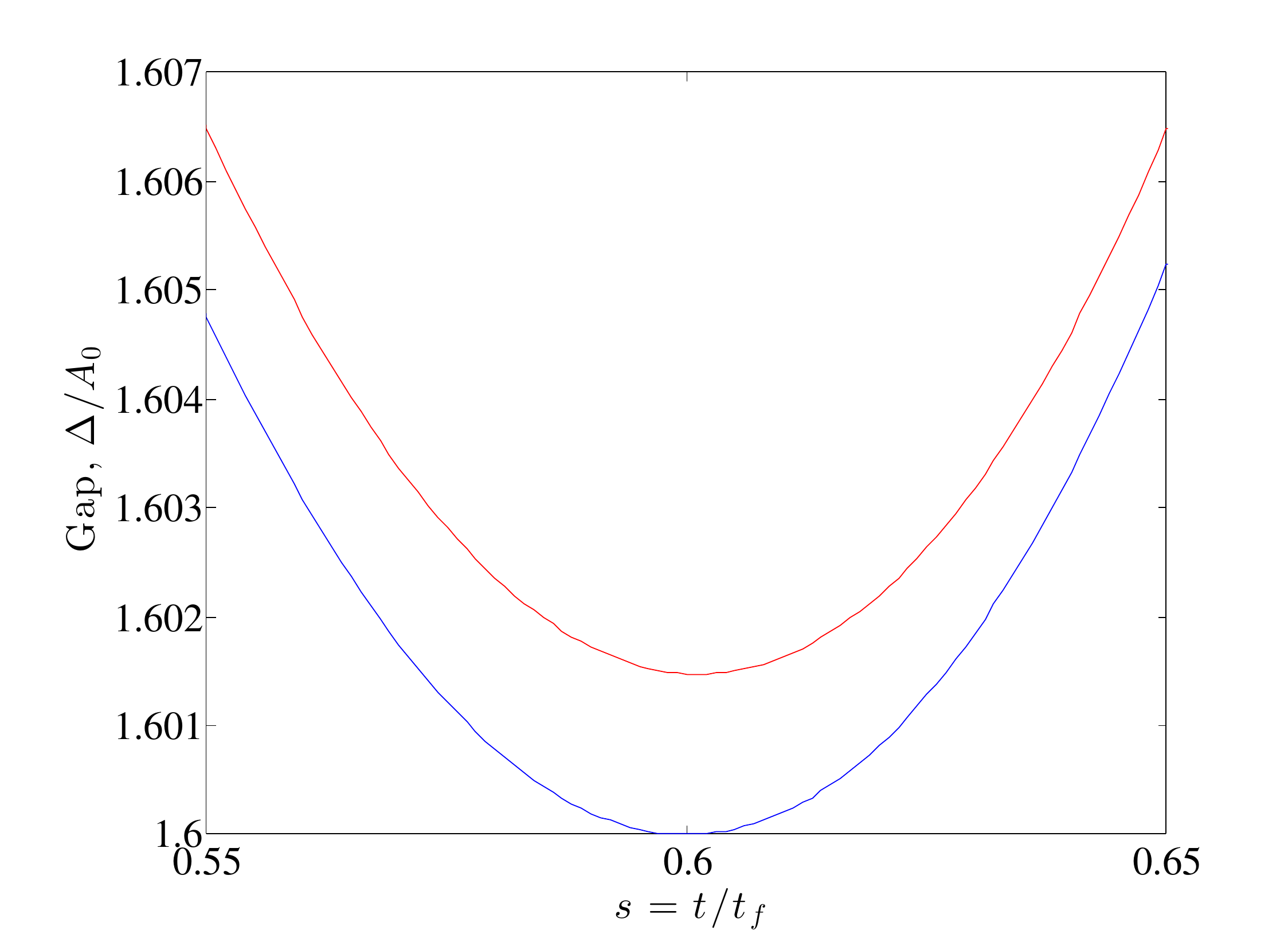} 
   \caption{The gap to the first relevant excited state for the unperturbed (blue curve) Hamiltonian and the perturbed (red curve) Hamiltonian with $\beta = 0.1$ and $\omega_0 = 0.001$.}
   \label{fig:2QubitBeta}
\end{figure}

\subsubsection{Three qubit pairs coupled to a penalty qubit}

As a proxy for the case of coupled logical qubits, which is difficult to analyze analytically, we consider three pairs of AF-coupled qubits, {with one triple coupled to a penalty qubit}. We start from a simple two-site chain, where for simplicity we set the AF-coupling to unity:
\beq \label{eqt:AF_H}
H(t) = A_0(1-s) \left( \sigma_1^x + \sigma_2^x \right) + A_0 s \sigma_1^z \sigma_2^z \ .
\eeq
The energy eigenvalues are given by:
\beq
\frac{\eps_0}{A_0}  = - \lambda  \ , \quad \frac{\eps_1}{A_0} = -s \ , \quad \frac{\eps_2}{A_0} = s \ , \quad  \frac{\eps_3}{A_0} = \lambda \ ,
\eeq
where $\lambda = \sqrt{4(1-s)^2 + s^2 }$, and the eigenvectors (in the computational basis $\uparrow \uparrow, \uparrow \downarrow, \downarrow \uparrow, \downarrow \downarrow$) are:
\bes
\begin{align}
\ket{\varepsilon_0(s) } =  & \frac{1}{c_-(s)} \left(1 - s, -\frac{s}{2} - \frac{\lambda}{2} , -\frac{s}{2} - \frac{\lambda}{2} , 1-s  \right) \\
\ket{\varepsilon_1(s) } = & \frac{1}{\sqrt{2}} \left( 0 , -1 , 1 , 0 \right) \\
\ket{\varepsilon_2(s) } = & \frac{1}{\sqrt{2}} \left( -1 , 0 , 0 , 1 \right) \\
\ket{\varepsilon_3(s) } = &  \frac{1}{c_+(s)} \left(1, -\frac{s}{2} + \frac{\lambda}{2}  , -\frac{s}{2} + \frac{\lambda}{2}, 1  \right) \ ,
\end{align}
\ees
where $c_\pm(s)$ are normalization functions.  Note that the Hamiltonian in Eq.~\eqref{eqt:AF_H} is invariant under $\sigma_i^z \to - \sigma_i^z$, so that the expectation value of $\sigma_i^z$ under any of the energy eigenstates is zero,
\beq \label{eqt:Invariance}
\bra{\varepsilon_i} \sigma_j^z \ket{\varepsilon_i} = 0 \ , \quad i = 0,1,2,3 \ , \ j = 1,2 \ .
\eeq
Furthermore, the only non-zero matrix elements of the sum of $\sigma_i^z$ in the instantaneous energy eigenbasis are:
\begin{eqnarray}
\bra{\varepsilon_0} \sigma_1^z  \ket{\varepsilon_2} &=& \frac{-\sqrt{2}(1-s)}{ c_-} \ , \quad \bra{\varepsilon_3} \sigma_1^z  \ket{\varepsilon_2} = \frac{-\sqrt{2}}{ c_+} \nonumber \\
\bra{\varepsilon_0} \sigma_1^z  \ket{\varepsilon_1} &=& \frac{s+\lambda}{ \sqrt{2} c_-} \ .
\end{eqnarray}
We now proceed as in the single qubit case: we make three copies of the AF chain and introduce a fourth qubit that does not interact with the three AF chains.  The ground state of this seven qubit system can be written as follows:
\beq
\ket{0} = \ket{\varepsilon_0} \otimes \ket{\varepsilon_0} \otimes \ket{\varepsilon_0} \otimes \ket{\epsilon_-} \ ,
\eeq
with energy
\beq
E_0 = 3 \varepsilon_0 + \epsilon_- \ . 
\eeq
The relevant excited state is the lowest excited state that does not decode to the correct ground state.  This is:
\beq
\ket{2} = \ket{\varepsilon_2} \otimes \ket{\varepsilon_0} \otimes \ket{\varepsilon_0} \otimes \ket{\epsilon_-} \ ,
\eeq
with a three-fold degeneracy because the state $\ket{\varepsilon_2}$ can be placed on any of the three chains, with energy
\beq
E_2 = 2 \varepsilon_0 +\varepsilon_2 + \epsilon_- \ . 
\eeq
We now again introduce the penalty as a perturbation:
\beq
H_P = - A_0 s \beta \sum_{i=1}^3  \sigma^z_{1_i}  \sigma^z_4 \ ,
\eeq
where $\sigma^z_{1_i}$ is the Pauli operator acting on the $i$th qubit of the first logical qubit of the AF chain.  From Eq.~\eqref{eqt:Invariance}, we obtain that the energy change vanishes to first order in perturbation theory, so we have to go to second order.  The change in the ground state energy is given by:
\onecolumngrid
\begin{eqnarray}
\frac{\delta E_0}{\beta^2 A_0^2 s^2} &=& 3 \frac{ | \bra{\varepsilon_0} \sigma_1^z \ket{\varepsilon_2}|^2}{\varepsilon_0 - \varepsilon_2}  | \bra{\epsilon_-} \sigma_4^z  \ket{\epsilon_-}|^2 
+3  \frac{ | \bra{\varepsilon_0} \sigma_1^z \ket{\varepsilon_2}|^2}{\varepsilon_0 - \varepsilon_2 + \epsilon_- - \epsilon_+}  | \bra{\epsilon_-} \sigma_4^z  \ket{\epsilon_+}|^2 
+ 3 \frac{ | \bra{\varepsilon_0} \sigma_1^z \ket{\varepsilon_1}|^2}{\varepsilon_0 - \varepsilon_1}  | \bra{\epsilon_-} \sigma_4^z  \ket{\epsilon_-}|^2 \nonumber \\ && 
+ 3  \frac{ | \bra{\varepsilon_0} \sigma_1^z \ket{\varepsilon_1}|^2}{\varepsilon_0 - \varepsilon_1 + \epsilon_- - \epsilon_+}  | \bra{\epsilon_-} \sigma_4^z  \ket{\epsilon_+}|^2 \ ,
\end{eqnarray}
while the change in the relevant excited state energy is:
\begin{eqnarray}
\frac{\delta E_2}{\beta^2 A_0^2 s^2} &=&  \frac{ | \bra{\varepsilon_2} \sigma_1^z \ket{\varepsilon_0}|^2}{\varepsilon_2 - \varepsilon_0}  | \bra{\epsilon_-} \sigma_4^z  \ket{\epsilon_-}|^2 
+  \frac{ | \bra{\varepsilon_2} \sigma_1^z \ket{\varepsilon_0}|^2}{\varepsilon_2 - \varepsilon_0 + \epsilon_- - \epsilon_+}  | \bra{\epsilon_-} \sigma_4^z  \ket{\epsilon_+}|^2 
+ 2\frac{ | \bra{\varepsilon_0} \sigma_1^z \ket{\varepsilon_2}|^2}{\varepsilon_0 - \varepsilon_2}  | \bra{\epsilon_-} \sigma_4^z  \ket{\epsilon_-}|^2 \nonumber \\ 
&&
+  2\frac{ | \bra{\varepsilon_0} \sigma_1^z \ket{\varepsilon_2}|^2}{\varepsilon_0 - \varepsilon_2 + \epsilon_- - \epsilon_+}  | \bra{\epsilon_-} \sigma_4^z  \ket{\epsilon_+}|^2 
+ \frac{ | \bra{\varepsilon_2} \sigma_1^z \ket{\varepsilon_3}|^2}{\varepsilon_2 - \varepsilon_3}  | \bra{\epsilon_-} \sigma_4^z  \ket{\epsilon_-}|^2 
+  \frac{ | \bra{\varepsilon_2} \sigma_1^z \ket{\varepsilon_3}|^2}{\varepsilon_2 - \varepsilon_3 + \epsilon_- - \epsilon_+}  | \bra{\epsilon_-} \sigma_4^z  \ket{\epsilon_+}|^2 \nonumber \\ &&
+ 2\frac{ | \bra{\varepsilon_0} \sigma_1^z \ket{\varepsilon_1}|^2}{\varepsilon_0 - \varepsilon_1}  | \bra{\epsilon_-} \sigma_4^z  \ket{\epsilon_-}|^2 
+  2\frac{ | \bra{\varepsilon_0} \sigma_1^z \ket{\varepsilon_1}|^2}{\varepsilon_0 - \varepsilon_1+ \epsilon_- - \epsilon_+}  | \bra{\epsilon_-} \sigma_4^z  \ket{\epsilon_+}|^2  \ .
\end{eqnarray}
\twocolumngrid
The perturbed gap is given by:
\beq
\Delta = E_2 - E_0 + \delta E_2 - \delta E_0 \ ,
\eeq
which is larger than the unperturbed gap, and moves slightly to the right, as shown in Fig.~\ref{fig:2QubitBeta}. Thus this incomplete model of coupled logical qubits captures the increase in the gap due to the penalty term, but not its shift to earlier time. However, we do observe both the increase in the gap and its shift to the left when we numerically compute the gap for AF chains, as can be seen in Fig.~\ref{fig:3a} in the main text.

\clearpage


%

\end{document}